\documentclass[12pt,preprint]{aastex}
\usepackage{graphics}
\usepackage{subfigure}

\shorttitle{XLFs and ULXs}
\shortauthors{Wang et al.}

\begin{document}

\title{CHANDRA ACIS SURVEY OF X-RAY POINT SOURCES IN NEARBY GALAXIES. II. X-RAY LUMINOSITY FUNCTIONS AND ULTRALUMINOUS X-RAY SOURCES}

\author{Song Wang\altaffilmark{1}, Yanli Qiu\altaffilmark{1,3},
Jifeng Liu\altaffilmark{1,2}, and Joel N. Bregman\altaffilmark{4}}
\altaffiltext{1}{Key Laboratory of Optical Astronomy, National Astronomical Observatories,
Chinese Academy of Sciences, Beijing 100012, China;\\
jfliu@bao.ac.cn, songw@bao.ac.cn}
\altaffiltext{2}{College of Astronomy and Space Sciences,
University of Chinese Academy of Sciences, Beijing 100049, China}
\altaffiltext{3}{University of Chinese Academy of Sciences, Beijing 100049, China}
\altaffiltext{4}{University of Michigan, Ann Arbor, MI 48109}


\begin{abstract}

Based on the recently completed {\it Chandra}/ACIS survey of X-ray point sources in nearby galaxies, we study the X-ray luminosity functions (XLFs) for X-ray point sources in different types of galaxies and the statistical properties of ultraluminous X-ray sources (ULXs). Uniform procedures are developed to compute the detection threshold, to estimate the foreground/background contamination, and to calculate the XLFs for individual galaxies and groups of galaxies, resulting in an XLF library for 343 galaxies of different types. With the large number of surveyed galaxies, we have studied the XLFs and ULX properties across different host galaxy types, and confirm with good statistics that the XLF slope flattens from lenticular ($\alpha\sim1.50\pm0.07$) to elliptical ($\sim1.21\pm0.02$), to spirals ($\sim0.80\pm0.02$), to peculiars ($\sim0.55\pm0.30$), and to irregulars ($\sim0.26\pm0.10$). The XLF break dividing the neutron star and black hole binaries is also confirmed, albeit at quite different break luminosities for different types of galaxies. A radial dependency is found for ellipticals, with a flatter XLF slope for sources located between $D_{25}$ and 2$D_{25}$, suggesting the XLF slopes in the outer region of early-type galaxies are dominated by low-mass X-ray binaries in globular clusters. This study shows that the ULX rate in early-type galaxies is $0.24\pm0.05$ ULXs per surveyed galaxy, on a $5\sigma$ confidence level. The XLF for ULXs in late-type galaxies extends smoothly until it drops abruptly around $4\times10^{40}$ erg s$^{-1}$, and this break may suggest a mild boundary between the stellar black hole population possibly including 30 $M_\odot$ black holes with super-Eddington radiation and intermediate mass black holes.

\end{abstract}


\section{INTRODUCTION}

X-ray observations of more than forty years have revealed a variety of X-ray
point sources beyond the solar system in the Milky Way and Magellanic Clouds.
%
Although the bright X-ray sources in the Milky Way are easily studied, many of
them are not observable due to the heavy obscuration of the galactic disk, and
one must correct for completeness when studying the statistical properties of
Galactic X-ray binaries \citep[e.g.,][]{Grimm2002}.
Studies of X-ray binaries in more distant galaxies may be free from the
obscuration problem, and more importantly, provide us uniform samples of
X-ray binaries in different environments.
%
%
Much work has been done to study the X-ray source populations in distant
galaxies since the lunch of {\it Einstein} \citep{Fabbiano1989} and {\it ROSAT} \citep{Roberts2000},
and a quantum leap in this field has been achieved with the {\it Chandra} mission
\citep[see][for a review]{FW2006, Fabbiano2006}.

New classes of X-ray sources not previously seen in the Milky Way have emerged
from studies of distant galaxies, such as the ultraluminous X-ray sources
(ULXs), which were first detected by {\it Einstein} \citep{Long1983, Fabbiano1989}.
ULXs are defined as extranuclear sources with an observed luminosity of at least 10$^{39}$ erg/s,
in excess of the Eddington limit of a neutron star.
A large amount of both observational and theoretical work has focused on the nature of ULXs,
including the high luminosity, the X-ray spectra and short time variability, the association with
active star-forming regions, and the formation and evolution of ULXs \citep[see][for a review]{Fabbiano2005, Fabbiano2006}.
Recently, it was generally believed that ULXs in nature appear in two categories:
the long-sought intermediate mass black holes (IMBHs) as the seeds of supermassive black holes \citep{Heger2002, Miller2004};
or stellar mass black holes in a new ultraluminous accretion state \citep{Makishima2000, King2001, Zezas2002}.


The X-ray luminosity function (XLF) is a powerful tool for the
characterization of the populations of discrete X-ray sources detected in
nearby galaxies.
\citet{Kilgard2002} studied three starburst and five spiral galaxies, and found that
starburst galaxies have flatter XLF slopes than do normal spirals.
\citet{Grimm2003} studied the X-ray sources (mainly high-mass X-ray binaries; HMXBs)
in a dozen of late-type and starburst galaxies, and found that the total X-ray
luminosity and the XLF scale with the star formation rate (SFR).
%
%
%
While HMXB XLFs are typically described by straight power-laws, XLFs are
typically described by broken or cut-off power-laws for low-mass X-ray binaries
in early-type galaxies or regions lacking current star formation.
\citet{Mineo2013} reported the numbers of ULXs (in one colliding galaxy pair)
are strongly correlated with the local SFR densities, but the luminosities
of these sources show weak correlation with SFR densities.

In the general picture, starbursting galaxies have flatter XLFs than spiral or early-type galaxies,
disagreements, however, exist for some important details of the XLFs.
For low mass X-ray binaries (LMXBs) in galaxies lacking current star formation, the XLFs are
typically described by broken or cut-off power-laws, with several possible but controversial knees:
($i$) The knees at a few $10^{37}$ erg s$^{-1}$ \citep{Grimm2003, Gilfanov2004} may be explained by accretion on neutron star from Roche lobe overflow,
which is driven by gravitational wave emission when below $\sim~2\times10^{37}$ erg s$^{-1}$, and by magnetic stellar winds at higher luminosities \citep{Postnov2005}.
Some studies proposed that systems below $10^{37}$ erg s$^{-1}$ may be transients \citep{Bildsten2004}. However, \citet{Kim2006} did not find such a break in the LMXB XLF in
two normal elliptical galaxies.
($ii$) The knees at a few $10^{38}$ erg s$^{-1}$ \citep{Sarazin2000,Sarazin2001} may be consistent with
the Eddington luminosity of normal neutron star (NS) binaries, the luminosity of
massive neutron stars (NSs), He-enriched NS binaries \citep{Ivanova2006},
or low-mass BH binaries.
($iii$) The knees at a few $10^{39}$ erg s$^{-1}$ \citep{Jeltema2003} could be attributable to ULXs or small number statistics.
In some sense, these discrepancies are expected, because those studies use
different galaxy samples that are usually small (on the order of a dozen), and
they adopt different methods to construct XLFs and different methods to correct for the survey incompleteness.


We have embarked on an effort to study the X-ray point sources, especially
ULXs, in nearby galaxies with uniform procedures using the wealth of {\it Chandra}
Data Archive after eight years' accumulation \citep[][hereafter Paper I]{Liu2011}.
As detailed in Paper I, 383 galaxies within 40 Mpc with isophotal major axis
above 1 arcminute have been observed by 626 ACIS observations, and our uniform
analysis of these observations has led to 11,824 point sources within the 2$D_{25}$
isophotes of 380 galaxies, by far the largest extragalactic X-ray point source
catalog of such.
There are a large number of galaxies observed for each galaxy morphological
type, making it possible to compare XLFs among galaxies of each type and
between galaxies of different types with good statistics.
Meticulous efforts have been made to identify the nuclear X-ray sources, so
that we can excise them from the catalog and study the remaining ordinary X-ray
binary populations.

In this paper, we study XLFs and statistical properties for ULXs for different
samples of galaxies with uniform procedures.
In section 2, we describe our treatments of the detection threshold and the
foreground/background contamination estimates for extragalactic point source
surveys.
In section 3, we describe our procedures to construct a stitched survey for
individual galaxies aided by an example, and present a library of XLFs for 343
galaxies.
In section 4, we describe how to construct surveys of a sample of galaxies, and
present the XLFs and statistical properties for different samples of galaxies.
We summarize our results and discuss the significance of the XLF break at
$4\times10^{40}$ erg s$^{-1}$ in section 5.

\section{TWO ISSUES FOR EXTRAGALACTIC SURVEYS}

In carrying out extragalactic point source surveys,
two issues are of particular importance: the incompleteness; and the
foreground/background contaminations.
Below we describe our methods to compute detection thresholds pixel by pixel
for ACIS observations used in this work, and the different ways to estimate the
foreground/background objects projected into the galaxies.

\subsection{Detection Threshold}

The sensitivity for an X-ray observation includes two issues, the detecting
limit and the incompleteness.
The first issue gives the minimum flux of a source that can be detected with a
specific detection algorithm such as {\tt wavdetect}, widely used for {\it Chandra}
observations.
The second issue gives the fraction of sources at certain fluxes that are
missed by the detection algorithm.
The XLF becomes artificially flattened at the faint end
due to the missing sources close to the detect limits, and the missing
source fraction must be known to correct for this flattening.
These issues are usually addressed by extensive simulations such as those for the
{\it Chandra} deep fields \citep{Giacconi2001} and for the ChaMP project \citep{Kim2004},
in which fake sources at different flux levels at different
positions on the CCD chips are simulated and detected.
As shown by these simulations, the detect limit and the missing source
fraction are very complicated functions of the off-axis angle and background
level, and there are no satisfying analytical forms for such functions.

\begin{figure}
\plotone{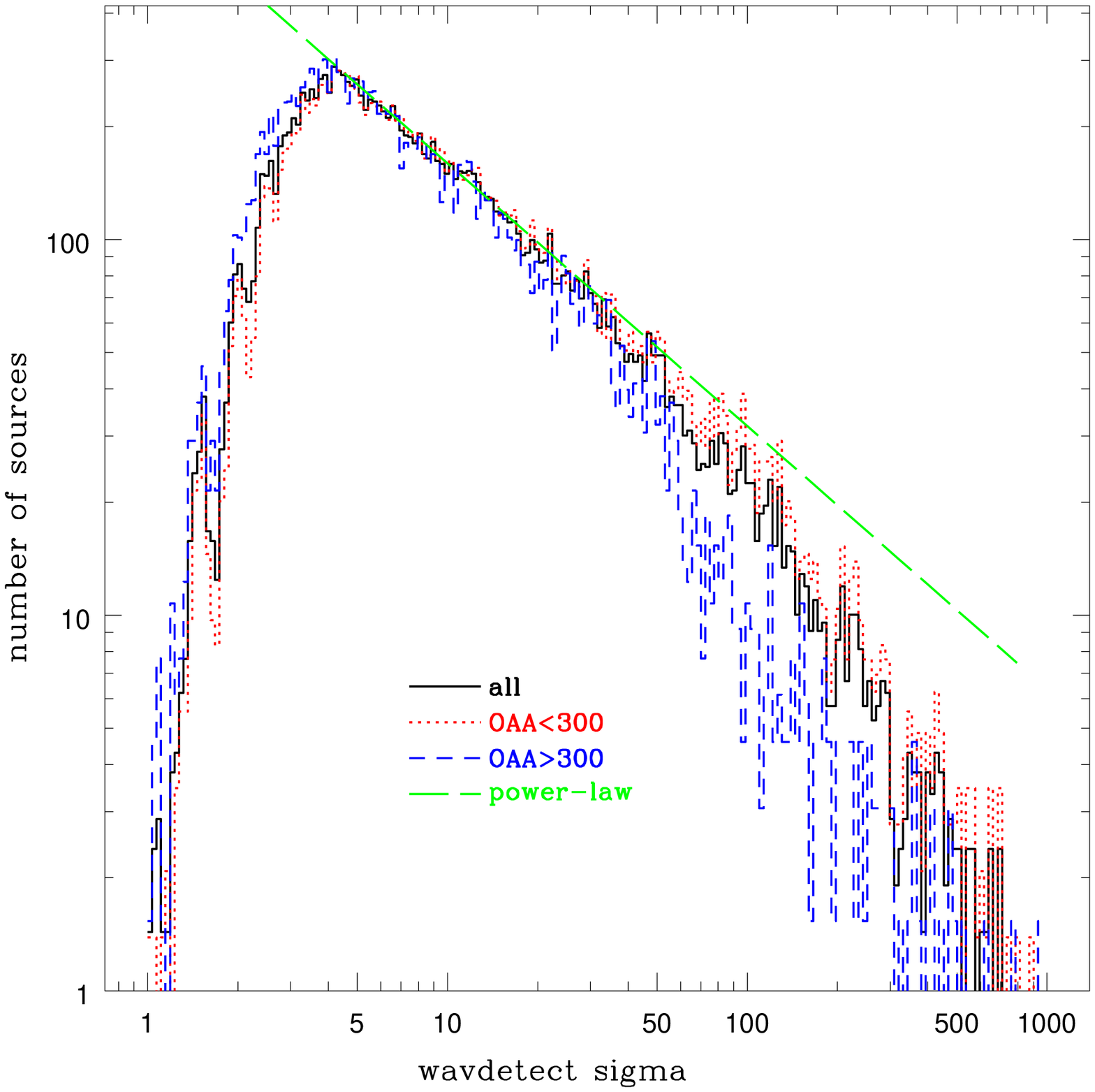}
\caption{Distribution of detection significance for all sources detected in 626
ACIS observations. As the detection significance decreases, the source number
increases in a power-law form till $\sigma \approx 4$, from where on the source
number drops sharply from the expected power-law form. The detected number is
about 55\% at $3\sigma$ and 15\% at $2\sigma$ of that from the expected
power-law.}
\label{fig.1}
\end{figure}

When constructing XLFs for this survey, we exclude the
flattened part at the faint end affected by the missing sources.
This ensures the XLFs are free of the artificial flattening problem by sacrificing
the sensitivity range from the detect limit to where incompleteness begins.
Then where does incompleteness come into play? We try to get the answer from
the survey itself following \citet{Kong2003}.
Figure \ref{fig.1} shows the distribution of detection significance for all sources
detected in 626 ACIS observations.
As the detection significance decreases, the source number increases in a
power-law form until $\sigma \approx 4$, beyond which the source number drops
sharply from the expected power-law form.
This indicates that incompleteness begins at $\sigma=4$, regardless of the
off-axis angles for the detection algorithm used for our survey.
As shown in Figure \ref{fig.1}, the missing source fraction increases for lower detection
significance, reaching 45\% at $3\sigma$ and 85\% at $2\sigma$.

\begin{figure}
\plottwo{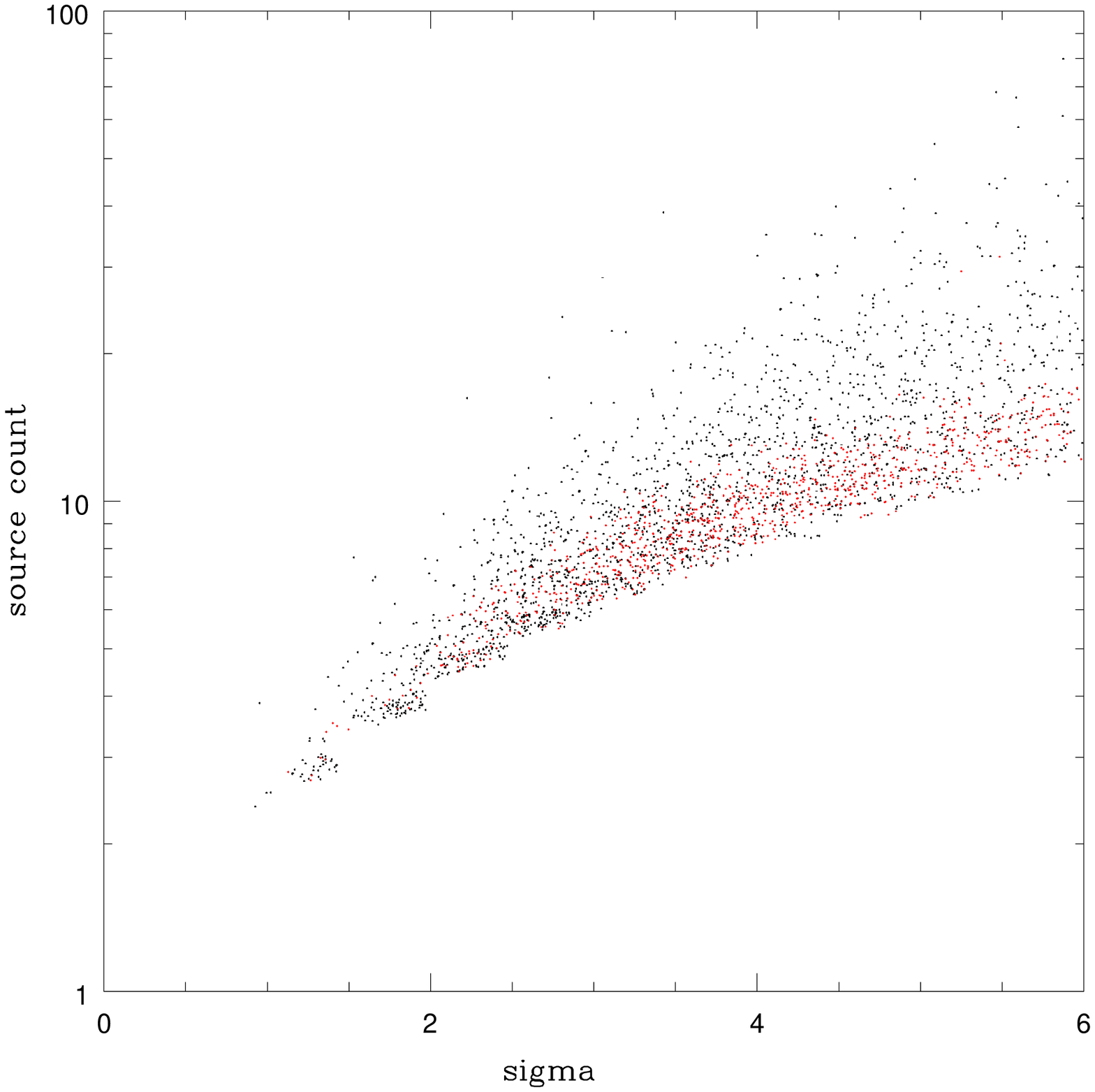}{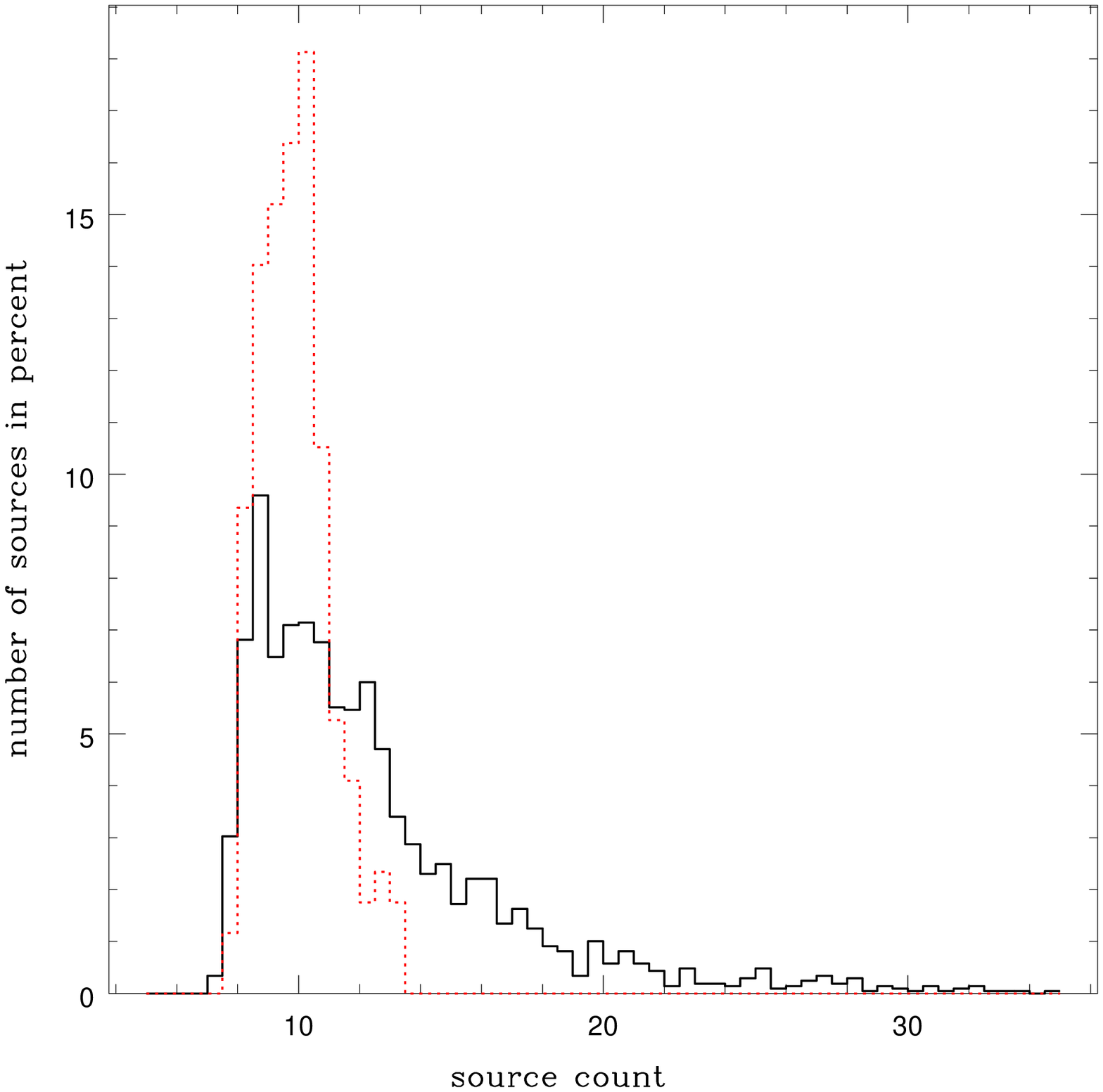}
\caption{(a) Source counts versus detection significance for all sources (black
dots) and for a subset of sources with OAA $\le2^{\prime}$ and
background level between 0.01 and 0.03 count pixel$^{-1}$ (red dots).  (b) The source
count histograms for sources with detection significance $3.8\le\sigma\le4.2$
(solid black) and for its subset with OAA $\le2^{\prime}$ and
background level between 0.01 and 0.03 count pixel$^{-1}$ (dotted red). }
\label{fig.2}
\end{figure}

The detection threshold is a complex
function of off-axis angles (OAA), background counts per pixel, and different source
counts \citep{KF2003}. In general,
the detection threshold increases as the effective background rate and the
OAA increase, and as the source count decreases.
The detection efficiency significantly decreases with the OAA,
meaning more source counts are needed to achieve a similar detection sensitivity
at larger off-axis radii compared to on-axis observations,
due to the variation of the {\it Chandra} PSF with the OAA \citep{Primini2011}.
As shown in \citet{Evans2010},
the PSF close to the optical axis of the telescope is approximately
symmetric with a 50\% enclosed energy fraction radius of $\sim~0.3''$ over a wide range of energies,
but the PSF at 15$''$ off-axis is asymmetric, significantly extended, and strongly energy dependent,
with a 50\% enclosed energy fraction radius of $\sim~13''$ at 1.5 keV.
In this paper, the detection threshold for certain detection significance is derived from the
survey itself.
As shown in Figure \ref{fig.2}a, the source counts increase monotonically with the
detection significance, albeit with large dispersions.
This dispersion, however, reduces greatly if we fix the OAA and
the background level.
In example, of the sources with $\sigma=3.8\sim4.2$, 90\% have
between 8 and 25 source counts, but 90\% have between 8 and 12 counts for the subset
with OAA less than 2$^{\prime}$ and background level between 0.01 and 0.03
count pixel$^{-1}$ (Figure \ref{fig.2}b).
To compute the detection threshold for $4\sigma$, we use sources with $\sigma$
between 3.8 and 4.2, and group them based on OAAs and background
levels.
For each group, we compute the average OAA, the average background
level, and the average source count as the detection threshold.
The resulting detection thresholds for $4\sigma$, as plotted in Figure \ref{fig.3}, are
used to calculate the detection threshold for any given OAA and
background level with interpolation or extrapolation.

\subsection{Estimates of Foreground and Background Objects}

When studying X-ray point source populations in nearby galaxies, one needs to
exclude the foreground stars and background QSO/AGNs projected into the host
galaxies by chance.
It would be best if we can identify the foreground/background objects with
multiwavelength observations, but is impractical given the huge telescope time
required to carry out such identifications for a large survey like this.
Fortunately, we do not really need to know which are foreground/background
objects for statistical work in this paper.
What we really need is a statistical estimate, as the number of contaminating
sources per flux interval, so that we can subtract these estimates from the
observed number of sources per flux interval.
Below we show two ways to make such statistical estimates for contaminating
sources.

One conventional way to estimate the contamination from the foreground/background
objects is to use the log$N$-log$S$ relation that predicts the number of X-ray
sources per deg$^2$ $N$ as a function of flux $S$.
\citet{Hasinger1993} derived a log$N$-log$S$ relation based on $ROSAT$
observations of the Lockman Hole region, where the differential form is
$dN/dS = N_1 S^{-\beta_1}$ for $S>S_b$ and $dN/dS = N_2 S^{-\beta_2}$ for $S<S_b$,
with $S$ in unit of $10^{-14}$ erg cm$^{-2}$ s$^{-1}$ in the 0.5--2 keV band,
$S_b=2.66\pm0.66$, $N_2 = 111\pm10$, $\beta_1 = 2.72\pm0.27$, $\beta_2 =
1.94\pm0.19$, and $N_1 = 238.1$.
\citet{Mushotzky2000} derived a log$N$-log$S$ relation based on a long {\it Chandra}
ACIS observation, where the number of sources over the flux range
(2.3--70)$\times10^{-16}$ erg cm$^{-2}$ s$^{-1}$ are given by $N(>S) = 185({S \over 7\times10^{-15}})^{-0.7\pm0.2}$,
with $S$ as the 0.5--2 keV flux.
\citet{Giacconi2001} derived another log$N$-log$S$ relation based on the {\it Chandra}
deep field source observations, where the number of sources over the flux range
(2--200)$\times10^{-16}$ erg cm$^{-2}$ s$^{-1}$ is given by $N(>S) = 370({S \over 2\times10^{-15}})^{-0.85\pm0.15}$, with $S$ as the 0.5--2 keV flux.

\begin{figure}
\plottwo{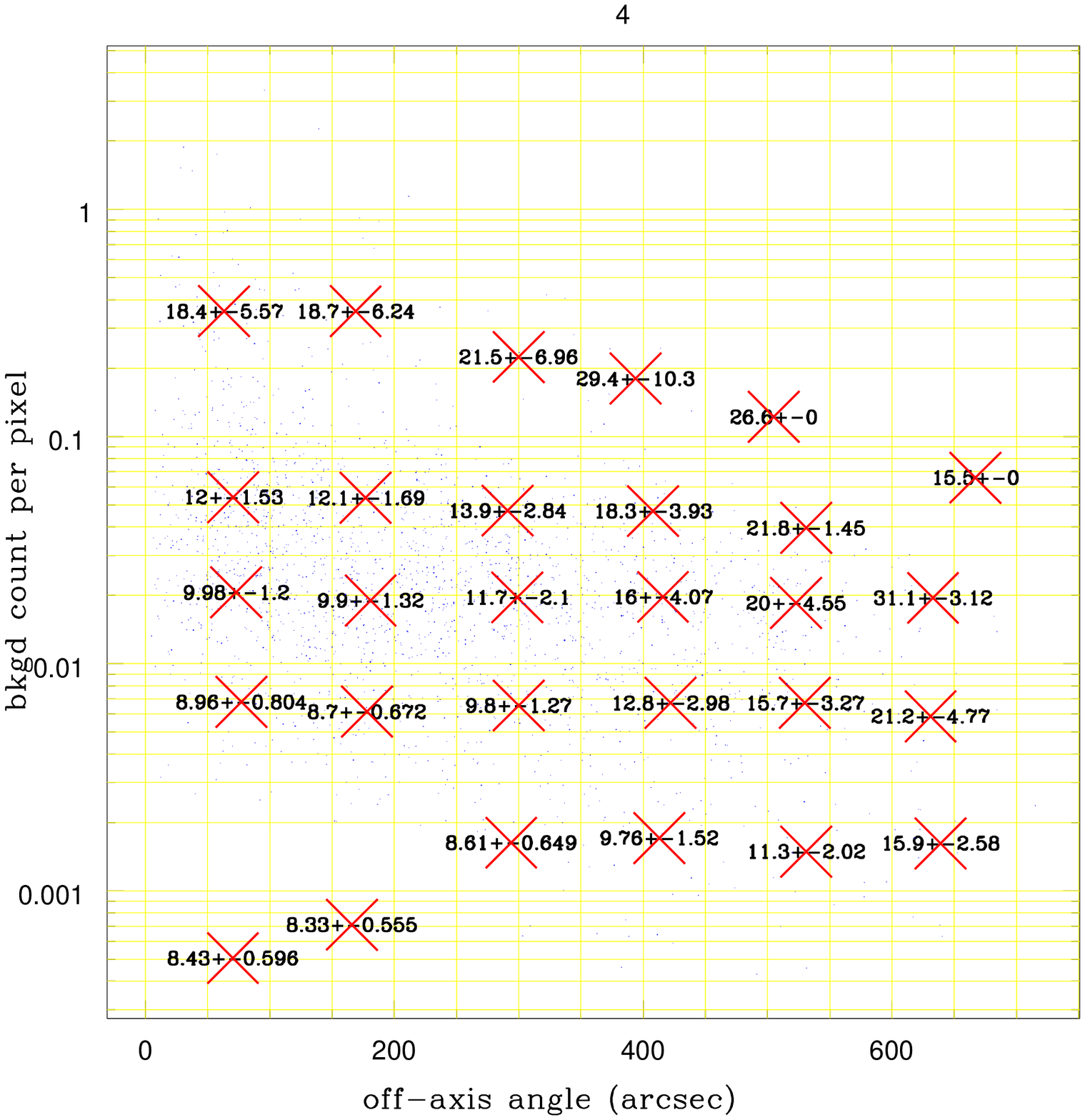}{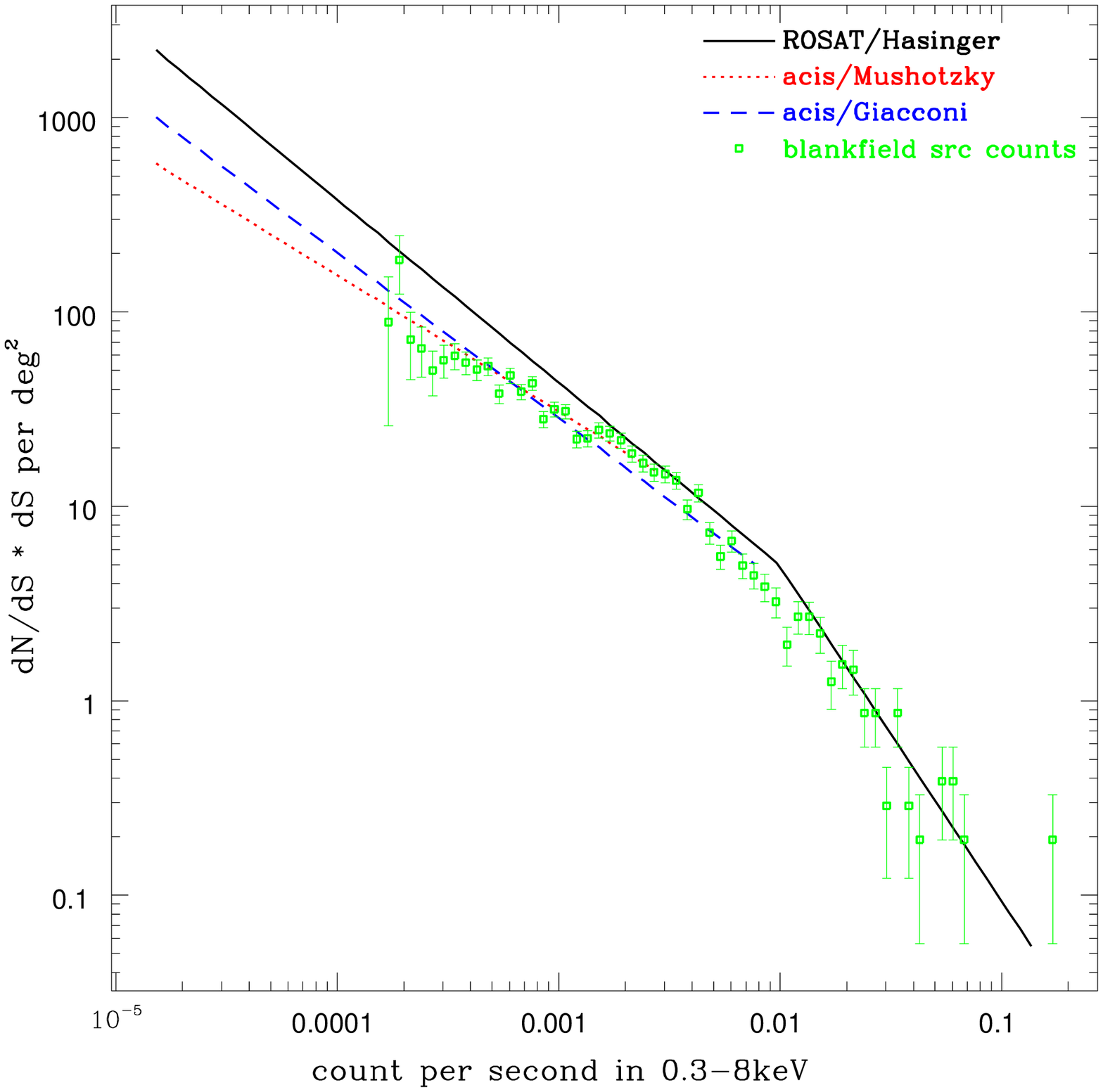}
\caption{The detection thresholds for $4\sigma$ for different combinations of OAAs and background levels.
The numbers are the detection thresholds and errors computed by averaging the detected sources with $3.8\le\sigma\le4.2$ (blue dots)
around the red crosses. }
\label{fig.3}

\caption{The numbers of foreground/background sources per deg$^2$ versus
equivalent ACIS-S3 count rates expected from the log$N$-log$S$ relations derived by
\citet{Hasinger1993}, \citet{Mushotzky2000}, and \citet{Giacconi2001}.
Overplotted are the detected sources outside 2$D_{25}$ isophotes of all
galaxies as a direct foreground/background estimate for our survey. }
\label{fig.4}
\end{figure}

Another way to estimate the contamination from the foreground and background
objects makes use of observed fields void of nearby galaxies.
From these blank fields outside 2$D_{25}$ isophotes of any nearby galaxies, X-ray
point sources should be detected and fluxes be computed from count rates with
the same procedures as from fields within 2$D_{25}$ isophotes of nearby galaxies.
Then the number of sources in a flux interval $\Delta N(F_1,F_2)$ can be
accumulated from these blank fields and serve as a template for contaminating
sources after normalized by the surveyed area of these blank fields.
This nearby galaxy survey, with its uniform analysis of 626 ACIS observations,
naturally provides a collection of blank fields, i.e., the regions outside of
2$D_{25}$ isophotes of galaxies.
Figure \ref{fig.4} shows the number of contaminating sources per deg$^2$ as a function of
flux derived from this collection of blank fields.

The contaminating sources from different estimates are compared in Figure \ref{fig.4}.
To enable the comparison between different contamination estimates, we plot
them in Figure \ref{fig.4} as a function of {\it Chandra} ACIS-S3 count rate.
Assuming $n_H = 2\times10^{20}$ cm$^{-2}$, one ACIS-S3 photon/count would
correspond to
the 0.5--2 keV flux $S$ of $2.57\times10^{-12}$ erg s$^{-1}$ used in the Hasinger
log$N$-log$S$ relation assuming a power law with $\Gamma=2$,
the 0.5--2 keV flux $S$ of $2.35\times10^{-12}$ erg s$^{-1}$ used in the Mushotzky
log$N$-log$S$ relation assuming a power law with $\Gamma=1.4$,
the 0.5--2 keV flux $S$ of $2.50\times10^{-12}$ erg s$^{-1}$ used in the Giacconi
log$N$-log$S$ relation assuming a power law with $\Gamma=1.7$,
and the 0.3--8 keV flux $F$ of $7.90\times10^{-12}$ erg s$^{-1}$ used in $\Delta
N(F_1,F_2)$ assuming a power law with $\Gamma=1.7$.
The fourth flux $F$ is the same as used for the source catalog in Paper I.
As shown in Figure \ref{fig.4}, the blank field estimate shows a knee around 0.01
count s$^{-1}$ like the Hasinger relation, but is significantly lower at fluxes
below 0.001 count s$^{-1}$ than the Hasinger estimate.
This is due to the poor extrapolation of the formula in Hasinger et al. (1993),
in which the lower flux limit of the used data points is $\sim$ 10$^{-15}$ erg cm$^{-2}$ s$^{-1}$
($\sim$ 0.002 count s$^{-1}$).
The blank field estimate is
roughly consistent with the Mushotzky and Giacconi estimates, but extends to
much higher flux levels.

In this work, the blank field estimate is adopted for the foreground/background
contamination.
It is advantageous to use this estimate because the contaminating sources are
detected and treated exactly the same way as the point sources associated with
the galaxies.
In addition, these blank fields have the same positions as the surveyed
galaxies, making the foreground/background estimate free of cosmic variance.
The other estimates are also computed for comparison and included in the XLF
library in Section 3.3.

\section{SURVEYS OF INDIVIDUAL GALAXIES}

Many galaxies in our archival survey have multiple {\it Chandra} ACIS observations
with different exposure times covering different parts of the galaxies.
To maximize the galaxy coverage, we stitch all available observations together
to construct surveys of individual galaxies.
That is, when observations block each other, pixels in longer observations take
precedence over pixels in shorter observations,
which means only the deepest single observations are used for a given area of the sky,
so as to make the stitched survey as deep as possible.
In such cases, we first order the observations by their exposure times, then
sum up pixels in each observation that are not blocked by pixels in any
preceding observations.
In the end, the stitched survey of a galaxy comes down to a collection of
pixels possibly from different observations for that galaxy.
We do not co-add/merge the observations with different exposures
to avoid additional uncertainties, such as different PSFs for same sources at different off-axis radii,
although this ``stitching'' method ``wastes'' some existing exposures.
As shown by recent studies of XLFs for NGC 3031, different observations provide
a consistent determination of its XLF despite measuring significant variability
in a considerable fraction of sources \citep{Sell2011}.
Thus, our stitched deep survey will yield XLFs similar to other surveys made by
changing the order of observations, yet maximizes the power to study XLFs at
the low luminosity end.
Below we describe the surveyed sky area curve and the surveyed blue light curve
computed from this collection of pixels in Section 3.1, show the construction of XLFs
by an example in Section 3.2, and present a library of XLFs for individual galaxies
in Section 3.3.

\subsection{Surveyed Sky Area and Blue Light}

Given a collection of pixels for a stitched survey, we first compute the
survey area curve $A($$>$$F)$ as the sky area in which an X-ray source above
the limiting flux $F$ can always be detected in an observation or a survey.
As discussed above, we compute $F$ from the $4\sigma$ detection threshold to
avoid incompleteness.
For each pixel in an ACIS observation, we compute its OAA and the
$4\sigma$ detection threshold in count as described above, where the background
level is taken as the average of the whole ACIS chip.
The detection threshold is converted to a count rate, given the exposure time and
the vignetting factor that is computed as an analytical function of the
OAA provided by the {\it Chandra} calibration team.
To convert the count rate to the limiting flux, we use the response matrix at
the chip center of that observation to account for {\it Chandra}'s response evolution
over the past years.
The limiting flux $F$ is then computed in 0.3--8 keV by assuming a power-law
with $\Gamma=1.7$ and Galactic absorption at the nominal pointing of that
observation.

The survey area curve $A($$>$$F)$ for a collection of pixels is computed by
summing the area of the pixels for which the detection thresholds correspond
to flux less than $F$.
To compute $A($$>$$F)$ for a galaxy, we compute for each pixel the separation
$\alpha$ between the galaxy center and the pixel and compare to the elliptical
radius $R_{25}$ of the $D_{25}$ isophotal ellipse along the great arc
connecting the galaxy center and the pixel.
A pixel is considered as in a galaxy if it is within the 2$D_{25}$ isophote of the
galaxy, or in a blank field if outside the 2$D_{25}$ isophotes of any galaxies.
For pixels within two overlapping galaxies, they belong to the closer galaxy in units of the elliptical radius.
The survey area curve $A($$>$$F)$ for a galaxy in this survey can be computed
by summing the area of the pixels within this galaxy for which the detection
thresholds correspond to flux less than $F$.
Similarly, the survey area curves for parts of a galaxy, e.g., the region
between $D_{25}$--2$D_{25}$, can be computed by considering only pixels within
those parts.

To compare X-ray point source populations in different types of galaxies, we
need to account for their variations in the stellar mass content,
which can be deduced from the total light of
the galaxy and the stellar mass-to-light ratio ($M/L$).
Although the $M/L$ varies significantly in the optical compared to the near-infrared \citep{Bell2001},
the blue light from RC3 is used because of the lack of a consistent compilation of near-infrared
magnitudes for these survey galaxies.
The blue and near-infrared light trace different star population and suffer different extinction,
which would result in different effective radii of galaxies.
This could affect the computed light curve $\pounds($$>$$F)$ and finally the $\pounds$-corrected XLF.
However, as shown by \citet{McDonald2011}, the radial distributions are quite similar for optical and near-infrared bands,
thus the discrepancy of the $\pounds$-corrected XLF due to the bulge light may be small.

The surveyed blue light curve $\pounds_B($$>$$F)$ for a galaxy gives the blue
light as a measure of the stellar contents in which X-ray sources above $F$ can
be detected in an observation.
This curve is calculated the same way as the surveyed area curve $A$($>$$F$),
except that we sum up the blue light in a pixel instead of the area of a pixel.
To compute the blue light in a pixel, we compute the light profile for the
galaxy from the total blue light $\pounds_B$ and the effective radius $R_e$
that encloses 50\% of the total light. This method is the same as used in our
previous work on a $ROSAT$/HRI survey of nearby galaxies \citep{Liu2006}, and
the procedures are also described in the appendix.
The survey area curve $A($$>$$L)$ and the surveyed blue light curve
$\pounds_B($$>$$L)$ on the luminosity intervals are calculated by converting
$F$ to $L$ with $L = 4\pi D^2 F$ using the distance to the surveyed galaxy.
To enable the summation of X-ray sources in different galaxies with the same
luminosities, we have adopted the same luminosity intervals for all
observations. This corresponds to different flux intervals for galaxies with
different distances.

%

\subsection{XLF for NGC 3031}

\begin{figure}
\plottwo{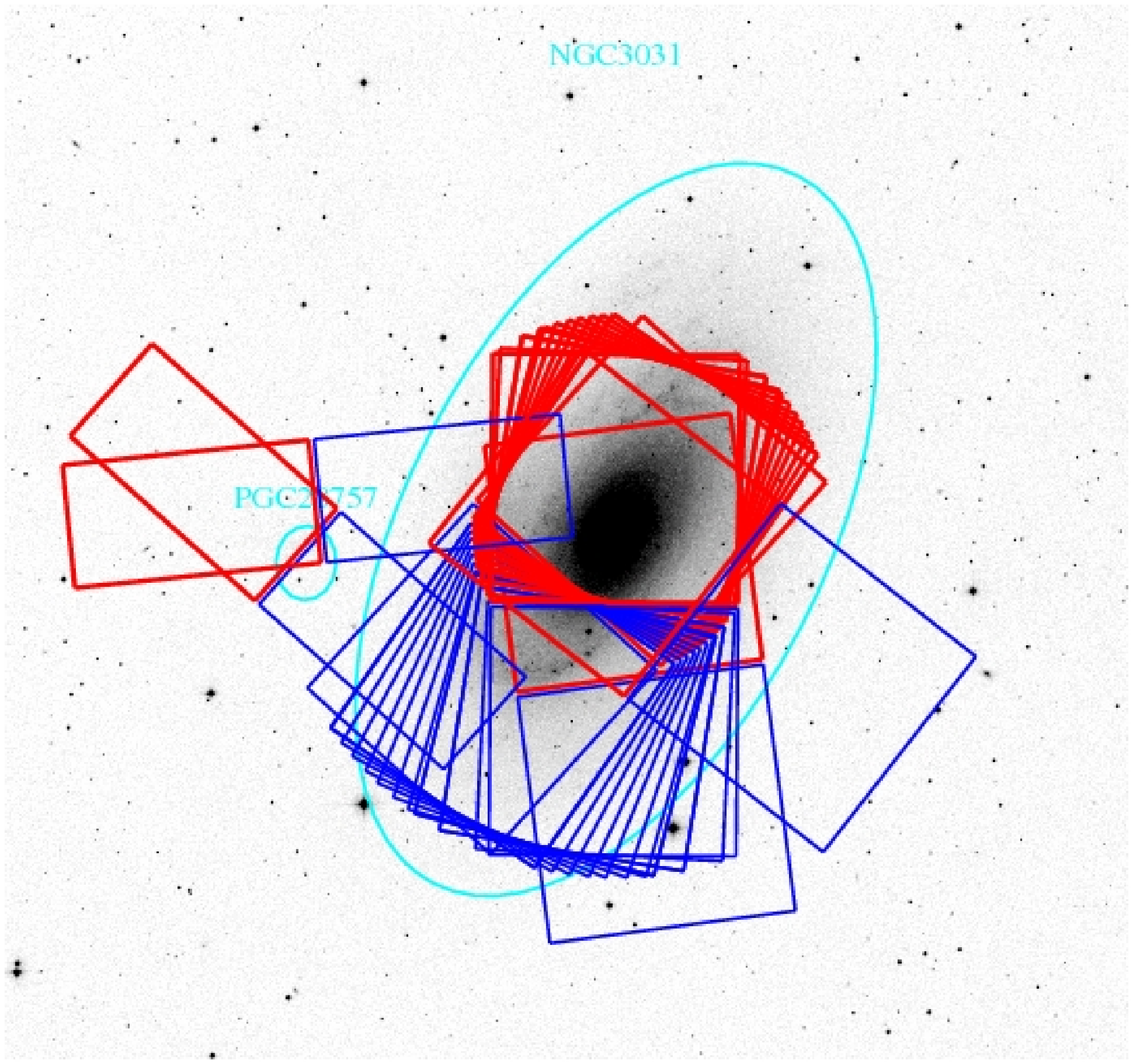}{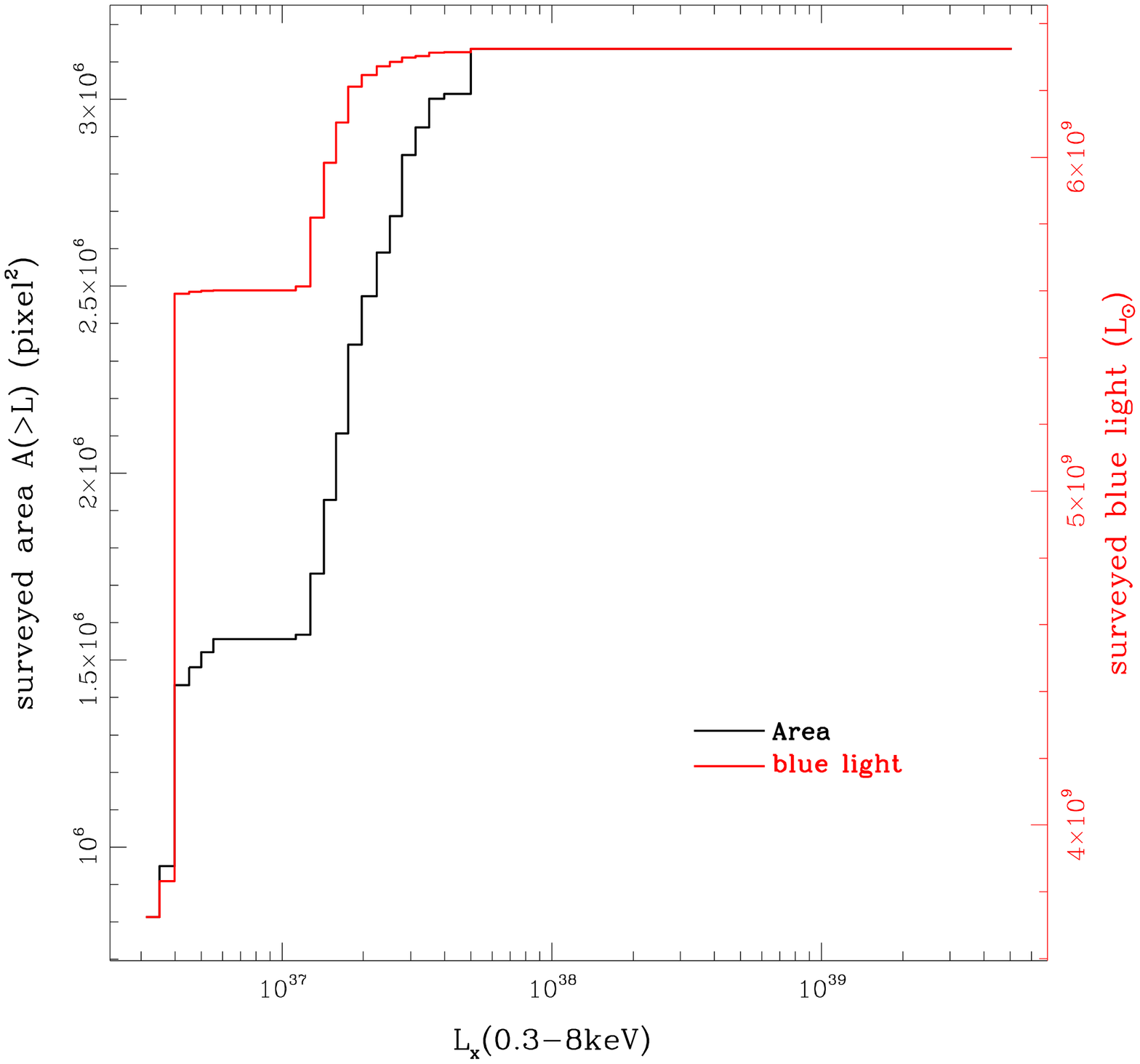}
\plottwo{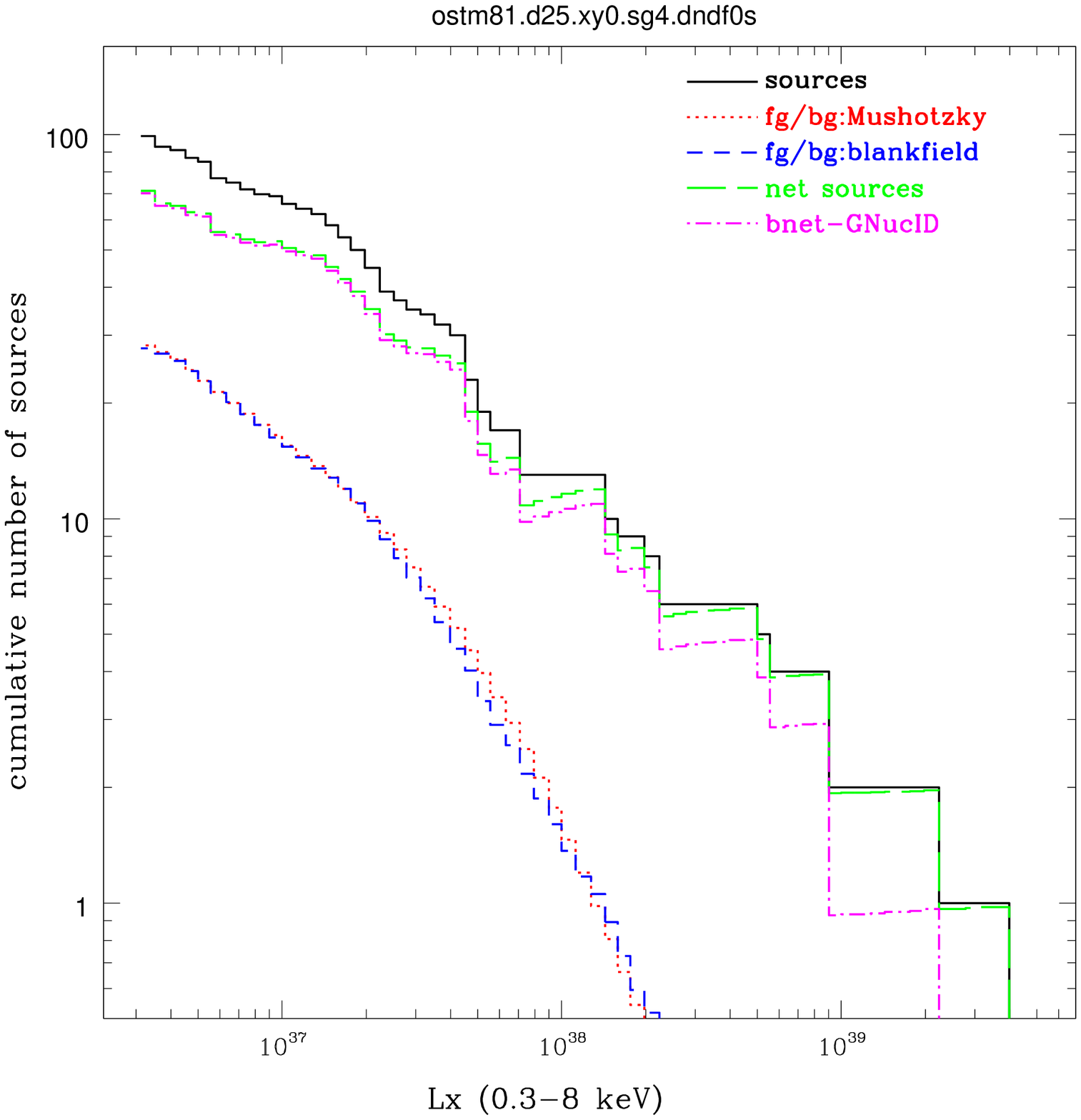}{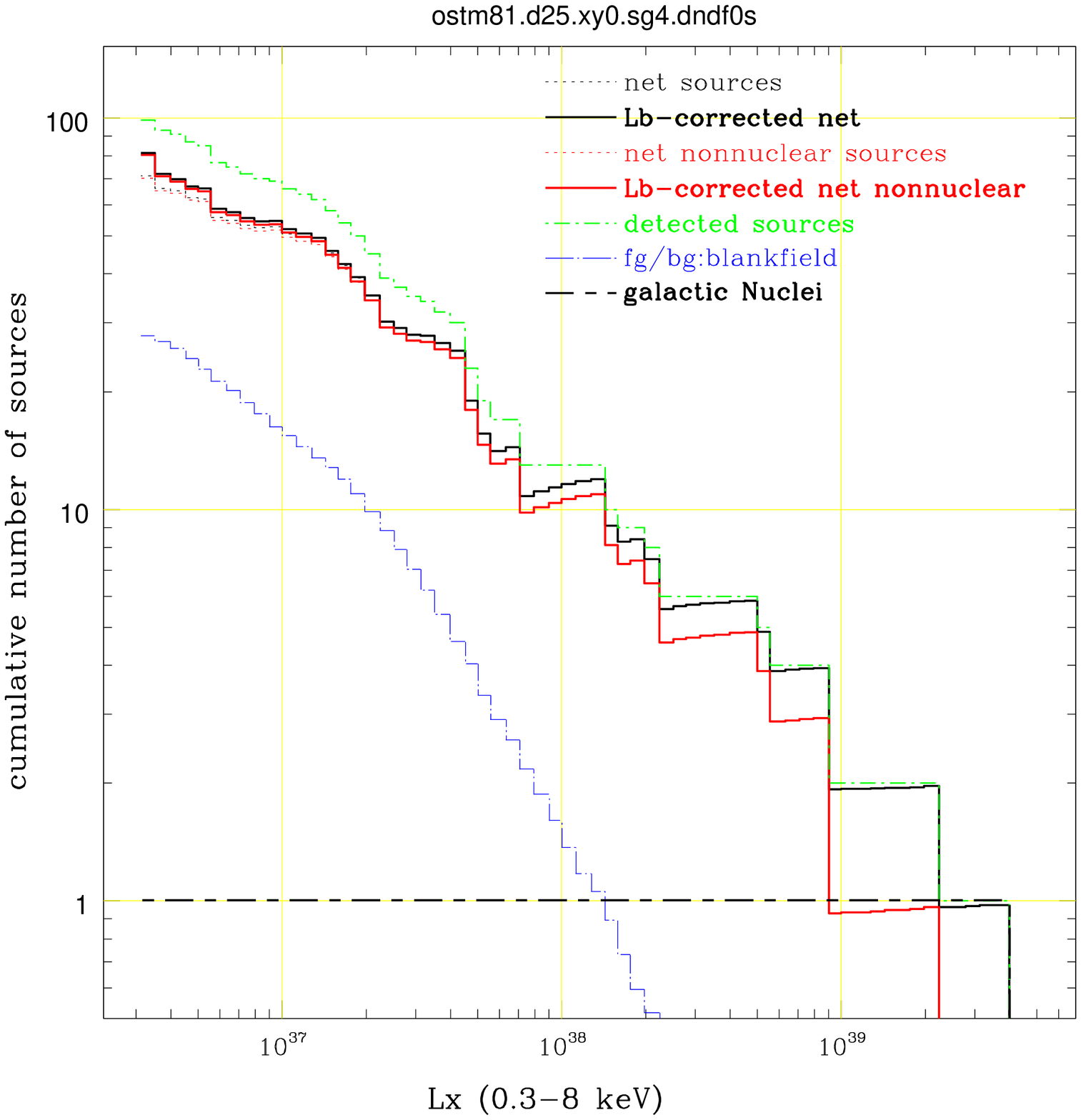}
\caption{(a) DSS image for NGC 3031 (M81) with the fields of view for 19 ACIS
observations overlayed. The red/blue squares are for S3/S2 chips, and the cyan
ellipses denote the isophotes for NGC 3031 and PGC 23757 (Holmberg IX). (b) The
surveyed sky area curve $A(>L)$ (black solid) and the surveyed blue light curve
$\pounds_B(>L)$ for the stitched deep survey out of 19 ACIS observations.  (c)
The cumulative curves for detected sources above $4\sigma$ (black solid), for
expected background/foreground objects based on the \citet{Mushotzky2000}
log$N$-log$S$ relation (red dotted) and based on the source counts in blank fields
(blue dashed), for ``net'' sources (green long-dashed), and for ``net''
non-nuclear sources (purple dot-dashed). (d) The cumulative curve for ``net''
sources in comparison to the $\pounds_b$-corrected ``net'' sources.  }
\label{fig.5}
\end{figure}

NGC 3031 (M81) is a well-studied nearby galaxy at 3.63$\pm$0.3 Mpc 
\citep[][using the Cepheid period-luminosity relationship]{Freedman1994}.  
It is a Sb-type
spiral with semi-major/minor axis of 13.5/7 arcmin, and has been observed with
19 {\it Chandra}/ACIS observations (Figure \ref{fig.5}a) with exposure times ranging from 10
ksec to 50 sec spanning five years from 2000 to 2005 as tabulated in \citet{Sell2011}.
We stitch a deep survey of NGC 3031 out of these 19 observations by using all
ACIS S3 and S2 pixels excluding those blocked by longer observations, which lead to a
total of $3.1\times10^6$ pixels, or 218 arcmin$^2$, covering 73\% of the $D_{25}$
isophote.
This ``stitched'' survey covers 50\% more sky area than any single ACIS
observation, which covers at most 146 arcmin$^2$ with S3 and S2 chips.

With these $3.1\times10^6$ pixels, the surveyed sky area curve $A($$>$$L)$ and
the surveyed blue light curve $\pounds_B($$>$$L)$ are computed following Section 3.1
and plotted on the luminosity grid in Figure \ref{fig.5}b.
The expected background/foreground sources for each luminosity bin $(L_1,L_2)$,
which corresponds to flux bin $(F_1,F_2)$,  is computed as $A($$>$$L_1)$ *
$\Delta N_{b/f}(F_1,F2)$, where $\Delta N_{b/f}(F_1,F2)$ is the number of
sources per square degree from the blank fields void of nearby galaxies.
X-ray point sources detected in these observations as listed in Table~3 of
Paper~I are assigned to the pixels where the source centroids are.
Point sources above $4\sigma$ associated with any of the $3.1\times10^6$ pixels
are considered detected in this stitched survey, and are assigned to
luminosity bins based on their 0.3--8 keV luminosity.
Of these detected, one point source is identified as the galactic nucleus of
NGC 3031.

The XLF curves are computed given the detected sources, identified galactic
nuclei, and the background/foreground estimates.
The cumulative curves are plotted in Figure \ref{fig.5}c, including the number of
detected sources $N($$>$$L)$, the numbers $N_{b/f}(>L)$ of expected
background/foreground objects, the number of net sources $N_{net}(>L)$ =
$N(>L) - N_{b/f}(>L)$, and the number of net non-nuclear sources
$N_{net,nG}(>L)$ = $N(>L) - N_{Gnuc} - N_{b/f}(>L)$ .
To correct the flattening of XLF at the low luminosity end due to less surveyed
blue light, we normalize the detected sources in each luminosity bin by the
surveyed blue light curve $\pounds_B(>L)$.
The $\pounds_B$-corrected number of net sources $N^\pounds_{net}(>L) =
N(>L) * \pounds_B^{tot} / \pounds_B(>L) - N_{b/f}(>L)$, and the
$\pounds_B$-corrected number of net non-nuclear sources
$N^\pounds_{net,nG}(>L)$ = $N(>L) * \pounds_B^{tot} / \pounds_B(>L) - N_{Gnuc}
- N_{b/f}(>L)$; both $\pounds_B$-corrected  curves are plotted in Figure \ref{fig.5}d.

This stitched survey is sensitive to luminosities as low as $2\times10^{36}$
erg s$^{-1}$, and detects 99 point sources above the limiting luminosity $L_{X,lim} =
4\times10^{36}$ erg s$^{-1}$, defined as the $L_X$ at which the surveyed blue light is
50\% of the total blue surveyed.
In comparison, there are 122 point sources from all observations with detection
significance above $4\sigma$ in at least one observation.
Out of the 99 detected sources above $L_{X,lim}$, about 27.8 (28\%) are
expected to be background/foreground objects.
There is one ULX detected above $2\times10^{39}$ erg s$^{-1}$, with 0.03
background/foreground objects expected, suggesting that the ULX is truly
associated with NGC 3031.
Recently, \citet{Sell2011} studied the XLFs in different regions and the entire
galaxy of NGC 3031. Although it is difficult to make direct comparisons between
their XLF and our XLF due to different detection thresholds and spectral models
used, our XLF does show a similar knee at $10^{37}$ erg s$^{-1}$ that also appears in
their XLF for the entire galaxy.
This knee in our XLF is a true feature instead of an artificial flattening due
to less surveyed blue light at the low luminosity end, because it is also
present in the $\pounds_B$ corrected XLF as shown in Figure \ref{fig.5}d.

\subsection{XLFs for 343 Galaxies}

\begin{figure}
\plottwo{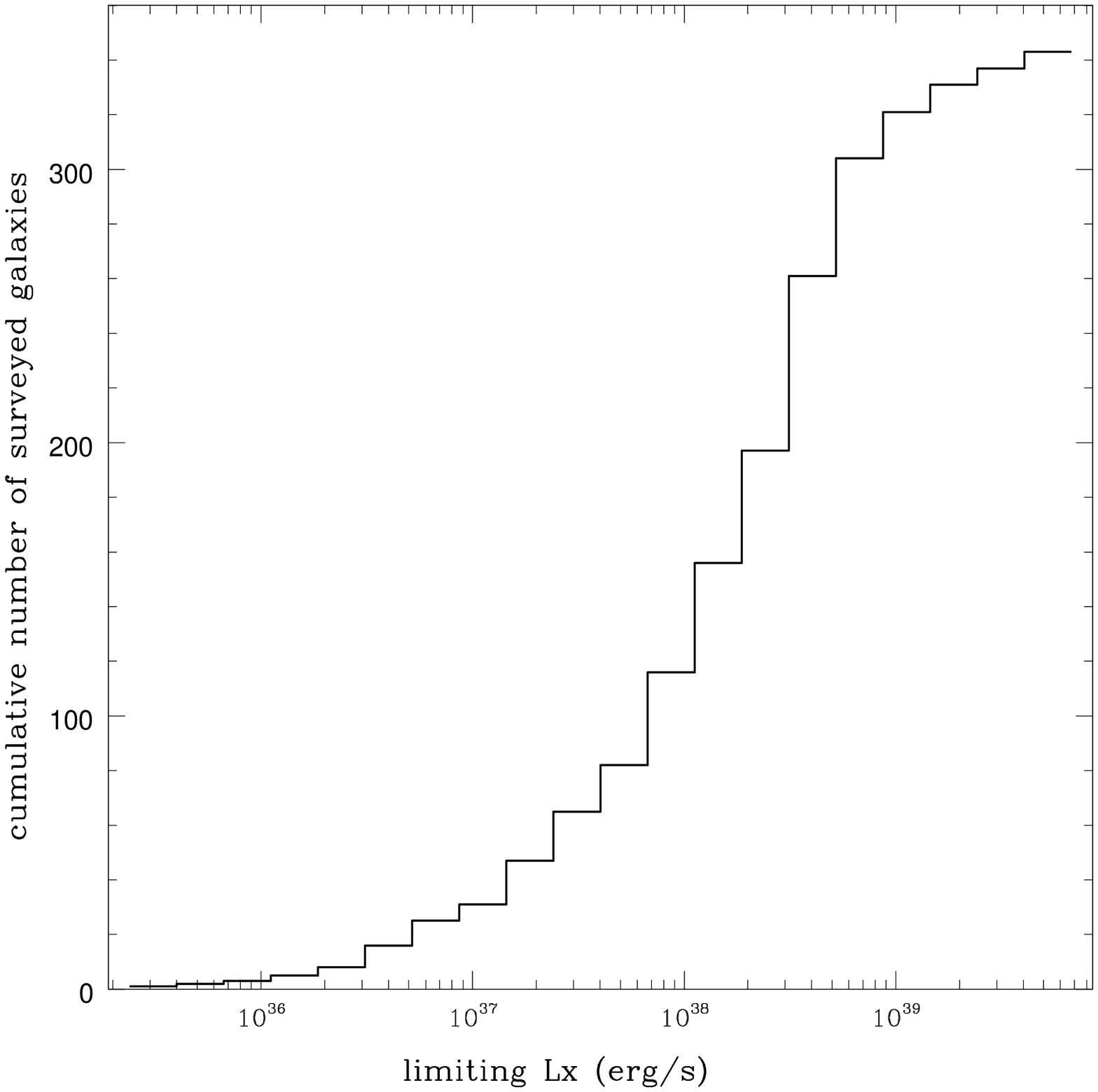}{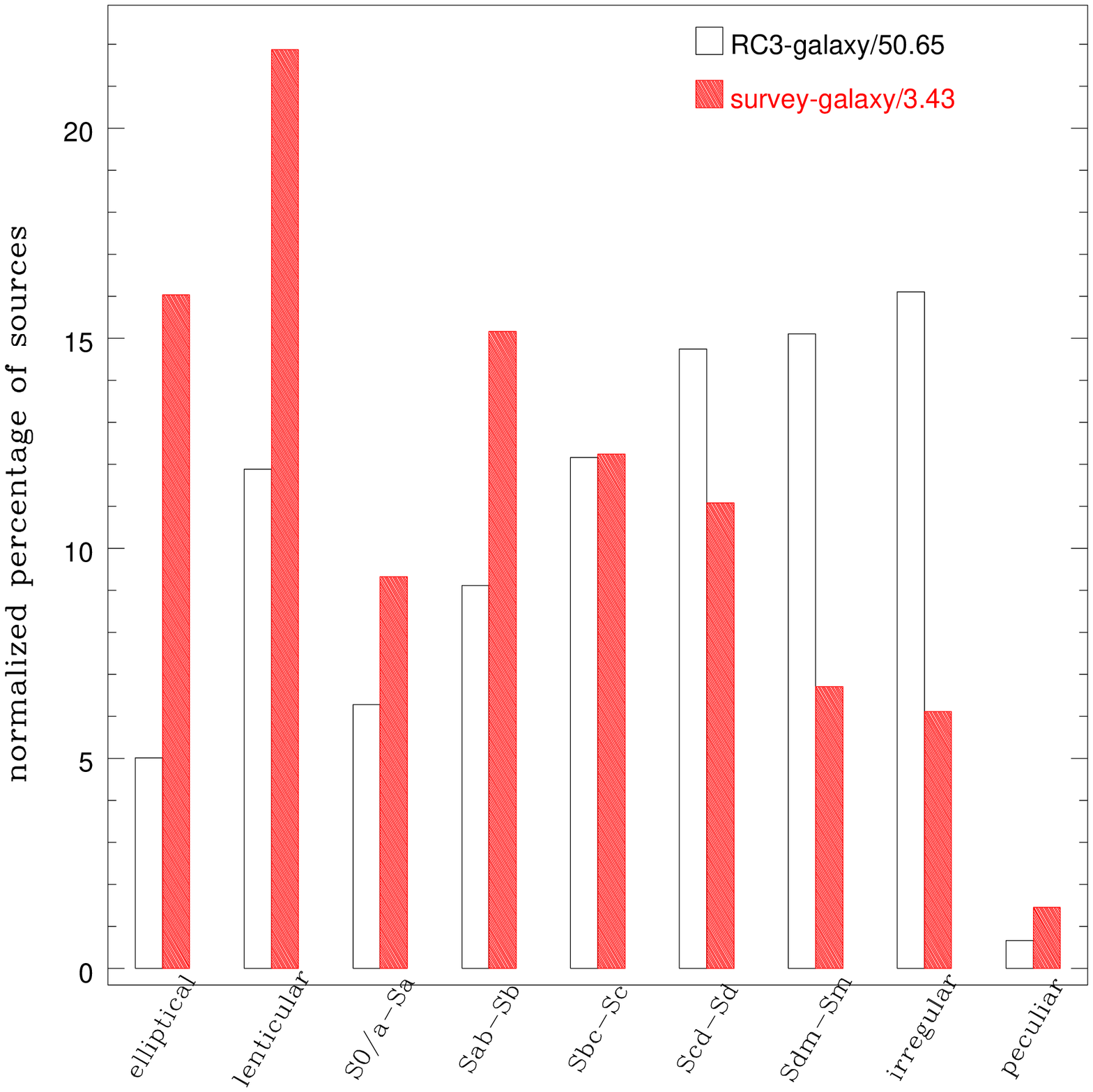}
\caption{(a) Histogram for the limiting $L_X$ of 343 surveyed galaxies. (b) The galaxy type distribution
of 343 surveyed galaxies in comparison to the RC3 galaxies within 40Mpc. }
\label{fig.6}
\end{figure}

XLFs for all surveyed galaxies are computed in the same way as for NGC 3031.
There are 343 galaxies with surveyed area more than 30\% of their $D_{25}$
isophotal sizes as listed in Table~1.
For each galaxy, listed are the number of ACIS observations utilized in Paper
I, the number of sources detected above $4\sigma$ from all observations, the
number of sources detected above $4\sigma$ in the stitched deep survey, the sky
coverage, the limiting luminosity $\log L_{X,lim}$, the number of sources
detected above $L_{X,lim}$, the expected background/foreground for sources
above $L_{X,lim}$, the number of ULXs from all observations, the net number
of ULXs from this stitched survey, the expected background/foreground for
sources above $2\times10^{39}$ erg s$^{-1}$, and the host galaxy properties.
The survey depths for these galaxies span 4 orders of magnitude ranging from
$3\times10^{35}$ to $6\times10^{39}$ erg s$^{-1}$.
As plotted in Figure \ref{fig.6}a, there are 27 galaxies with a limiting luminosity
$L_{X,lim}$ below $10^{37}$ erg s$^{-1}$, and 107 galaxies with limiting luminosity
$L_{X,lim}$ below $10^{38}$ erg s$^{-1}$.
%
These 343 galaxies include 130 early-type galaxies, 187 spiral galaxies, 5
peculiars and 21 irregular galaxies.  As compared to the RC3 \citep{Vaucouleurs1991}
galaxies within 40 Mpc as shown in Figure \ref{fig.6}b, our survey oversamples
early galaxies, S0/a-Sb galaxies and peculiars, and undersamples Sbc-Sm and
irregular galaxies.
A query against the NED service finds 15 starburst galaxies, and 121 AGNs of
different kinds.

XLFs for the above 343 galaxies are created on the luminosity
bins from $\log L=34$ to $\log L=42$ with $\Delta\log L=0.05$.
For each luminosity bin, we include the corresponding flux bin, the surveyed
sky area, the surveyed blue light, the numbers of background/foreground objects
estimated based on the Hasinger log$N$-log$S$ relation, the Mushotzky log$N$-log$S$
relation and the blank fields , the number of detected sources $N_{src}$, the
number of galactic nuclear sources $N_{Gnuc}$, the number of net sources
$N_{net} = N_{src} - N_{b/f}$, the number of net non-nuclear sources
$N_{net,nG} = N_{src} - N_{Gnuc} - N_{b/f}$ , and the $\pounds_B$-corrected
numbers of net sources $N^\pounds_{net} = N_{src} * \pounds^{tot}_B /
\pounds_B(>L) - N_{b/f}$ and net non-nuclear sources $N^\pounds_{net,nG} =
N_{src} * \pounds^{tot}_B / \pounds_B(>L) - N_{Gnuc} - N_{b/f}$.  For each
quantity, we include both the differential value for the luminosity bin and the
cumulative value.
%
%
For each galaxy, we present XLFs for sources within $D_{25}$ isophotes, which are used
by studies in Section 4. We also present XLFs for sources within elliptical annuli of
the galaxy from galactic center to 2$D_{25}$ in steps of 0.1 elliptical radii, which
can be used to study XLFs for sources in various parts of galaxies.

\section{SURVEYS OF GALAXY SAMPLES}

Studies of X-ray point source populations with individual galaxies are often
limited by the small number of sources at high luminosities. For better
statistics, it is conventional to combine X-ray sources from a sample of
galaxies with certain properties. Our XLF library provides a large pool of
galaxies to construct samples of galaxies.
In the following, we describe the construction of XLFs from the total survey of
all 343 galaxies and compute the statistical properties in Section 4.1, compare the
XLFs and statistical properties for galaxy samples across different
morphological types in Section 4.2, and compare the XLFs and statistical properties
for galaxy samples with different SFRs in Section 4.3.
To maximize the number of sources yet minimize the contamination of
background/foreground objects, we compute XLFs and corresponding statistics for
sources within $D_{25}$ isophotes unless specifically stated.

\subsection{Total Survey of All Galaxies}

\begin{figure}
\plottwo{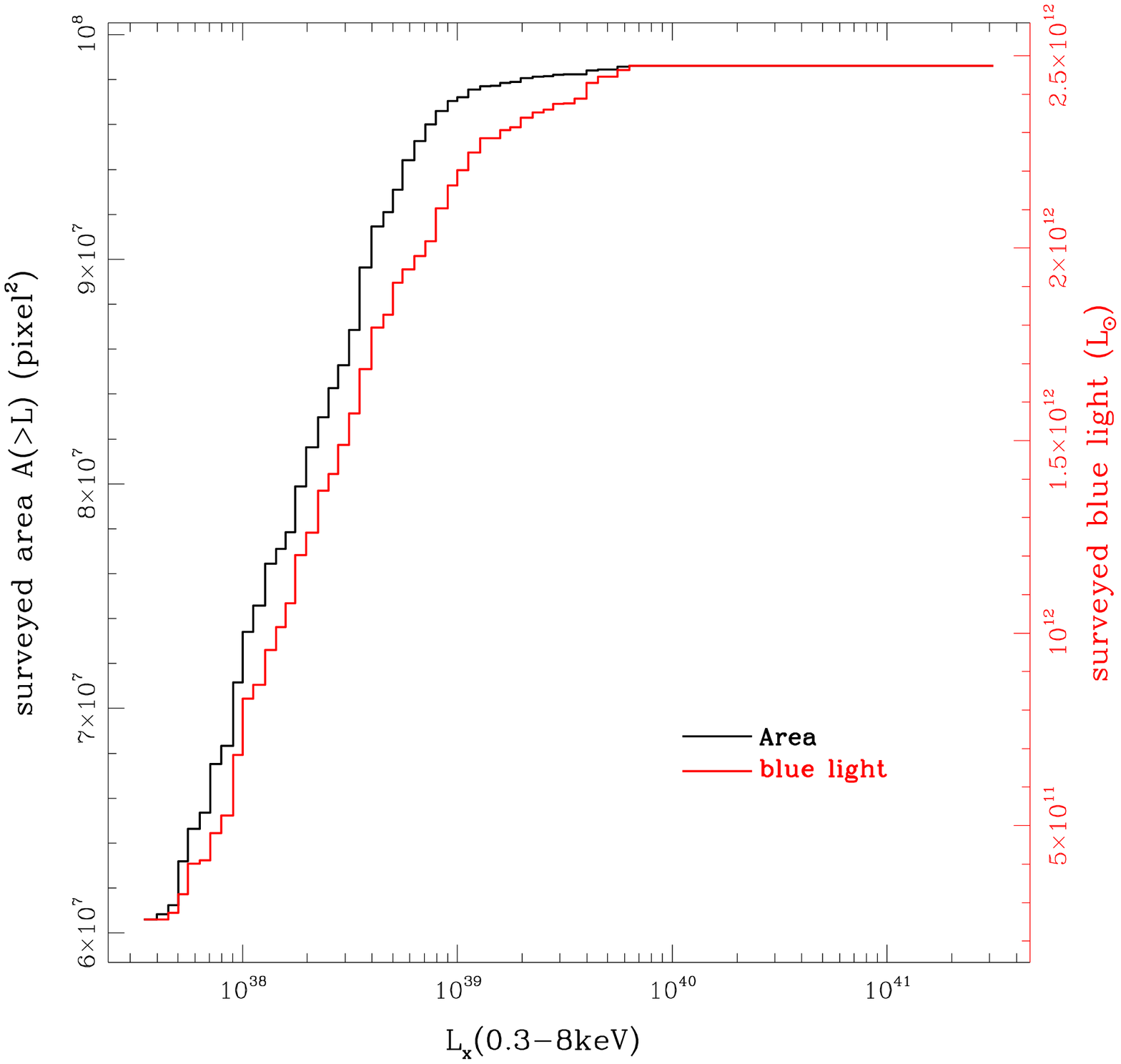}{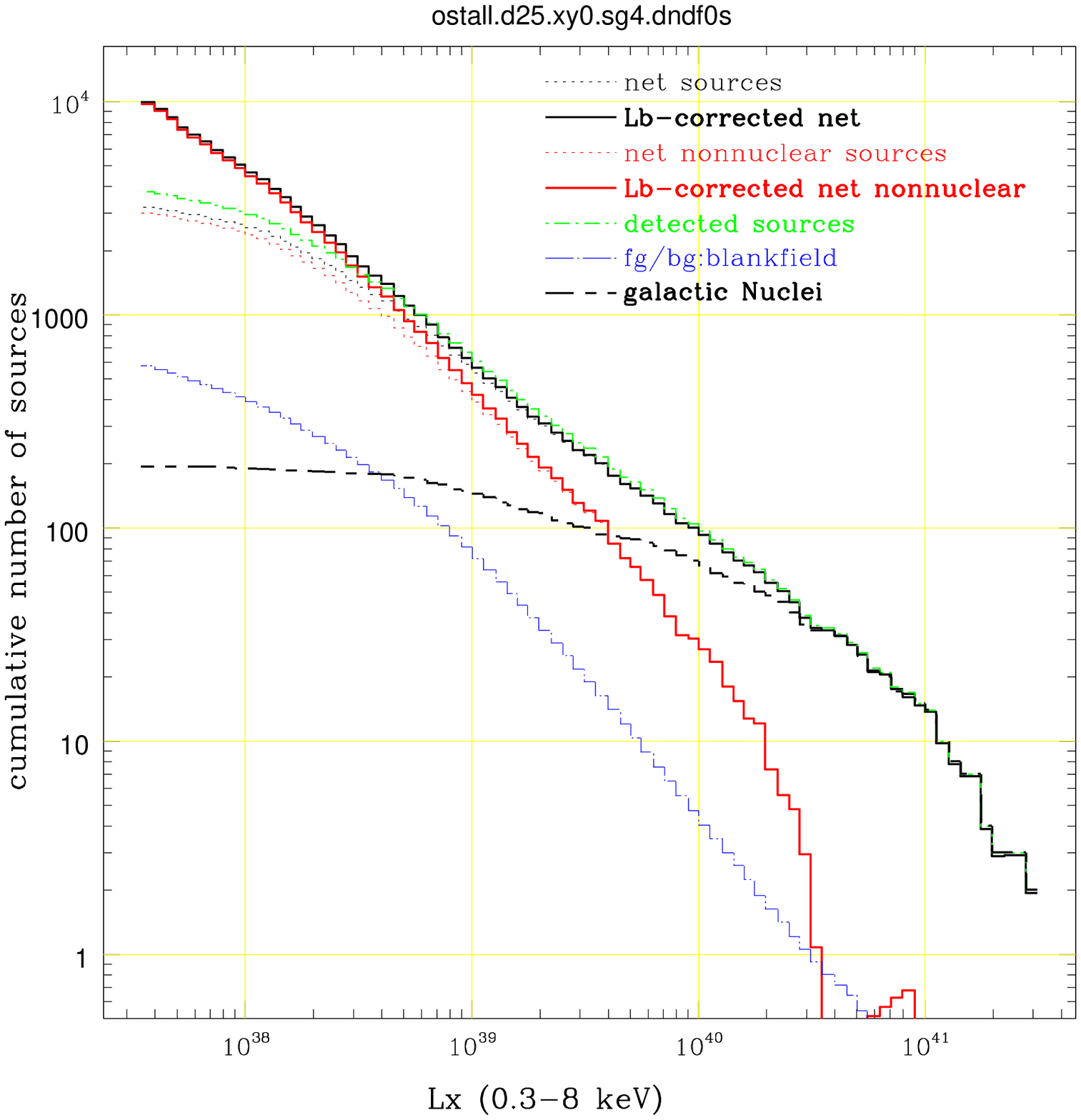}
\caption{(a) The
surveyed sky area curve $A(>L)$ (black solid) and the surveyed blue light curve
$\pounds_B(>L)$ for the stitched deep survey of all 343 galaxies. (b) Cumulative histograms for the
numbers of the detected sources, the estimated
background/foreground objects, the ($\pounds_B$-corrected) net sources, the
identified galactic nuclei, and the ($\pounds_B$-corrected) net nonnuclear
sources. }
\label{fig.7}
\end{figure}

A total survey of all 343 galaxies is constructed by summing up the stitched
deep surveys of individual galaxies.
As listed in Table~2, the combined survey detects 4970 sources above $4\sigma$
while all observations of these 343 galaxies detected 5920 X-ray point sources
(including 282 ULXs above $2\times10^{39}$ erg s$^{-1}$).
The combined XLF curves can be computed by directly summing up (from individual
galaxies in each luminosity bin): the surveyed sky area; the surveyed blue light;
the estimated numbers of background/foreground objects; the number of
detected sources; the number of identified galactic nuclei; the number of net
sources; the number of net non-nuclear sources.
The $\pounds_B$-corrected numbers of net sources and net non-nuclear sources
can then be computed based on the combined surveyed blue light curve and the
combined numbers.
As shown in Figure \ref{fig.7}a, the combined sky area coverage amounts to $\sim10^8$
ACIS pixels (1.9 square  degree) within the $D_{25}$ isophotes of 343 galaxies, and
the combined blue light amounts to $2.5\times10^{12} L_\odot$ with 40\% of the
blue light above $2\times10^{38}$ erg s$^{-1}$.
Figure \ref{fig.7}b shows the cumulative curves for the detected sources, the estimated
background/foreground objects, the ($\pounds_B$-corrected) net sources, the
identified galactic nuclei, and the ($\pounds_B$-corrected) net nonnuclear
sources.
%

The expected background/foreground objects are an increasing fraction of the
detected sources toward lower luminosities.
At luminosities above $10^{40}$/$10^{39}$/$10^{38}$/$10^{37}$ erg s$^{-1}$, there
are 4.1/71.9/390.5/803.0 background/foreground objects expected, about
4.2\%/11.9\%/13.2\%
/18.3\% of the 97/605/2955/4383 detected sources.
This fraction goes to its maximum of 23.6\% (1170.6/4970) at the lowest
luminosity of $1.2\times10^{35}$ erg s$^{-1}$ that the combined survey can detect.
Galactic nuclear sources are considered as another ``contamination'' for the
normal X-ray binary population in nearby galaxies. As shown in Figure \ref{fig.7}b, about
half of the detected sources above $4\times10^{39}$ erg s$^{-1}$ are identified as
galactic nuclei, while all detected sources above $10^{41}$ erg s$^{-1}$ are
identified as galactic nuclei.
We have listed in Table~2 the numbers of nonnuclear sources (or ULXs), of
identified galactic nuclei, and of expected background/foreground objects for
luminosities above $2\times10^{39}$/$4\times10^{39}$/$10^{40}$ erg s$^{-1}$
respectively.  Also listed are the numbers with errors of ULXs per surveyed
galaxy as an indicator for ULX frequency, and of ULXs per $10^{10} L_\odot$
surveyed blue light with error as a normalized indicator.

\begin{figure}
\plottwo{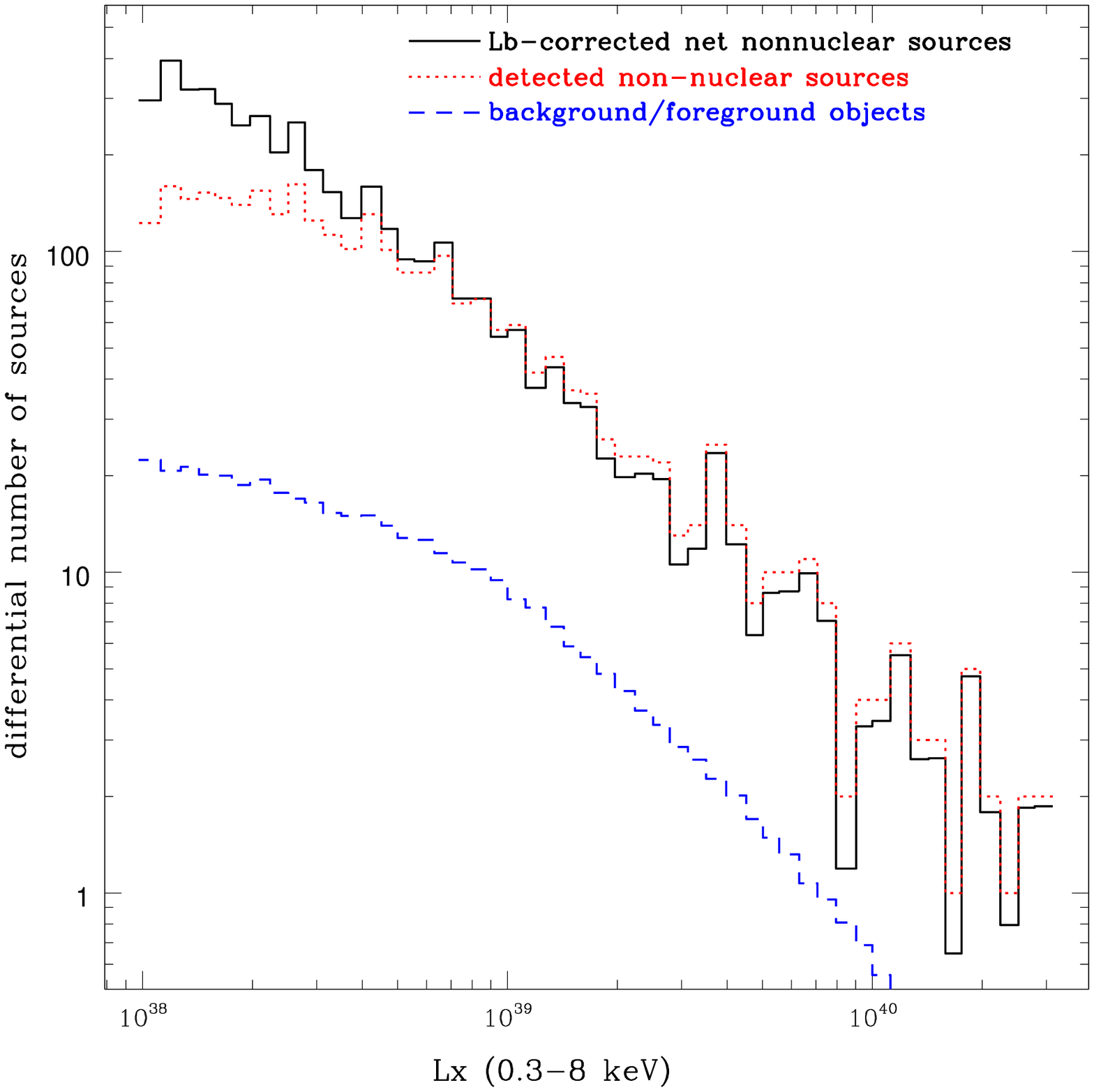}{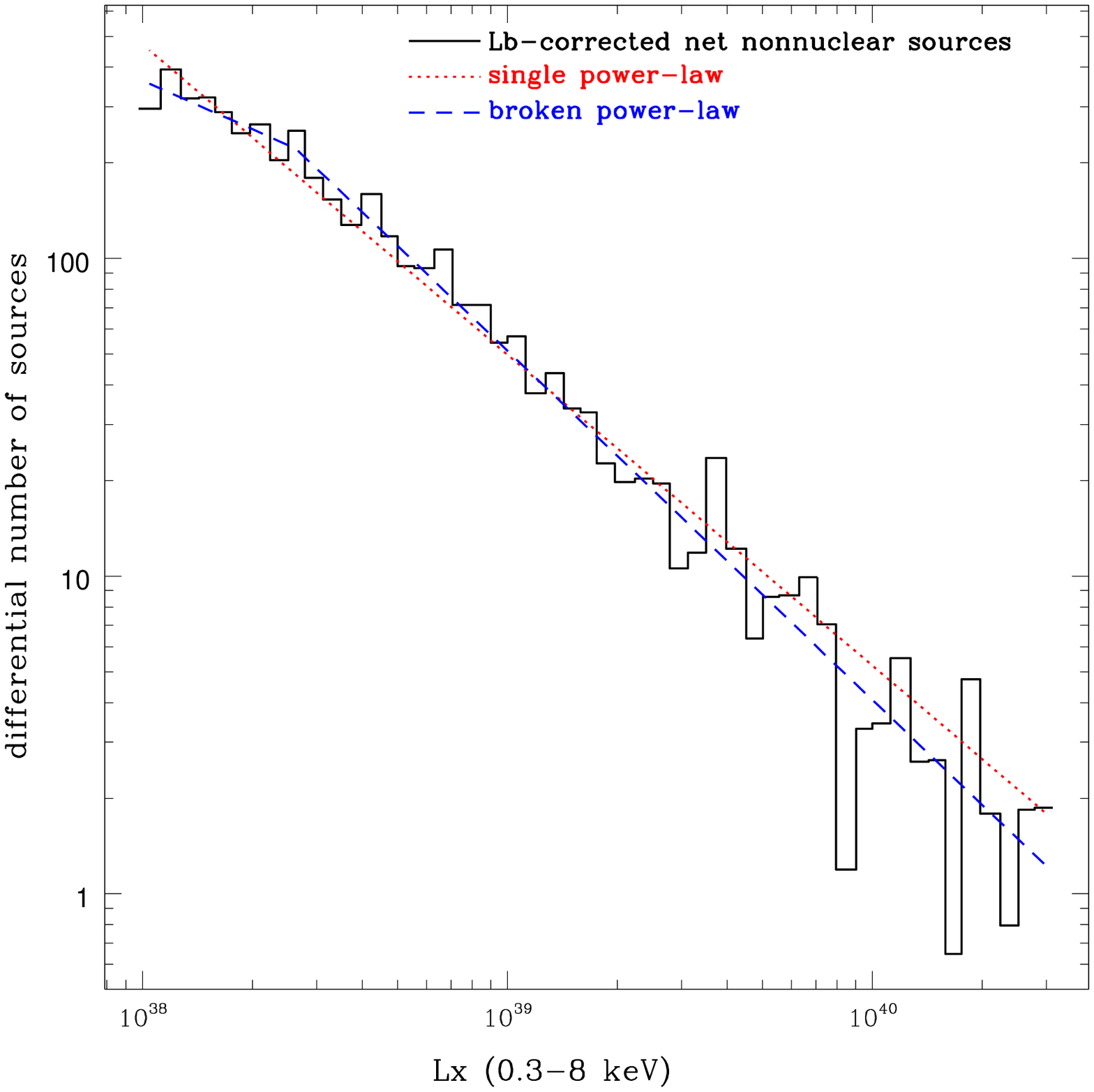}
\caption{(a) The differential curves for the numbers of detected non-nuclear
sources, predicted background/foreground objects, and the $\pounds_B$-corrected
net nonnuclear sources for the total survey of 343 galaxies. (b) The
differential curve for the $\pounds_B$-corrected net nonnuclear sources, the
single power-law fit, and the broken power-law fit. }
\label{fig.8}
\end{figure}

The normal X-ray binary population in nearby galaxies can be described by the
net nonnuclear sources.
At the low luminosity end, the net nonnuclear curve flattens due to less
surveyed blue light.
After correction by the combined surveyed blue light curve, the net nonnuclear
source curve can be described by a power-law between a few $\times10^{38}$
erg s$^{-1}$ and $\sim10^{40}$ erg s$^{-1}$.
The $\pounds_B$-corrected curve below $10^{38}$ erg s$^{-1}$ appears to have a
flatter power-law slope. This may reflect a true feature of the X-ray binary
population, or it could be caused by the insufficient correction for the
surveyed blue light.
As shown in Figure \ref{fig.8}a, the ``raw'' differential histogram of net non-nuclear
sources can be described by a power-law until it bends over with a sharp break
around $2\times10^{38}$ erg s$^{-1}$. After corrected by the surveyed blue light
curve, the curve is less bent and the break moves toward slightly higher
luminosities.
There are only two sources detected above $2\times10^{40}$ erg s$^{-1}$ as shown in
the differential histogram; this leads to a cut-off around this luminosity in
the cumulative curve in Figure \ref{fig.7}b.

Throughout the work, the differential histogram of net non-nuclear sources are
fitted with a single power-law of $dN(>L)/dL = A L_{39}^{-\alpha}$, or a broken
power-law of $dN(>L)/dL = A (L/L_{b})^{-\alpha}$ for $L\le L_{b}$ and
$dN(>L)/dL = A (L/L_{b})^{-\alpha2}$ for $L\ge L_{b}$ if the reduced
$\chi^2_\nu>1$ for the single power-law fit.
Furthermore, The $\chi^2$ test is used to determine which is statistically acceptable.
Here A is normalized by the total
surveyed blue light to enable comparison across different galaxy samples.
The parameters are obtained through minimizing $\chi^2 = \sum_i (N(L_i) - N_i)^2/e_i^2$, with
$L_{39}$ as the luminosity in unit of $10^{39}$ erg s$^{-1}$,
$L_i$ as the central luminosity of a luminosity bin, $N_i$ as the number of net
nonnuclear sources in the luminosity bin, and $e_i$ as the Poisson error for the
number of detected sources in the luminosity bin.
To reduce the effects of possible insufficient $\pounds_B$ corrections, we only
fit the luminosity range above which $\pounds_B(>L)$ exceeds 30\% of the total
surveyed blue light.
We further restrain our fit to below the luminosity above which there are only
two ULXs found.
As shown in Figure \ref{fig.8}b, the differential histogram for this total survey can be
fitted by a single power law of $A = 0.2$ and $\alpha = 0.98$ with $\chi^2/{\rm dof}$ = 86.8/49,
or fitted with a broken power law of $A = 0.21$, $\alpha=0.51$,
$\log L_b = 38.4$, $\alpha_2 = 1.1$ with $\chi^2/{\rm dof}$ = 48.8/47 (Table~3).
When the significance level is set as $\alpha = 0.05$,
the critical value of $\chi^2$ distribution for the broken power law (dof $= 47$) is 64.001, greater than 48.8;
the critical value for the single power law (dof $= 49$) is 66.339, less than 86.8.
Therefore, not only is the broken power law is a better fit, it is
statistically acceptable while the single power-law fit must be rejected.
The 1$\sigma$ statistical uncertainty is determined during the fitting.


\subsection{Galaxies of Different Types}

The above total survey uses all galaxy types known to have a mixture of
different X-ray binary populations, and the explanation of the XLF is rather
difficult.
Here we try to isolate different X-ray binary populations by studying them in
different types of galaxies. We first divide them into early-type galaxies and
late-type galaxies, then we divide them further into ellipticals and
lenticulars for early-type galaxies, and S0/a-Sa, Sab-Sb, Sbc-Sc, Scd-Sd,
Sdm-Sm for spiral galaxies, and irregular and peculiar galaxies.
For each sample of galaxies, the survey statistics and the ULX statistics are
computed as for the above total survey  and listed in Table~2.
The cumulative and differential XLF curves are computed for all galaxy samples
as for the total survey.
%
The differential curves of the net non-nuclear sources are fitted with a single
power-law, and a broken power-law if a single power-law does not fit well.
The resulted  parameters are listed for the single power law and
for the broken power law in Table~3.

\begin{figure}
\plottwo{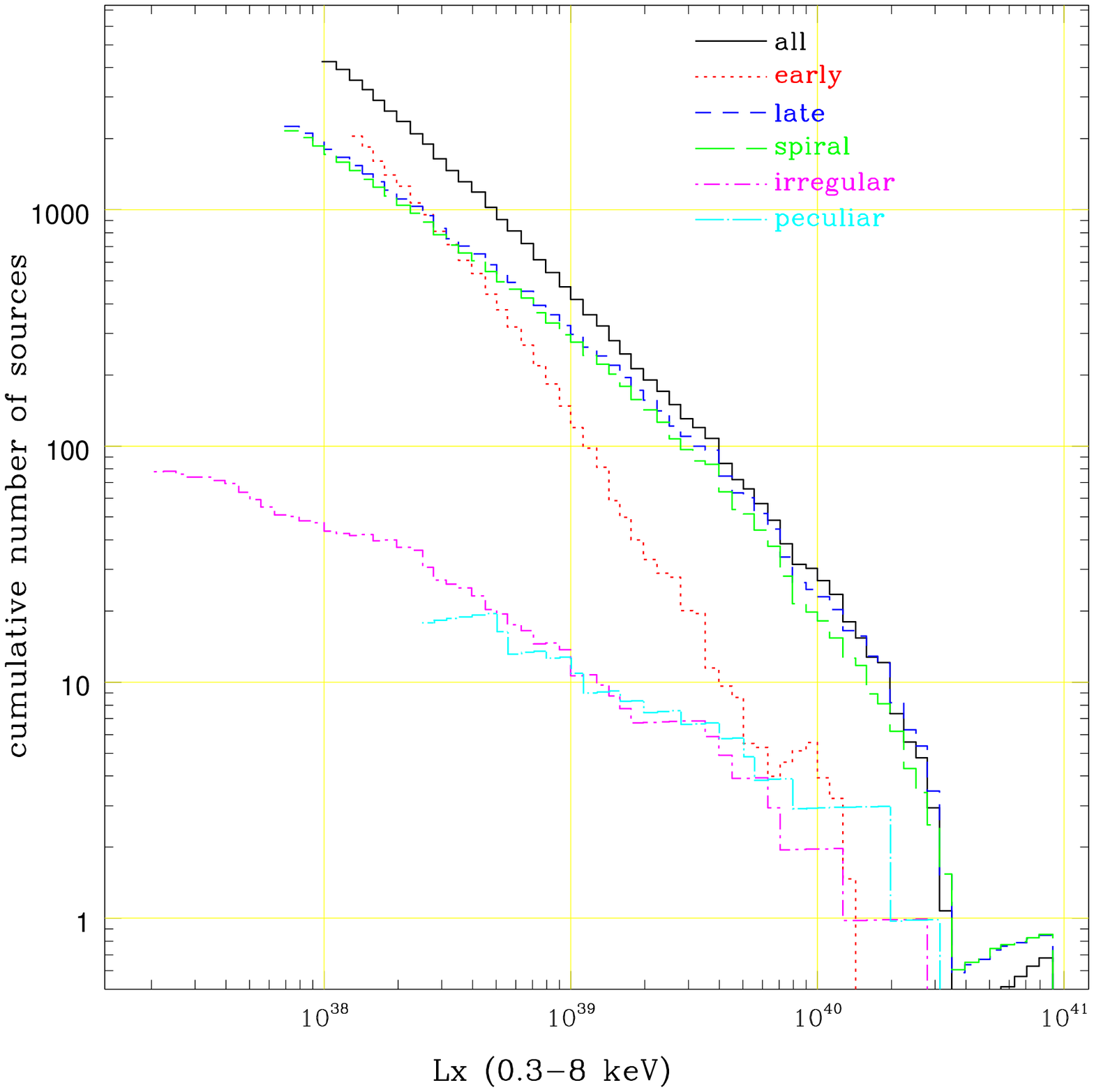}{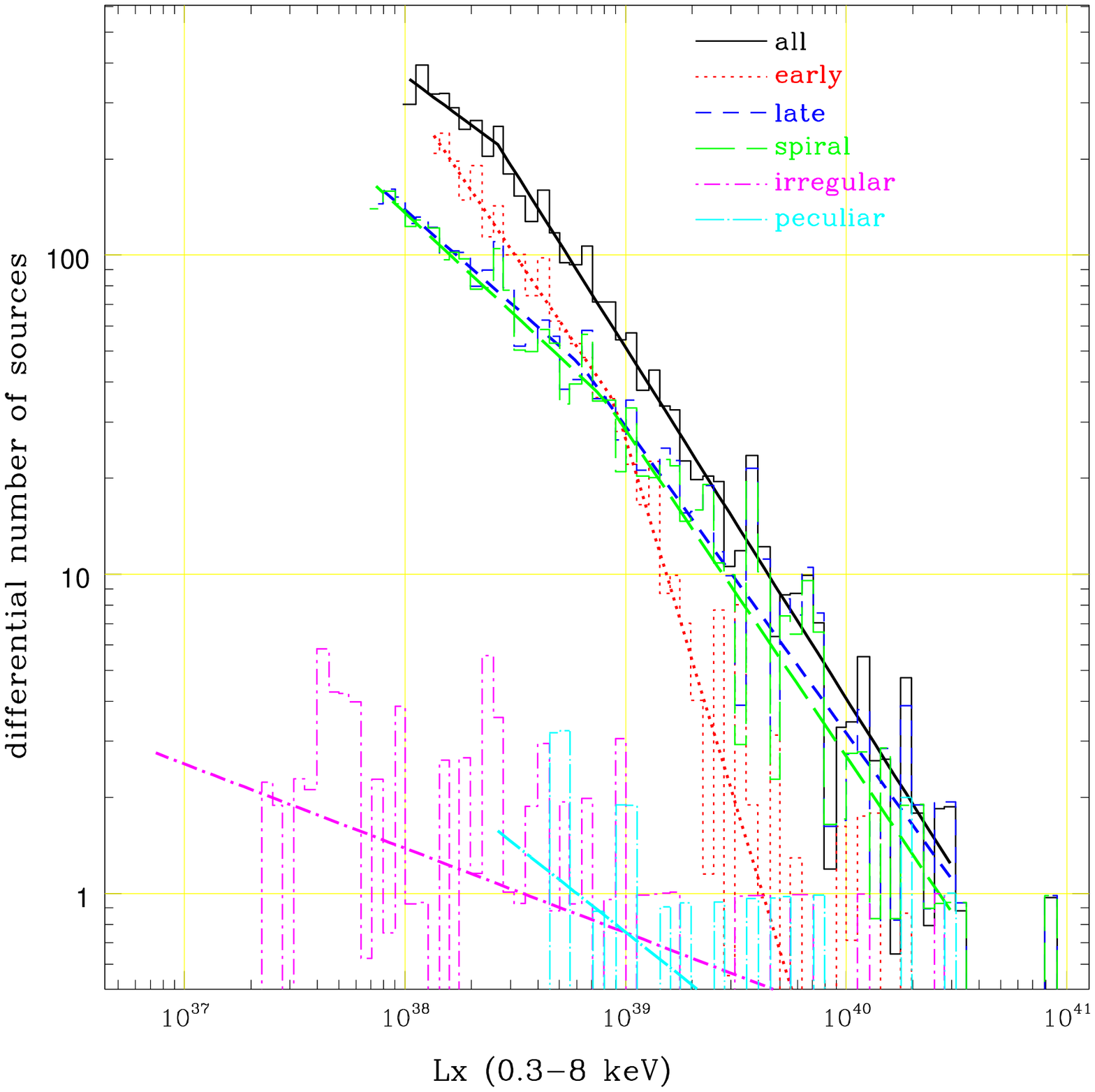}
\caption{(a) the cumulative curves for the $\pounds_B$-corrected net nonnuclear sources for samples of early-type
galaxies, late-type galaxies, spiral galaxies, irregular galaxies and peculiar galaxies. The curves are all above
the luminosity at which $\pounds_B$ is 50\% of the total surveyed blue light for the sample. (b) The differential curves
and the best-fit XLF models for the $\pounds_B$-corrected net nonnuclear sources. 
It is obvious that the XLF is much steeper for early-type galaxies than for late-type galaxies.}
\label{fig.9}
\end{figure}

The cumulative and differential net non-nuclear source curves for different
galaxy samples are overplotted to facilitate comparison between different
galaxy types.
As is obvious from Figure \ref{fig.9}a,b, the XLF is much steeper for early-type galaxies than
for late-type galaxies.
X-ray sources from late-type galaxies dominate the mixed total survey above a
few $10^{38}$ erg s$^{-1}$, while X-ray sources from early-type galaxies
dominate at lower luminosities.
The late-type galaxy sample is dominated by spiral galaxies, with very few
X-ray sources from the irregular galaxy sample and the peculiar galaxy sample.
A single power-law can fit the XLFs for irregular/peculiar galaxy samples well,
and no broken power-law are fitted given the small number of X-ray sources from
these two samples.
Despite the huge difference in the number of X-ray sources in different
samples, it is clear that the slopes for XLFs change significantly for
different galaxy types, becoming flatter and flatter in the order of early-type
galaxies, late-type galaxies, spiral galaxies, peculiars and irregulars.

\begin{figure}
\plottwo{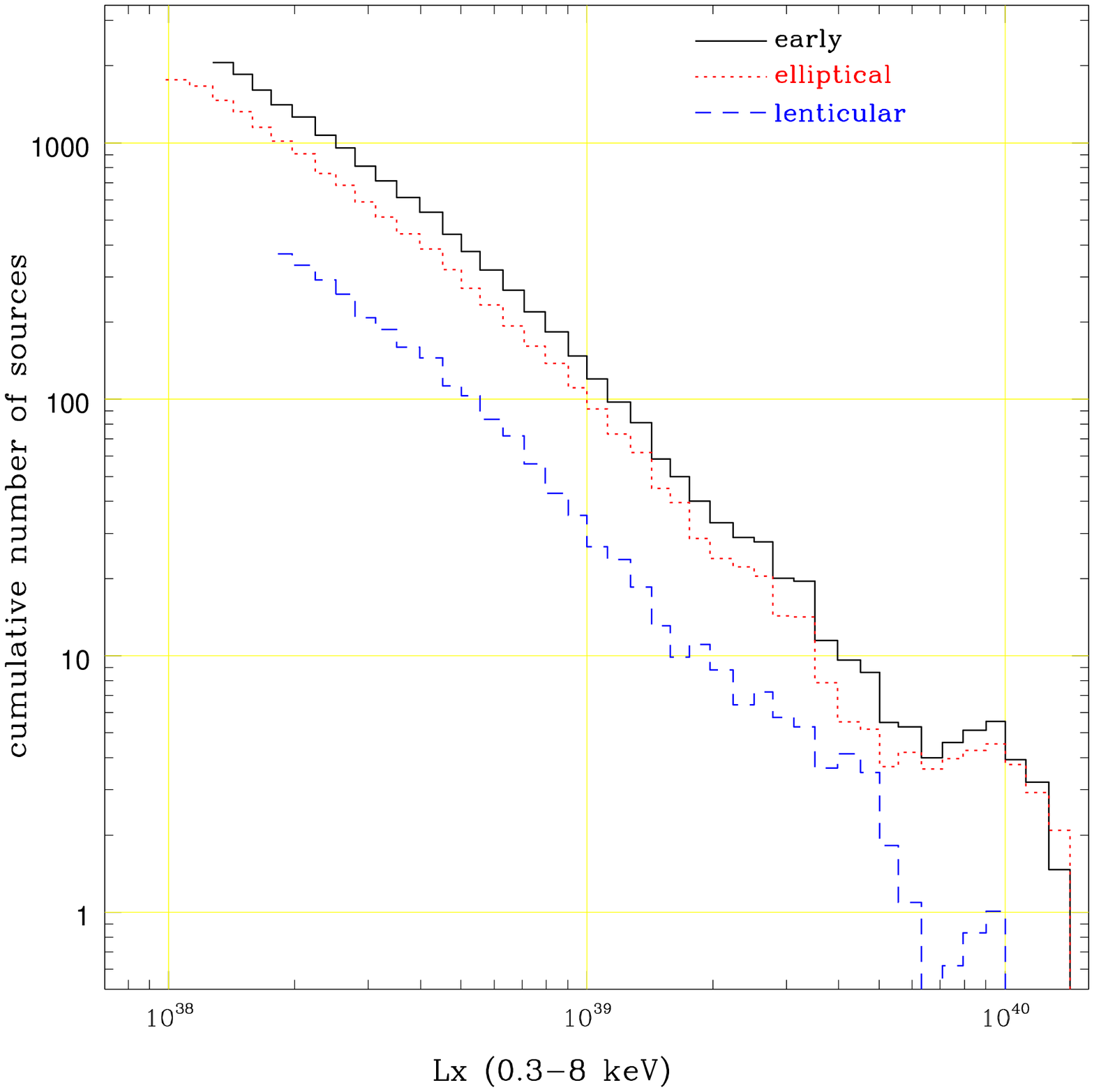}{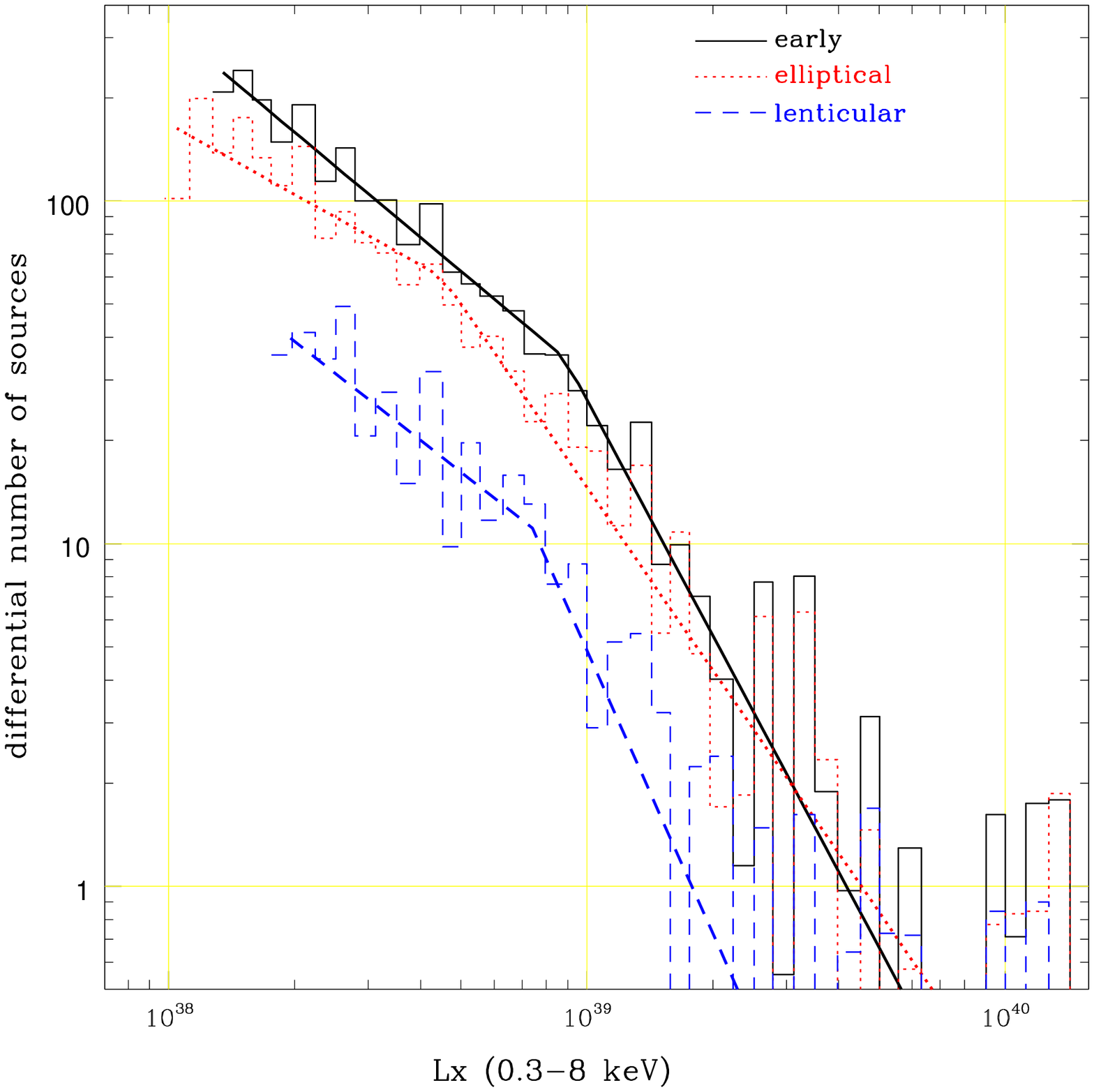}
\caption{(a) the cumulative curves for the $\pounds_B$-corrected net nonnuclear sources for samples of early-type
galaxies, elliptical galaxies and lenticular galaxies. (b) The differential curves and the best-fit XLF
models for the $\pounds_B$-corrected net nonnuclear sources. }
\label{fig.10}
\end{figure}

The cumulative and differential XLF curves for early-type and subtype (elliptical and lenticular) galaxies
are plotted in Figure \ref{fig.10}.
While earlier works suggest that ULXs in early-type galaxies are mostly
background/foreground objects \citep[e.g.,][]{Irwin2004}, this {\it Chandra} ACIS
survey does find dozens of ULXs above $2\times10^{39}$ erg s$^{-1}$ that are
significantly more than predicted by the background/foreground estimate.
Indeed, the net non-nuclear source curve extends smoothly down to a few
$10^{39}$ erg s$^{-1}$, with a possible bend down around $10^{39}$ erg s$^{-1}$.
The quantitative fitting results prove that a broken power-law fit
(as overplotted in Figure \ref{fig.10}b) is indeed better than a single power-law on a
98.5\% confidence level (F test).
Above $6\times10^{39}$ erg s$^{-1}$, there appears to be an extra population of a
dozen of ULXs in addition to the power-law fit.
Inspection of the sub-type samples shows that this extra population mainly
comes from elliptical galaxies, while the XLF curve for lenticular galaxies
show a cut-off at $6\times10^{39}$ erg s$^{-1}$.
There are 2--9 times more ULXs above $2\times10^{39}$ erg s$^{-1}$ per elliptical
galaxy than per lenticular galaxy.
When normalized by the surveyed blue light, the ULX density in the elliptical
galaxies is  1--4 times higher than that in the lenticular galaxies.
For less luminous sources, the $\pounds_B$ normalized source density in
elliptical galaxies is about 2 times higher than that in the lenticular
galaxies.

\begin{figure}
\plottwo{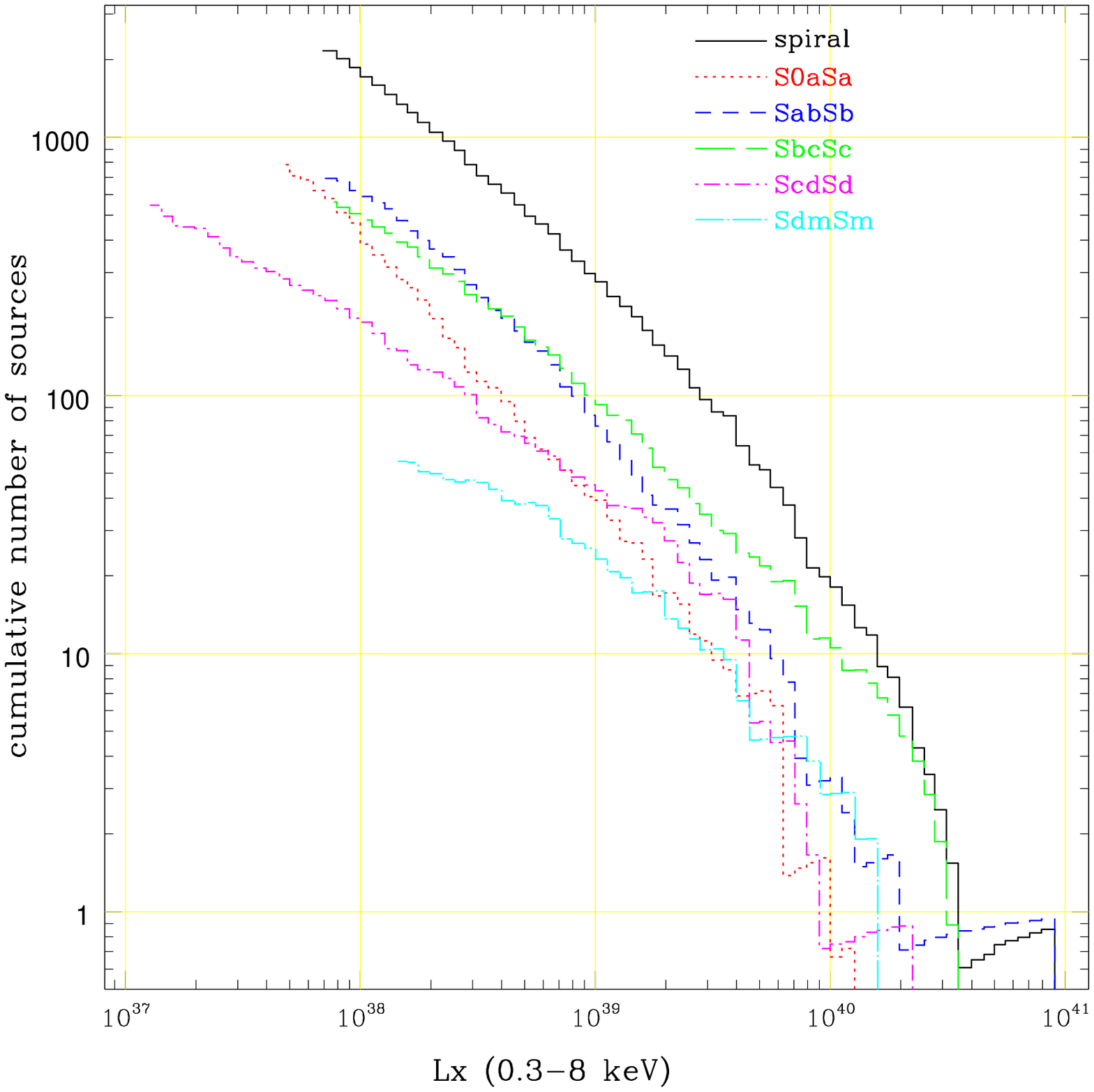}{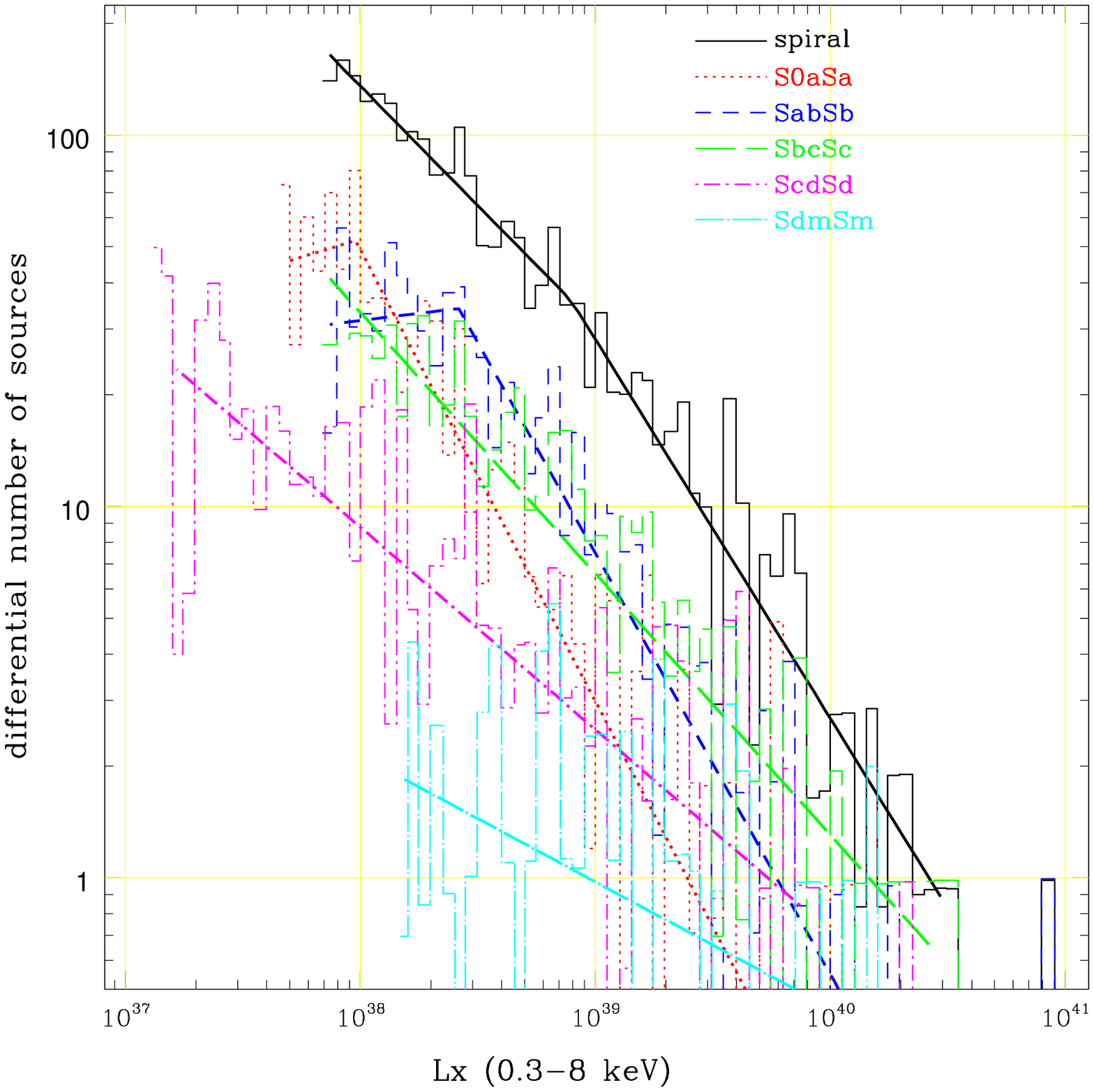}
\caption{(a) the cumulative curves for the $\pounds_B$-corrected net nonnuclear
sources for samples of spiral and subtype galaxies.  (b) The differential
curves and the best-fit XLF models for the $\pounds_B$-corrected net nonnuclear sources. }
\label{fig.11}
\end{figure}

The XLF curves for spiral and subtype galaxies are plotted in Figure \ref{fig.11}.
The net non-nuclear source curve for the spiral galaxy sample extends down
smoothly with a gradual bend down until it cuts off at $3\times10^{40}$ erg s$^{-1}$.
Quantitative fitting to the differential curve shows that a broken power-law (as
overplotted in Figure \ref{fig.11}b) is slightly better than a single power-law on a 69\%
confidence level.
For the subtypes, a broken power law fit is a significant improvement than a
single power-law for Sab-Sb galaxies, while only a slight improvement for
S0/a-Sa galaxies. However, it is not clear whether the power-law
break is caused by the insufficient $\pounds_B$ correction.
A single power law is a good fit for Sbc-Sc, Scd-Sd and Sdm-Sm galaxies, and no broken
power law fits are tried due to the small number of X-ray sources for the latter two
subtypes.
For $L_X>10^{39}$ erg s$^{-1}$, the XLF slopes are similar for S0/a to Sc galaxies, but
becomes flatter as the subtype goes later as shown in Figure \ref{fig.11}b.
Not surprisingly, there are more ULXs in late-type galaxies than in early-type
galaxies.
With better statistics from our large sample, we find that the number
of ULXs per late-type galaxy is 2--4 times higher than the number per early-type
galaxy for ULXs above  $2\times10^{39}$ erg s$^{-1}$, and is 4--10 times higher for
ULXs above  $4\times10^{39}$ erg s$^{-1}$.
The number of ULXs per surveyed galaxy peaks at Sbc-Sc galaxies, and the number of
ULXs per $10^{10}L_\odot$ surveyed blue light peaks at later galaxy types,
consistent with previous findings \citep{Liu2006, Swartz2008}.

\begin{figure}
\plottwo{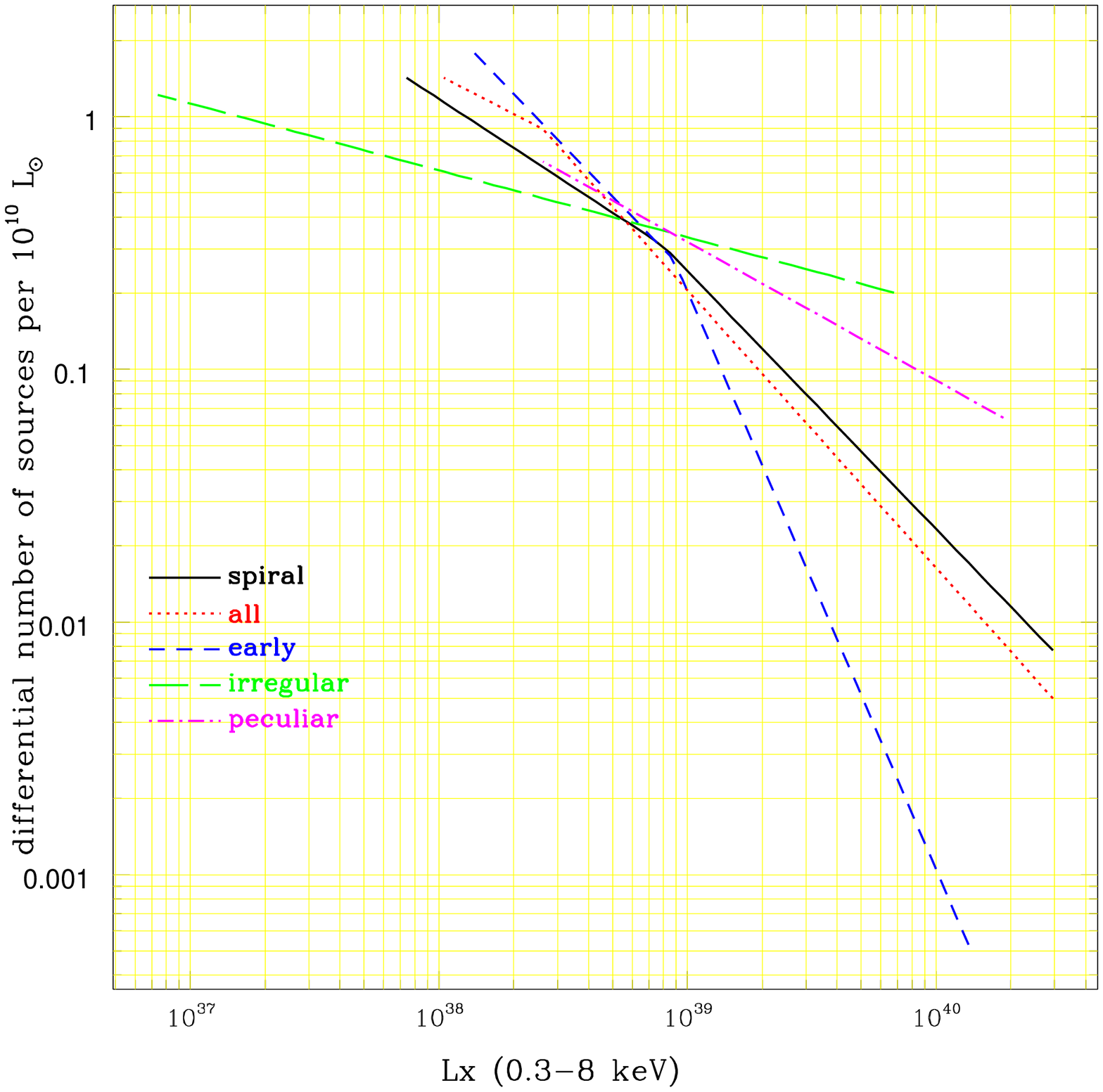}{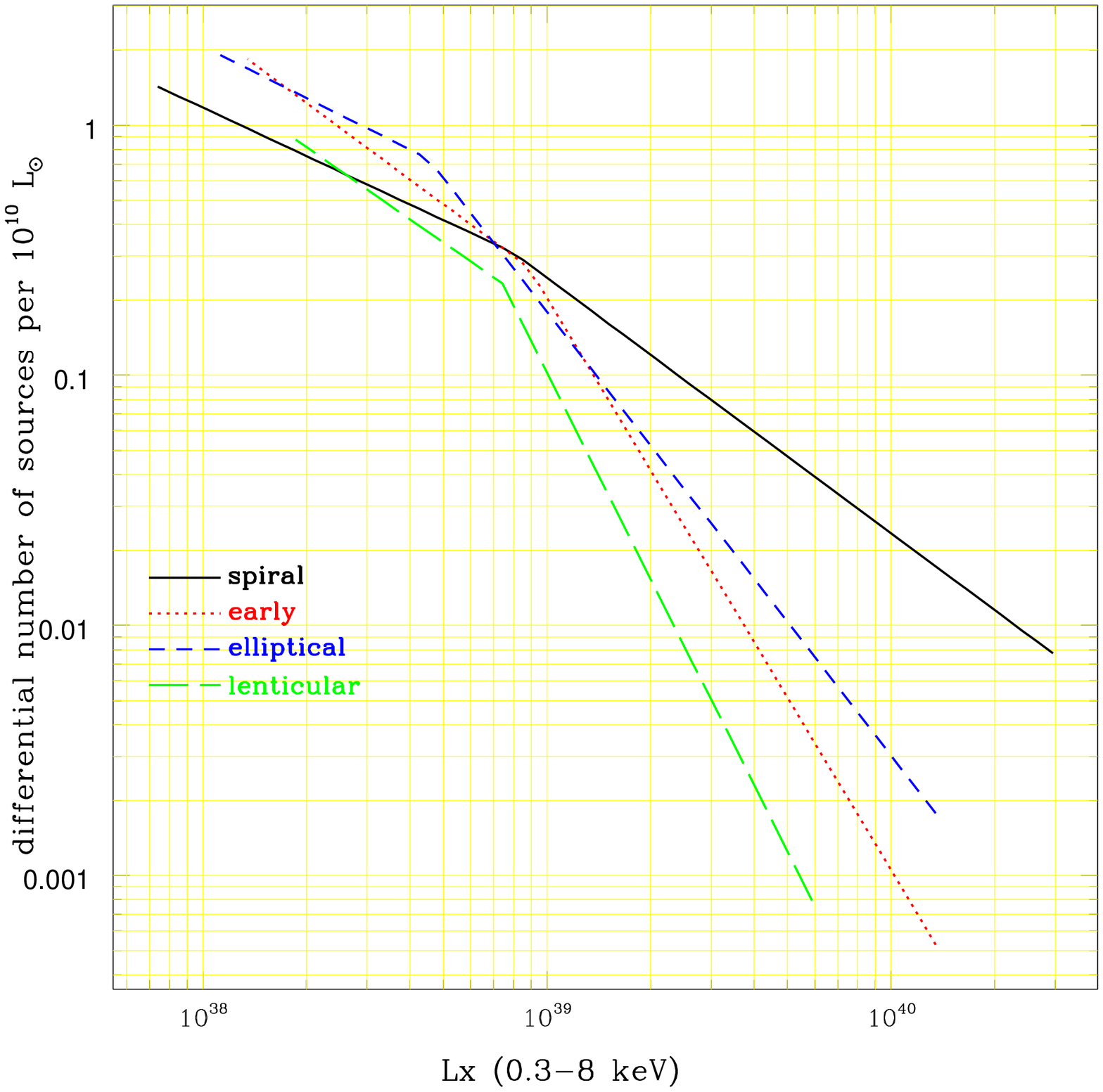}
\plottwo{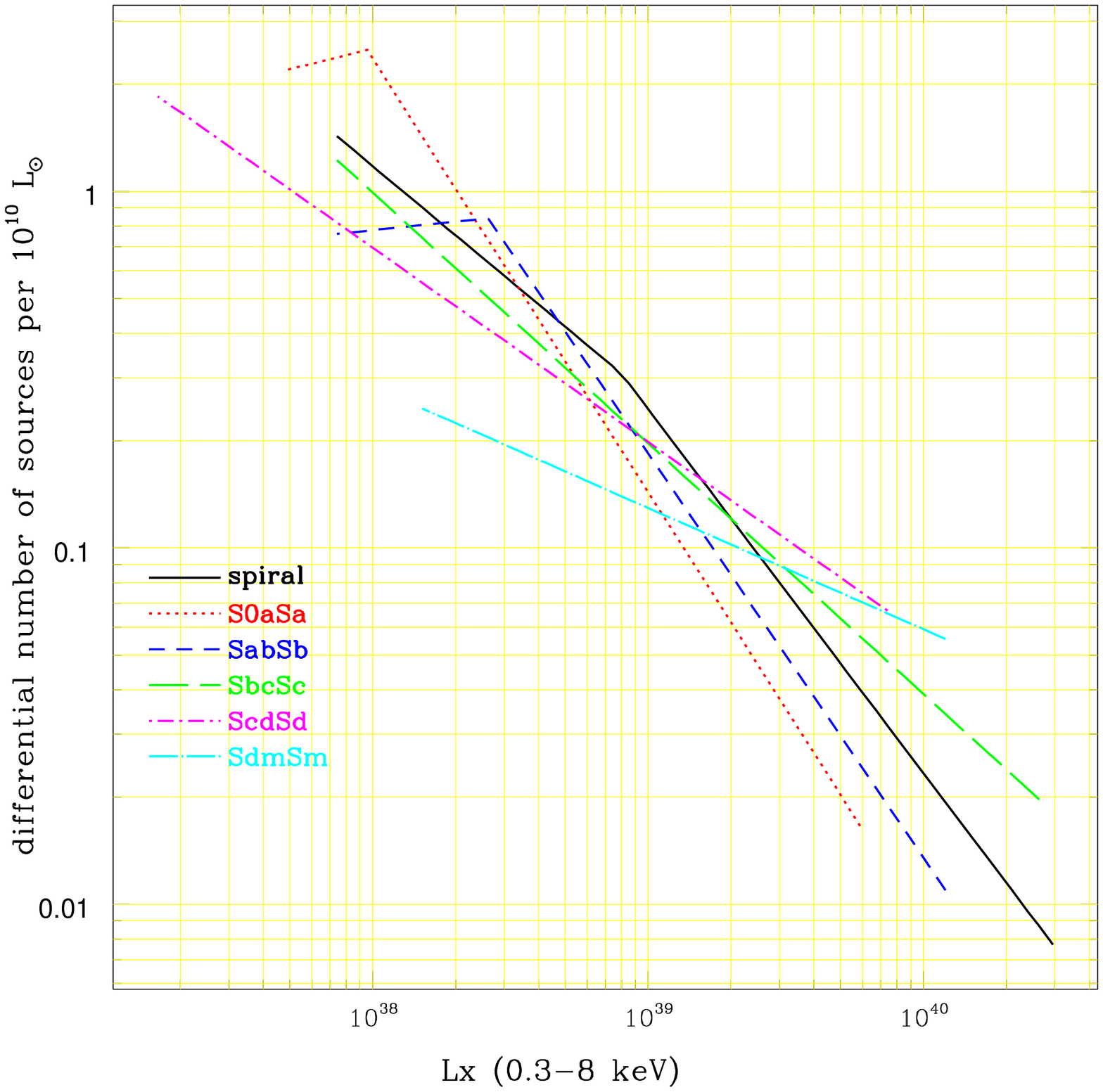}{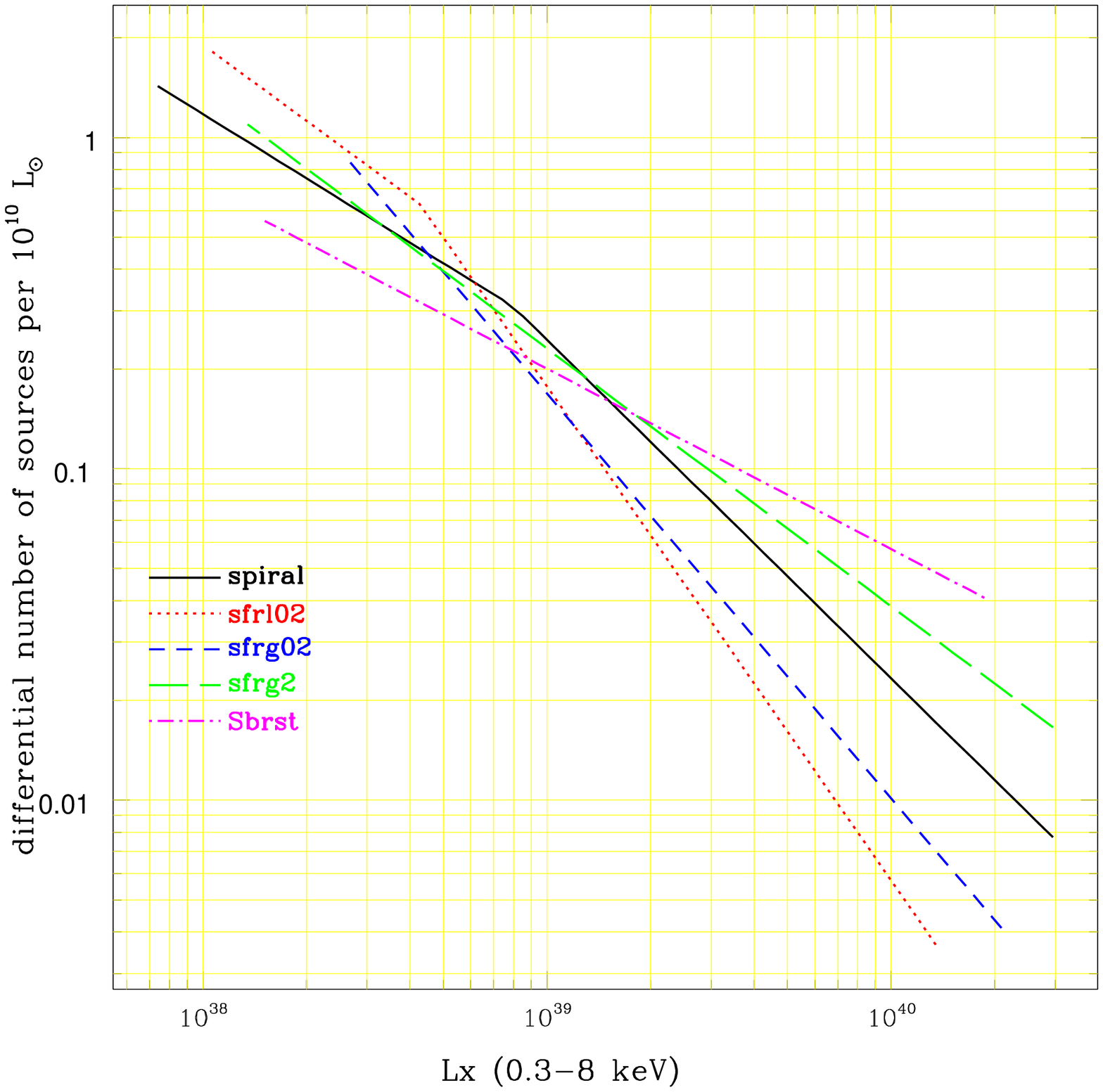}
\caption{The $\pounds_B$ normalized power-law fits to the net nonnuclear
sources in different galaxy samples. (a) Fits for all, early, spiral, peculiar,
and irregular galaxies. (b) Fits for early and subtype galaxies. (c) Fits for
spiral and subtype galaxies. (d) Fits for spiral, starburst galaxies and
galaxies with different SFRs. The same black solid line is used
for the power-law fit for spiral in all four panels. }
\label{fig.12}
\end{figure}

\begin{figure}
\plottwo{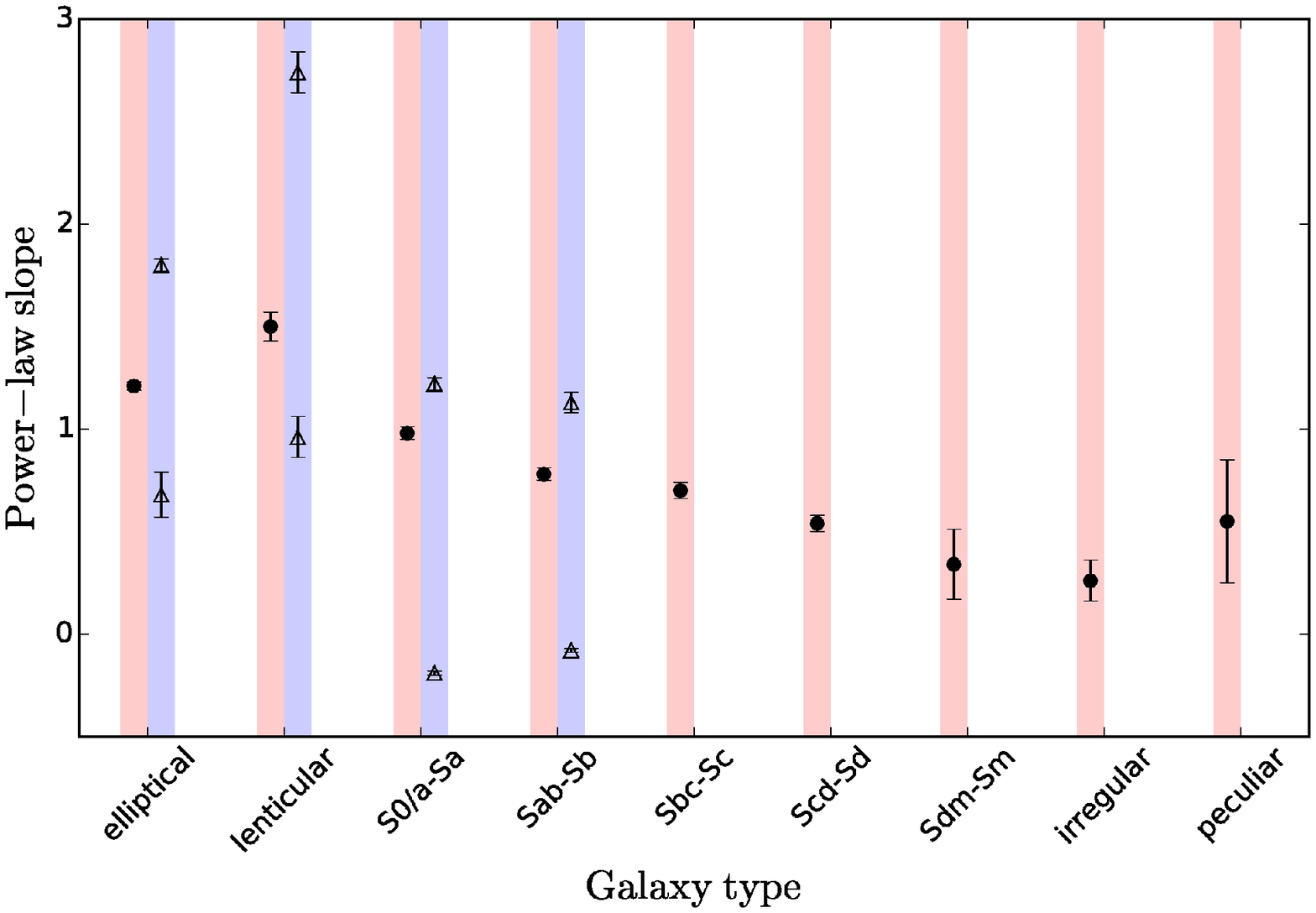}{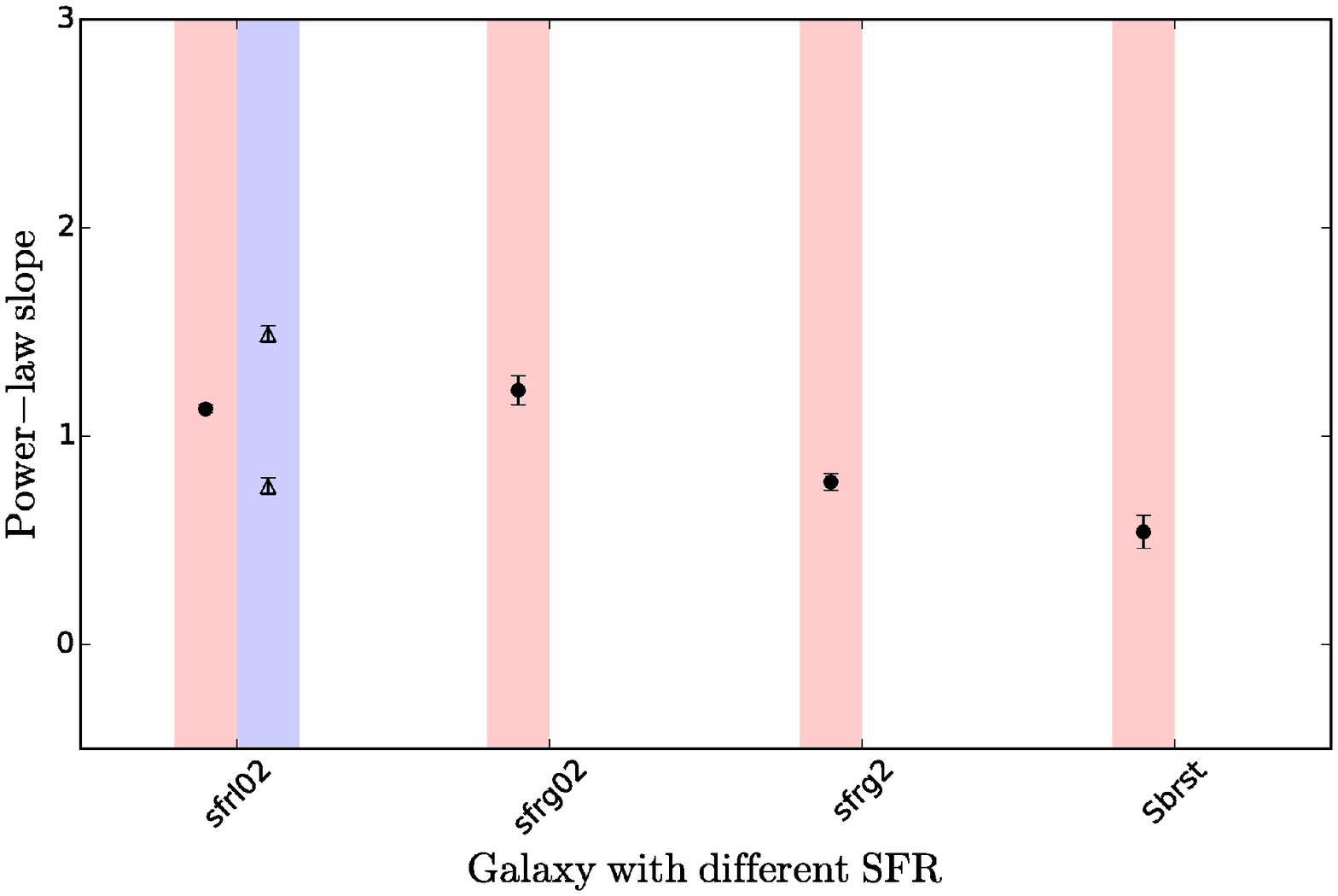}
\caption{(a) The fitted XLF slopes for galaxies with different types.
The solid circles in the red bar shows the results from single power fits,
while the open triangles in the blue bar shows the results from broken power fits.
It is obvious that the XLF slopes become flatter from early-type to late-type galaxies.
(b) The fitted XLF slopes for galaxies with different SFRs.
The solid circles in the red bar shows the results from single power fits,
while the open triangles in the blue bar shows the results from broken power fits.
It also shows the flatting trend of the XLF slopes with increasing SFRs.}
\label{fig.gtype}
\end{figure}

The X-ray source density per surveyed stellar mass is compared among different
host galaxy types.
For ULXs above $2\times10^{39}$ erg s$^{-1}$, the cumulative source density ($U/L_B$
as listed in Table~2) is the highest for peculiars and irregulars, and
decreases along the sequence of spirals, ellipticals and lenticulars.
The highest cumulative source densities in peculiars and irregulars clearly indicate
a link to their high SFR and young stellar populations.
This cumulative source density is not always in agreement with the differential
source density, because it also depends on the brightest source detected in the galaxy sample.
To estimate the differential source density, we normalize the
best-fit XLF models by the surveyed blue light, overplotted in
Figure \ref{fig.12}.
At luminosities around $2\times10^{39}$ erg s$^{-1}$, the differential source density
in irregular galaxies is the highest, which is 1.3 (2.4, 5.6, 20) times higher
than in peculiar (spiral, elliptical, lenticular) galaxies.
The trend is reversed at luminosities around $2\times10^{38}$ erg s$^{-1}$, and the
differential source density in irregular galaxies becomes the lowest, while it
is 1.37 (1.45, 1.55, 2.47) times higher in spiral (peculiar, lenticular,
elliptical) galaxies.
For spiral subtypes, the differential source density is lowest in Sdm-Sm
galaxies at $2\times10^{38}$ erg s$^{-1}$, and increases for earlier and earlier
subtypes.
The trend is reversed at luminosities around $2\times10^{39}$ erg s$^{-1}$, except
for that the source density in Sdm-Sm galaxies is lower than in Scd-Sd
galaxies.
However, the source density in Sdm-Sm galaxies does exceed that in Scd-Sd
galaxies at luminosities above $7\times10^{39}$ erg s$^{-1}$.
Note that the levels of significance for these differences are low due to the large uncertainties.
Figure \ref{fig.gtype}a displays the fitted XLF slopes in a quantitative way, which shows
the flatting trend from early-type to late-type galaxies.

\subsection{Galaxies with Different SFRs}

The correlation between the XLF slope (also ULX density) and the galaxy type
sequence suggests that X-ray source populations at high luminosities are linked
to star formation activities \citep{Ranalli2003,Grimm2003,Gilfanov2004a,Mineo2012},
which are believed to be more prominent for later galaxy types.
To quantify the relation between star formation and the ULX phenomenon, we
group the galaxies based on their SFRs.
The SFR is calculated with ${\rm SFR}(M_\odot/{\rm yr}) = 4.5\times10^{-44}\times L$
(60$~\mu$m) \citep{Rosa-Gonzalez2002}. The flux
densities at 60 $\mu$m are taken from the IRAS point source catalog (IPAC,
1986), with some nearby galaxies from \citet{Rice1988}. For galaxies that are
not detected, the $3\sigma$ upper limit is calculated by adopting noise levels
of 8.5 mJy/arcmin$^2$ \citep{Rice1988}.
The calculated rates are compared to the compilation of \citet{Grimm2003} for
11 galaxies, and are consistent with their rates within 50\% without systematic
biases.
We put the galaxies into three groups based on the SFR
from 60 $\mu$m emission, {\tt sfrl02} for 208 galaxies with SFR upper limits
or SFR $<$ 0.2 $M_\odot$/yr, {\tt sfrg02} for 90 galaxies with SFR $>$ 0.2
$M_\odot$/yr but $<$ 2 $M_\odot$/yr and an average of 0.76 $M_\odot$/yr, and
{\tt sfrg2} for 45 galaxies with SFR $>$ 2 $M_\odot$/yr and an average of
6.6 $M_\odot$/yr.
Note that a galaxy in {\tt sfrg2} does not necessarily have a truly high star
formation rate because the 60 $\mu$m emission may come from the AGN instead of
the dust associated with star formation; indeed, 27 out of 45 galaxies in this
group host AGNs.
To construct a sample of galaxies with truly high SFRs, we
build a group {\tt Sbrst} of 15 galaxies identified as starburst galaxies in
NED with an average SFR of 8.5 $M_\odot$/yr.

\begin{figure}
\plottwo{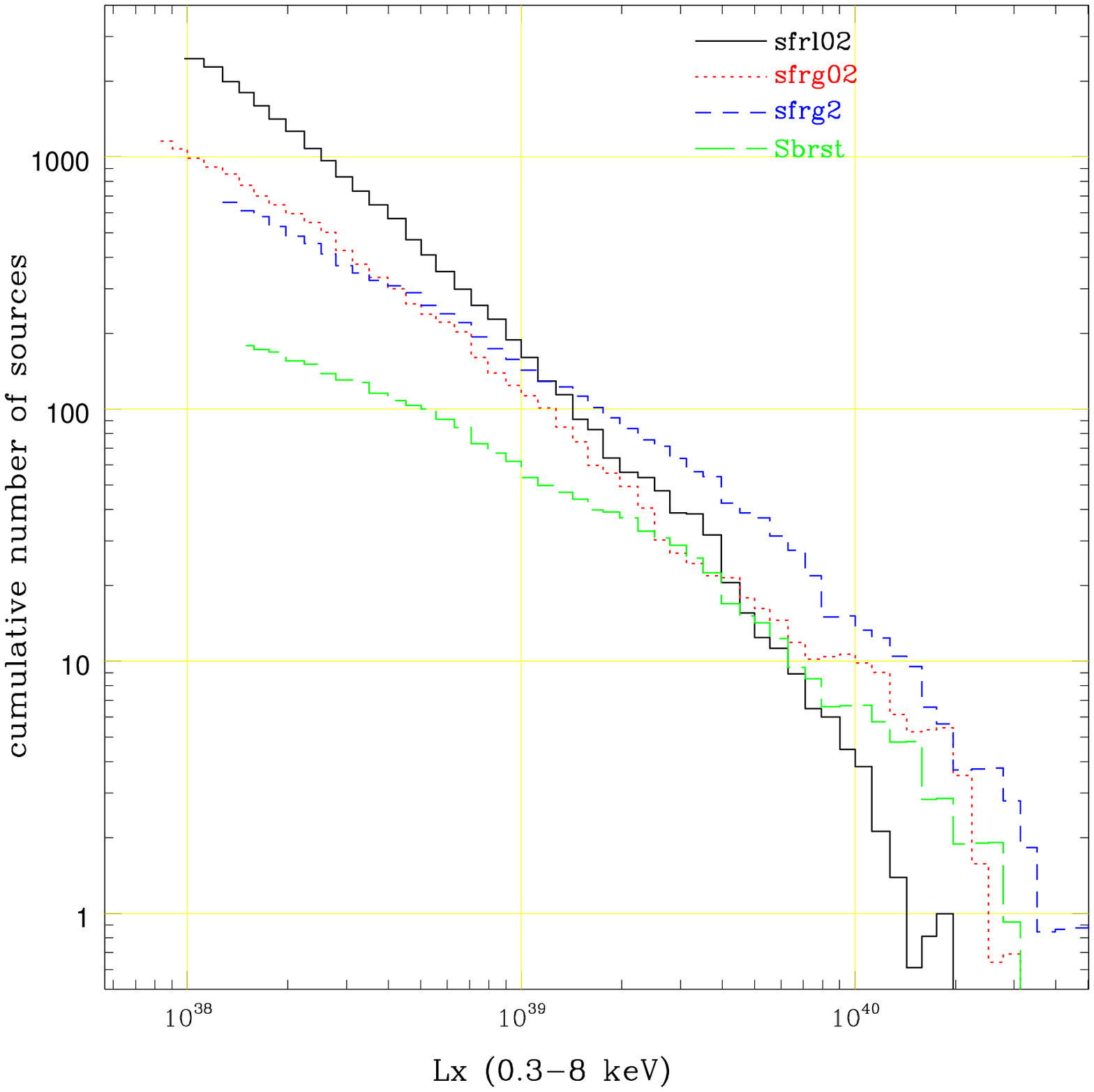}{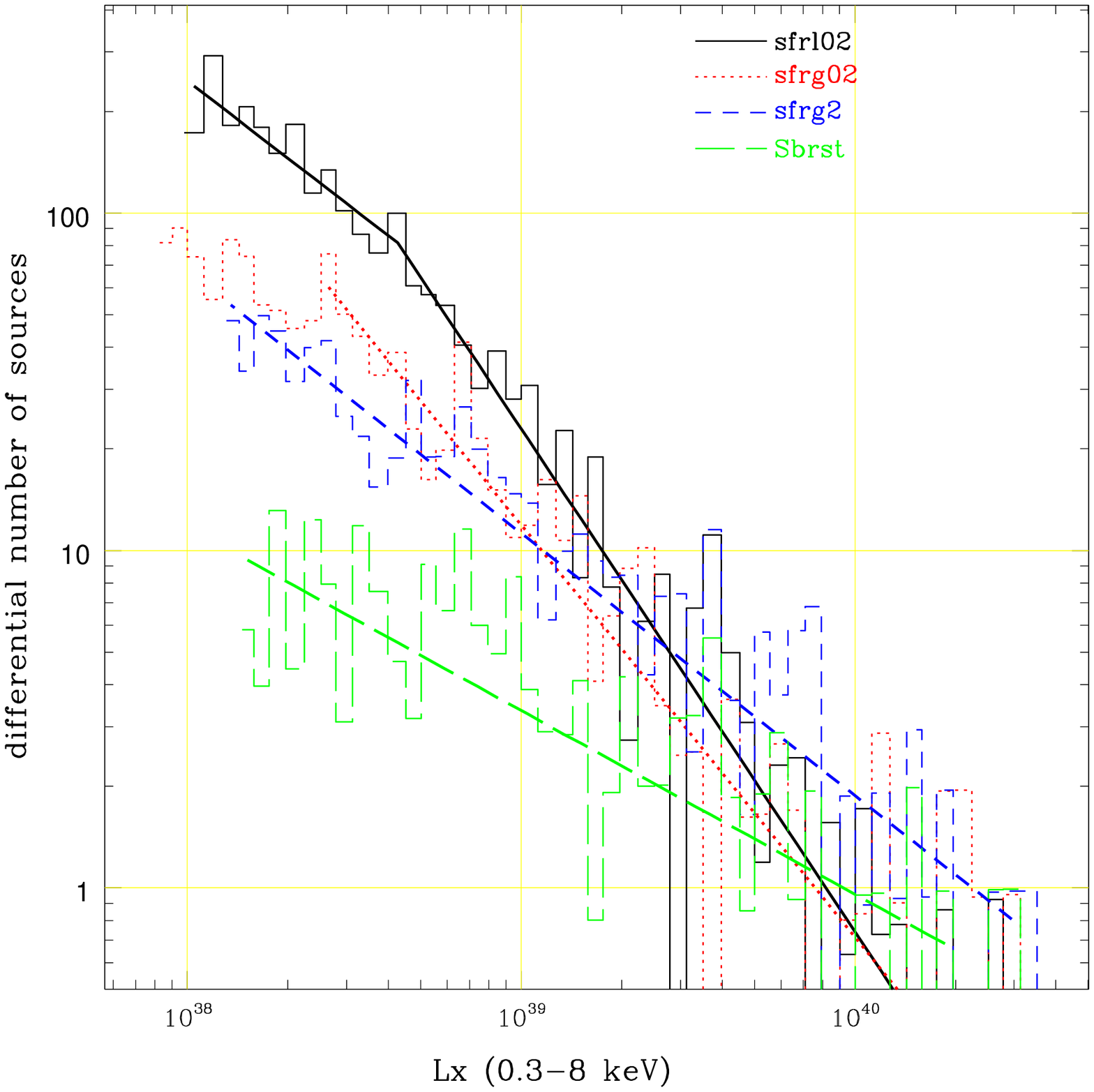}
\caption{(a) The cumulative curves for the $\pounds_B$-corrected net nonnuclear
sources for galaxies with different SFRs.  (b) The differential
curves and the best-fit XLF models for the $\pounds_B$-corrected net nonnuclear sources. }
\label{fig.13}
\end{figure}

The cumulative and differential XLF curves are plotted in Figure \ref{fig.13} for galaxy
samples with different SFRs.
For galaxy samples with gradually higher SFRs, the luminosity
of the most luminous sources becomes higher, and the XLF slope becomes
flatter.
For the {\tt sfrl02} sample, a broken power-law with a break around
$4\times10^{38}$ erg s$^{-1}$ fits the differential XLF better than a single power-law
at a 90\% confidence level.  For all other three samples with higher star
formation rates, the differential XLF curves can be well fitted with a single
power-law with $\chi^2_\nu < 1$, with the slope decreasing from $\alpha\sim1.2$
to $\alpha\sim0.5$.
As shown in Figure \ref{fig.12}d, the $\pounds_B$-normalized XLFs reveal a higher source
density above a few $10^{39}$ erg s$^{-1}$ for samples with gradually higher
SFRs; the trend gets reversed at luminosities of a few
$10^{38}$ erg s$^{-1}$ for all except for the {\tt sfrg2} sample which lacks
sources below $10^{39}$ erg s$^{-1}$.
As listed in Table~2, the cumulative ULX rates become larger for samples with
higher SFRs, albeit not in a linear way.
For example, the SFR for {\tt Sbrst} is 11 times higher than
for {\tt sfrg02}, but the number of ULXs above $2\times10^{39}$ erg s$^{-1}$ per
surveyed galaxy is 4.5 times higher, and the $\pounds_B$-normalized ULX rate is
3.2 times higher.
Figure \ref{fig.gtype}b displays the fitted XLF slopes for galaxies with different SFRs,
which clearly shows the flatting trend with increasing SFRs.

\section{DISCUSSION}

Some often-debated questions of XLFs for nearby galaxies could be addressed,
with the large number of surveyed galaxies and resulting X-ray point sources,
combined with the uniform procedures devised for this survey.

\subsection{XLF Breaks}
\label{sec.break}

Different break luminosity may reflect the differences in the binary formation mechanisms \citep{Voss2009}, the binary
configurations and consequently the accretion rates of the X-ray binaries in
different types of galaxies.
Some studies of X-ray point sources in early-type galaxies have identified an
XLF shape that required two power-laws with a break around $3\times10^{38}$
erg s$^{-1}$ \citep{Sarazin2000}. However, such a break may result from biases
affecting the detection threshold of the data \citep{KF2003}.
\citet{KF2004} analyzed 14 early-type galaxies with varying
sizes of point source samples and found that, although XLFs of individual
galaxies may not require a broken power law, the combined differential XLF
shows a statistically significant break around $L_b=(5\pm1.6)\times10^{38}$ erg s$^{-1}$
with a slope of $\alpha=1.8\pm0.2$ below $L_b$ and $\alpha=2.8\pm0.6$ above $L_b$.
Using a {\it Chandra} survey of LMXBs in 24 early-type galaxies,
\citet{Humphrey2008} acquired a best-fit power law for the composite XLF,
with a break at ($2.21^{+0.65}_{-0.56}$)$\times10^{38}$ erg s$^{-1}$,
and slopes being $1.40^{+0.10}_{-0.13}$ and $2.84^{+0.39}_{-0.30}$ below and above it.
It is suggested that the lower end of the XLF is mainly populated by neutron
star X-ray binaries \citep[e.g., ultracompact binaries that form predominantly in
star clusters;][]{Bildsten2004},
while the upper end of the XLF is
mainly populated by black hole X-ray binaries \citep{Ivanova2006}.
The break would correspond to the Eddington luminosity of massive neutron stars;
if always true for individual galaxies, locating this break in the XLF for an
early-type galaxy may allow us to determine the distance of the galaxy.

This survey confirms that the break is present in the composite XLFs for both
early-type galaxies and late-type galaxies \citep[e.g.,][]{KF2004, Humphrey2008}.
The composite differential XLF for 130 early-type galaxies shows a significant
break around $L_b=(8.9\pm0.2)\times10^{38}$ erg s$^{-1}$ (or $4.6~L_{\rm Edd}$
of a 1.5 $M_\odot$ neutron star) with a slope of $\alpha=1.0\pm0.03$ below $L_b$
and $\alpha=2.3\pm0.2$ above $L_b$.
The break is also present in the XLFs for the sample of 55 elliptical galaxies
at a break luminosity of $L_b=(4.5\pm0.2)\times10^{38}$ erg s$^{-1}$ and for the sample of 75
lenticular galaxies at a break luminosity of $L_b=(7.4\pm0.3)\times10^{38}$ erg s$^{-1}$.
The composite XLF for early-type galaxies in this survey are flatter than those
in \citet{KF2004} and \citet{Humphrey2008}, because we have excluded the background/foreground
contaminating sources that are present in larger numbers at lower luminosities.

\begin{figure}
\plotone{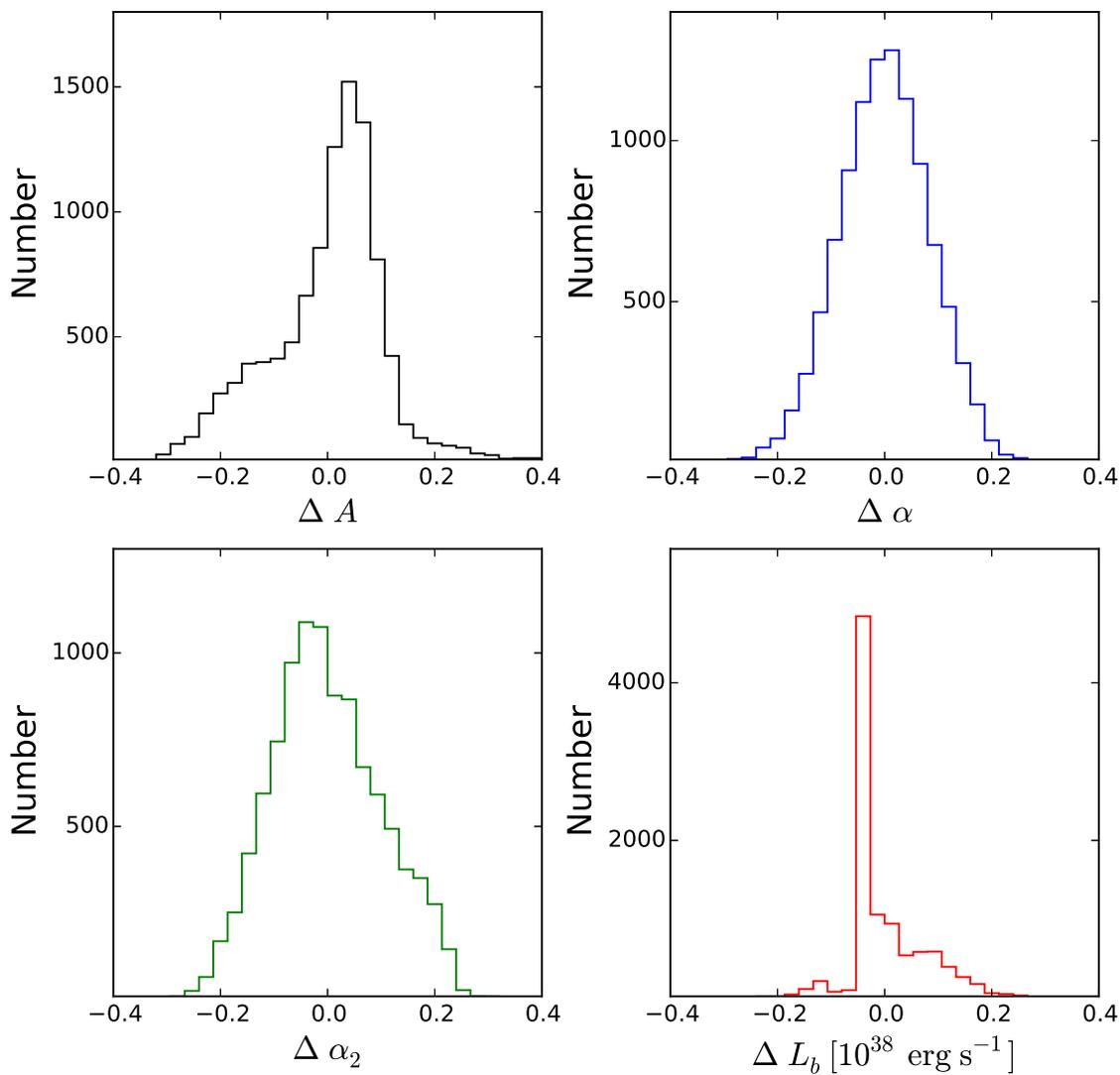}
\caption{The distribution of the fitting parameters ($A$, $\alpha$, $\alpha_2$ and $L_b$) compared to the results in this paper.
The simulation ($10^4$ runs) is performed for early-type galaxies
and the composite XLF is fitted with a broken power law.}
\label{fig.test}
\end{figure}

The higher break luminosity of early-type galaxies in this paper is examined.
First, we study the systematic effect due to the observation bias
since we are not dealing with a truly serendipitous survey.
A simulation ($10^4$ runs) is performed that 100 galaxies are randomly selected from the 130 early-type galaxies
and the composite XLF is fitted with a broken power law.
Figure \ref{fig.test} shows the distribution of the fitting parameters compared to the results in this paper.
The 1$\sigma$ systematic uncertainty is 0.08$\times10^{38}$ erg s$^{-1}$ for the break luminosity,
which is too small to affect the fitting results.
Another possible bias is due to that there may be more nearby lower optical luminosity galaxies,
which preferentially populate the low $L_X$ part of the XLF,
while the higher optical luminosity galaxies contribute more higher $L_X$ objects.
For distant galaxies ($D \gtrsim$ 30--50 Mpc),
sometimes $Chandra$ resolution becomes insufficient and only the total luminosity of the galaxy can be measured \citep{Gilfanov2004c}.
To examine this bias, we divide the early-type galaxies into three samples as 0-20 Mpc (68 galaxies),
10-30 Mpc (95 galaxies), and 20-40 Mpc (62 galaxies), then perform the XLF fittings.
The fitted breaks are 8.1$\times10^{38}$, 9.8$\times10^{38}$, and 1.3$\times10^{39}$ erg s$^{-1}$, respectively.
A small bias does exist, however, it can not explain the discrepancy between this paper and previous studies.
Actually, both the samples in \citet{Kim2004} and \citet{Humphrey2008} included galaxies with different distances,
and these studies may also suffer from this bias.
Therefore, we think the higher break luminosity in this paper is real.
Finally, we investigate the bias induced by the $B$-band luminosity correction.
One elliptical sample
(NGC1399, NGC1407, NGC3379,  NGC4365, NGC4374, NGC4472, NGC4621, NGC4636, NGC4649, and NGC4697)
and one spiral sample
(NGC253, NGC628, NGC891, NGC2403, NGC3031, NGC3628, NGC4631, NGC4945, NGC6946, and NGC7793)
are chosen to produce a composite XLF, respectively.
Both of the XLFs are corrected by the $B$-band and $K$-band luminosity, and then fitted with a broken power law.
The $K$-band luminosity for these galaxies are from \citet{Jarrett2003} and \citet{KF2004}.
The fitted break luminosity is $2.1\times10^{38}$ erg s$^{-1}$ for the ellipticals with $B$-band correction,
below that ($2.7\times10^{38}$ erg s$^{-1}$) with $K$-band correction.
For the spirals, the break luminosities are $7.2\times10^{38}$ and $1.2\times10^{39}$ erg s$^{-1}$ for
the XLFs with $B$- and $K$-band luminosity correction, respectively.
These suggest that the blue light correction may be slightly insufficient for correcting the XLF in the low luminosity side.

The dependency of the break luminosity on galaxy age is then examined.
We collect (averaged) ages of elliptical galaxies
from previous studies \citep{Trager2000, Terlevich2002, Thomas2005, Sanchez2006, Annibali2007},
and obtain seven young ($<$ 5 Gyr) and twenty old ($\ge$ 5 Gyr) elliptical samples
with detection thresholds below $10^{38}$ erg s$^{-1}$.
The fitted break luminosity is 7.08$\times10^{38}$ and 4.67$\times10^{38}$ erg s$^{-1}$
for the young and old elliptical samples, respectively.
The age effect indicates possible presence of binaries with massive neutron stars or
He-enriched neutron stars \citep{Podsiadlowski2002, KF2004} in young elliptical galaxies.
This may account for the higher break luminosity in this paper,
since there are a group of young early-type galaxies in our sample,
while in previous studies, few such galaxies were included.
In addition, \citet{Fragos2008} reported at early times ($<$ 5--6 Gyr) of one galaxy,
the XLF receives notable contributions from intermediate-mass X-ray binaries,
and that more luminous sources at earlier times makes the shape of the XLF flatter.

The composite differential XLF for 213 late-type galaxies shows a break,
although less significant, around $L_b=(6.3\pm0.3)\times10^{38}$ erg s$^{-1}$ with a slope
of $\alpha=0.6\pm0.03$ below $L_b$ and $\alpha=1\pm0.05$ above $L_b$.
Note that these breaks are the results of combining galaxies with different
XLFs, and are not necessarily present in individual galaxies,
even if the break is universal and the composite XLF is representative of the sample XLF.
For instance, a galaxy may not sample the population enough (i.e., too few sources) to unveil the break.
In this sense, this break can not be used to determine the distance of host
galaxies.

\subsection{XLF Normalizations and Slopes}

The large number of surveyed galaxies in this study allows to reveal how the
XLFs change over different galaxy types of different galaxy properties.
In Section 4.2, we normalize the best-fit XLF models by the surveyed blue light
to estimate the differential source density, and one obvious trend
shown in Figure 12 is that the $\pounds_B$ normalized X-ray point source density
at luminosities below $10^{38}$ erg s$^{-1}$ decreases from elliptical to lenticular,
to spiral, to peculiar and to irregular galaxies.
For instance, the source density around $10^{38}$ erg s$^{-1}$ are
1.99$\pm$0.35/1.51$\pm$0.65/1.07$\pm$0.14/
1.06$\pm$0.75/0.60$\pm$0.14 for these galaxies.
While the blue light $\pounds_B$ is intended as a mass indicator, there is a
trend for the mass-to-blue light ratio to decrease along the sequence of galaxy
types because the stars in these galaxies become progressively younger, bluer,
and brighter on average.
In addition, the same amount of mass corresponds to decreasing number of
stars/binaries along the sequence because the stars in these galaxies become
progressively more massive on average.
Another contributing factor is the globular cluster (GC) specific frequency, which
decreases from elliptical to spiral galaxies. This leads to less X-ray binaries
along the sequence because large fractions (20\%-70\%) of X-ray binaries are
formed through stellar interactions in GCs as confirmed by
observations \citep[e.g.,][]{Sarazin2003}.

Another obvious trend revealed in this study is that the XLF slopes become
flatter from early-type to late-type galaxies, as already revealed by previous
studies of smaller samples.
For example, \citet{Kilgard2002} presented a comparison of XLFs of four nonstarburst spiral galaxies
and three starburst galaxies, with the slopes determined from $\sim$1.3 to 0.5.
Using $Chandra$ observations of 32 nearby spiral and elliptical galaxies,
\citet{Colbert2004} determined the power-law index as $\sim$1.4 for elliptical galaxies,
and 0.6-0.8 for spiral and starburst galaxies.
The large number of galaxies in this study also allows to confirm this trend
for subtypes of galaxies, that is, the XLF slope flattens from lenticular
($\alpha\sim1.50\pm0.07$) to elliptical ($\sim1.21\pm0.02$), to spirals ($\sim0.80\pm0.02$) of S0/a-Sa
($\sim0.98\pm0.03$), Sab-Sb ($\sim0.78\pm0.03$), Sbc-Sc ($\sim0.70\pm0.04$), Scd-Sd ($\sim0.54\pm0.04$) to
Sdm-Sm ($\sim0.34\pm0.17$) subtypes, to peculiars ($\sim0.55\pm0.30$), and to irregulars
($\sim0.26\pm0.10$).

A flatter XLF slope means a larger population of X-ray sources at high
luminosities (i.e., a larger ULX population), and such an XLF slope evolution
suggests a connection between the ULX population and the star formation
activities, which are believed to be more prominent in later type/subtype
galaxies \citep{Zezas1999, Roberts2000, Fabbiano2001, Colbert2004}.
Most ULXs found in regions of star formation are HMXBs \citep{King2004},
and a significant correlation has been found between these ULXs and young OB associations \citep{Swartz2009}.
Indeed, studies of galaxy samples with different SFRs do show a
monotonic increase of the ULX rate with the SFR \citep{Mineo2013}, and the XLF
slope becomes monotonically flatter for higher SFRs.
Different shape of XLFs (e.g., breaks) of HMXBs and LMXBs can be used to diagnose on-going star formation,
while non-detection of luminous sources immediately constrains the star formation rate of the galaxy \citep{Gilfanov2004}.
The XLF evolution with different galaxy types may reflect a flatter mass distribution for the compact
objects in later type galaxies, meaning more massive black holes in later type
galaxies.
Alternatively, this may reflect more binaries at higher Eddington luminosities
in later type galaxies, due to higher accretion rates from more massive and
younger secondaries in later type galaxies.

\subsection{Radial Dependency}

As mentioned in Section 3.3, for each galaxy, we determine XLFs for sources within elliptical annuli of
the galaxy from galactic center to 2$D_{25}$ in steps of 0.1 elliptical radii, which
can be used to study the radial dependency of XLF for different galaxy types.

\begin{figure}
\plottwo{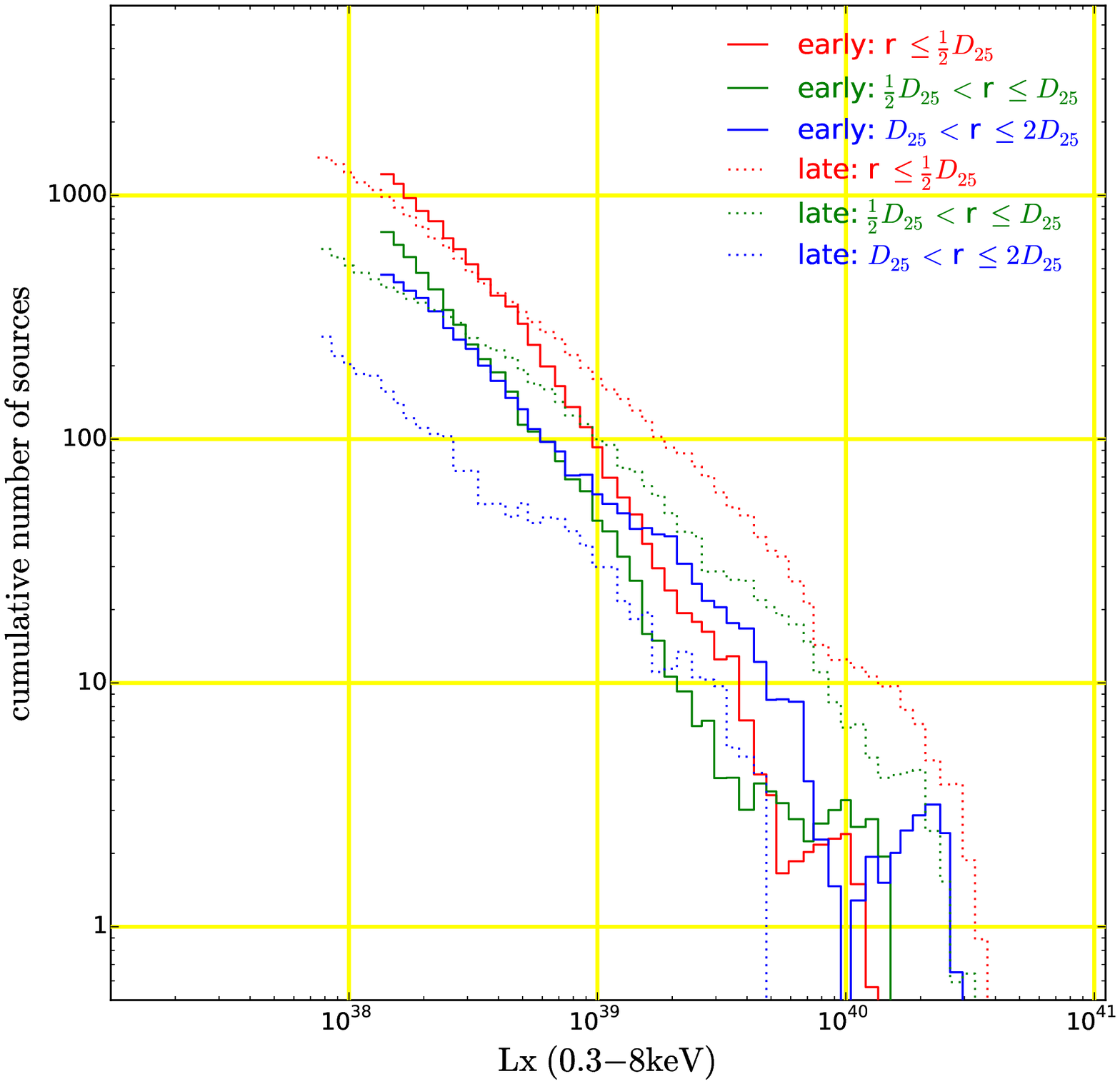}{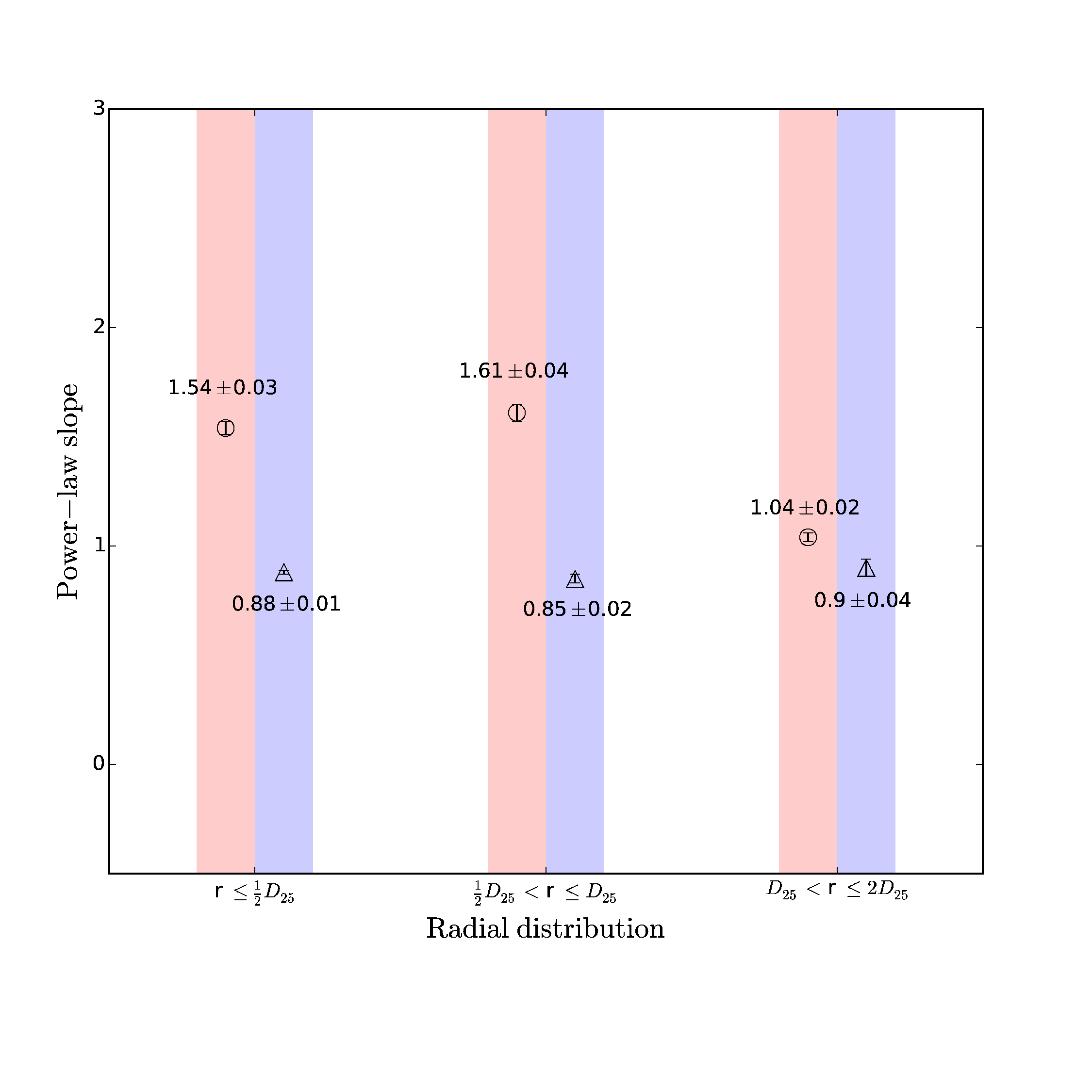}
\caption{(a) The cumulative curves for sources in various parts of early-type and late-type galaxies.
(b) The fitted XLF slopes for sources in various parts of early-type (circles) and late-type galaxies (triangles).
It seems that the XLF slopes of late-type galaxies are similar for different parts of galaxies, 
while the XLF slopes of early-type galaxies become flatter for sources between $D_{25}$ and $2D_{25}$ isophote.
}
\label{fig.Cr}
\end{figure}

Figure \ref{fig.Cr} presents the cumulative curves for sources in various parts of early-type and late-type galaxies.
Only galaxies with most of the 2$D_{25}$ region having been observed are used,
and this leads to a sample of 101 early-type and 189 late-type galaxies.
It seems that the XLF slopes of late-type galaxies are similar for different parts of galaxies.
In contrast, The XLF slopes of early-type galaxies become flatter for sources between $D_{25}$ and 2$D_{25}$,
indicating a larger proportion of bright sources in the outer part of elliptical galaxies.
The cumulative curves are fitted with a single power-law of $N(>L) = A L_{39}^{-\alpha}$,
with results shown in Figure \ref{fig.Cr}b.

\begin{figure}
\plotone{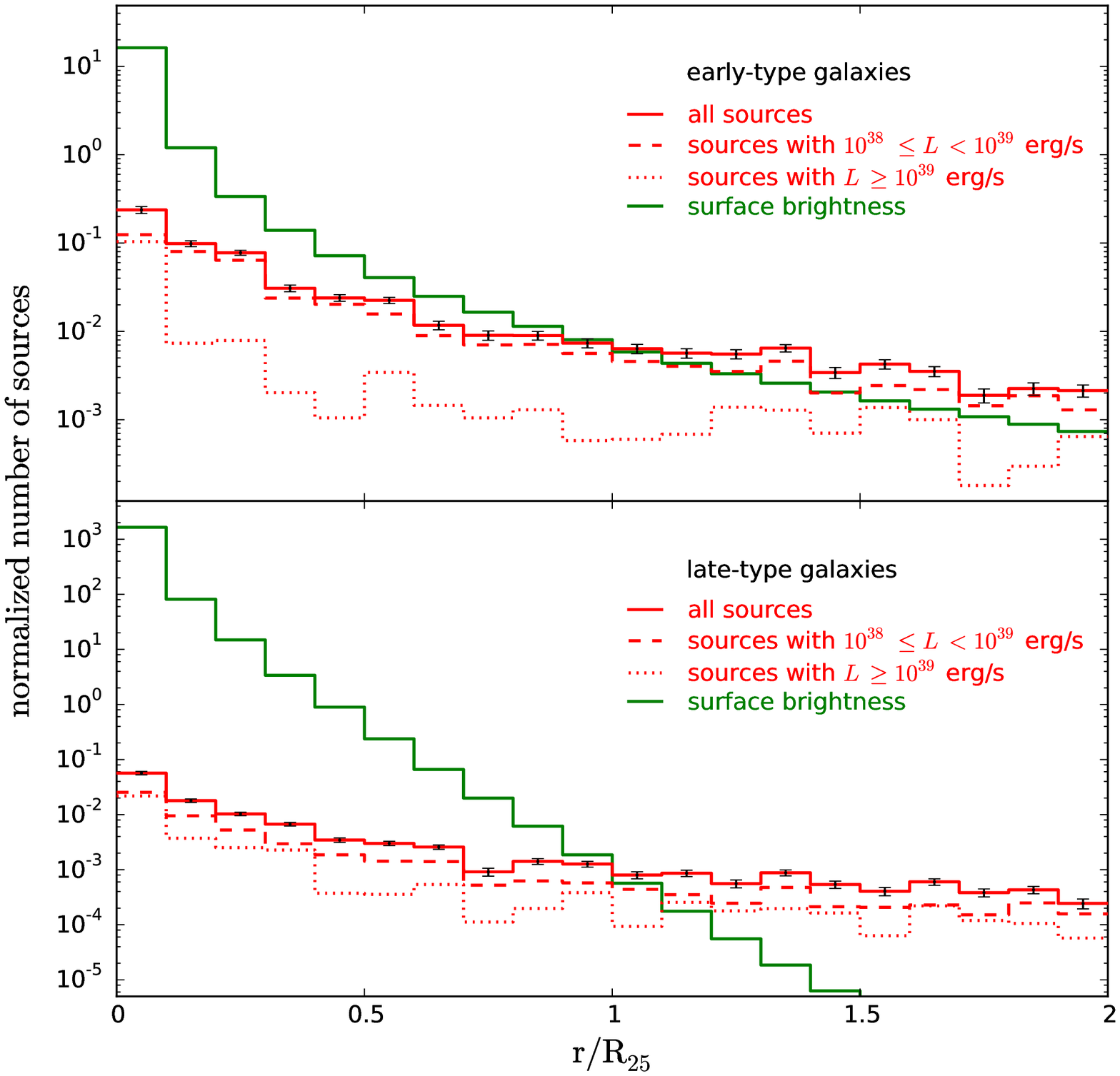}
\caption{The radial distribution of normalized number of sources for early-type and late-type galaxies.
The green line reprensents the radial distribution of B-band surface brightness, 
while the red lines represent the radial distribution of normalized number of X-ray sources.
}
\label{fig.radius}
\end{figure}

The radial distribution of sources,
which is normalized to the actually observed area of each individual annulus in unit of pixel,
is determined for different galaxy types (Figure \ref{fig.radius}).
Here we select the galaxies with detection thresholds below $10^{38}$ erg s$^{-1}$,
and use the sources brighter than $10^{38}$ erg s$^{-1}$ to avoid selection effects.
Also, only galaxies with most of the 2$D_{25}$ region having been observed are used,
and this leads to a sample of 66 early-type and 120 late-type galaxies.
The radial distribution of X-ray sources in early-type galaxies is slightly flatter than the $B$-band surface brightness.
Previous studies showed that in elliptical galaxies, the spatial distribution of GC LMXBs is more extended than field LMXBs
\citep{Kim2006, Kundu2007}, and the surface density of (blue) GC LMXBs exceeds the stellar surface brightness
when the radius increases \citep{Paolillo2011}.
Therefore, the slight excess of X-ray surface brightness in Figure \ref{fig.radius}a may suffer from the GC LMXBs \citep{Mineo2014a}.
Another feature is that the distribution of luminous sources
($10^{39} \leq L_{X} < 10^{49}$ erg s$^{-1}$) is nearly uniform, which is quite different with less luminous ones.
Because GCs host relatively more bright sources than filed LMXBs \citep{Kim2009, Peacock2016},
we conclude the XLF slopes in the outer region of early-type galaxies are dominated by GC LMXBs \citep{Irwin2005}.
For late-type galaxies, the X-ray surface brightness significantly extends past the optical, suggesting
strong effects from bright young populations.


\subsection{Age and Luminosity Dependency of Luminous X-ray Sources}

\begin{figure}
\plottwo{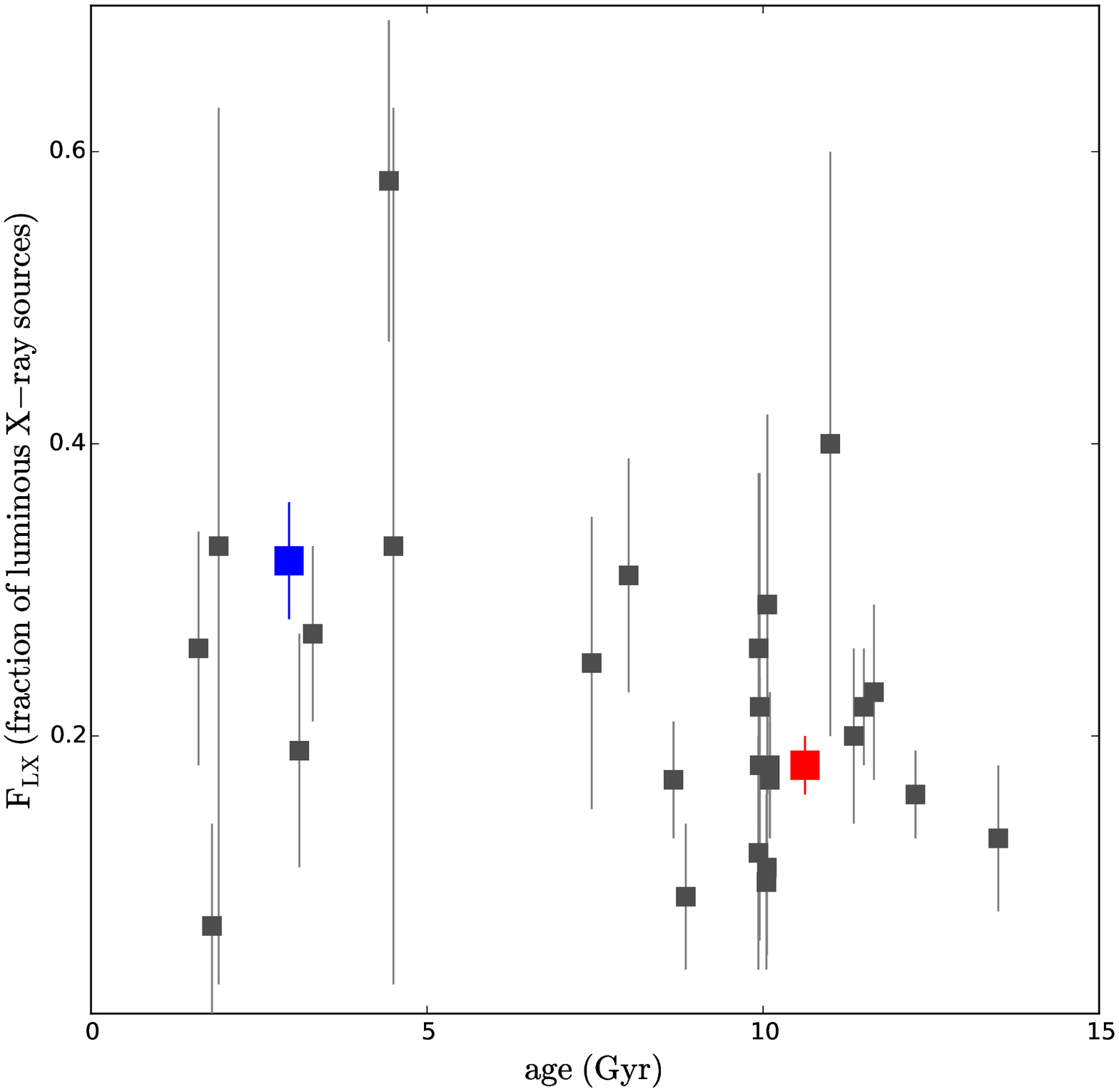}{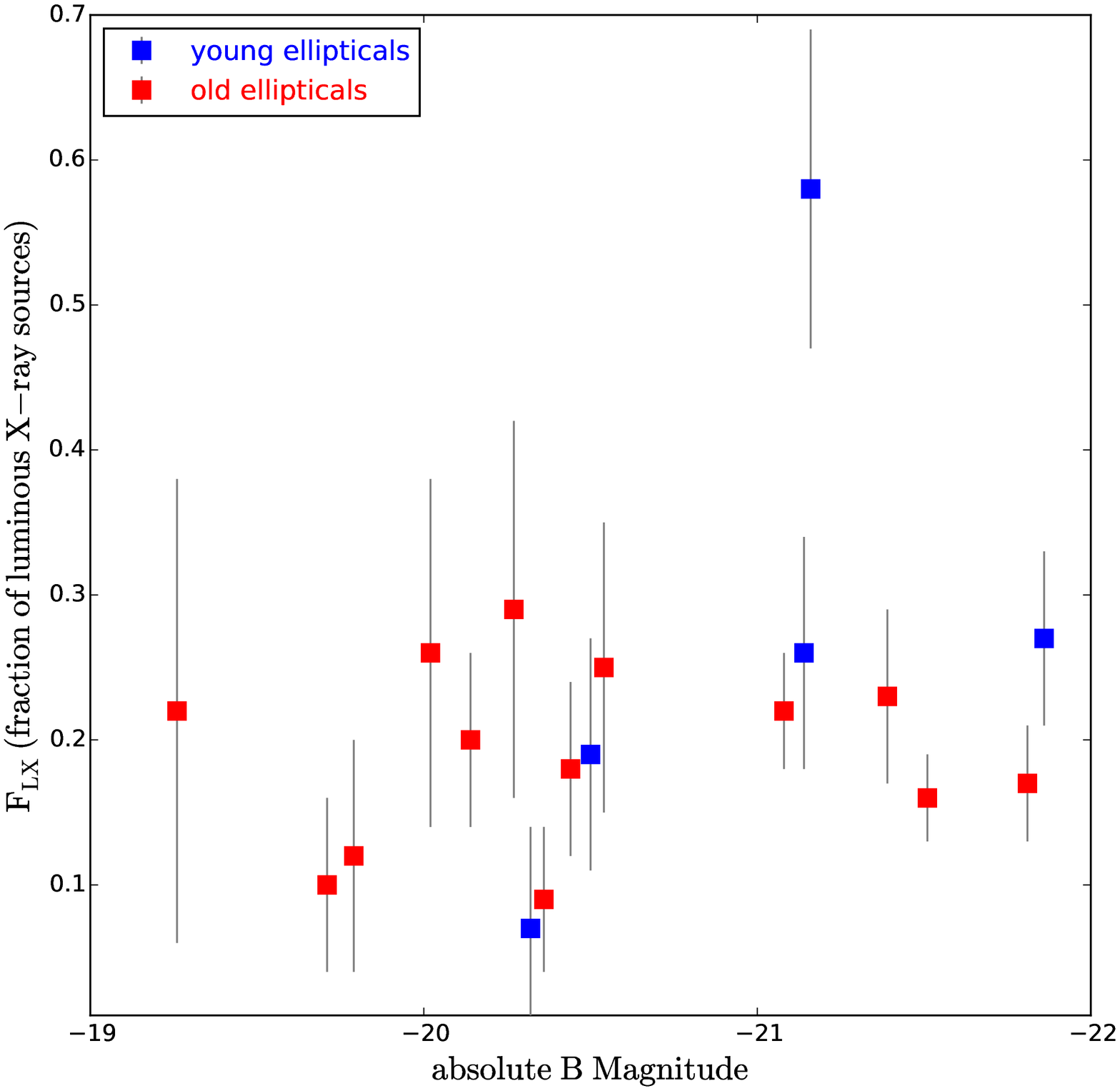}
\caption{Fraction ($F_{LX}$) of luminous ($L_{X} > 5\times10^{38}$ erg s$^{-1}$) X-ray sources
against age and absolute $B$-band magnitude.
The number of luminous X-ray sources is higher by a factor of $\sim$ 1.8 in the young sample compared to the old sample,
while no dependency of $F_{LX}$ on the stellar luminosity of the galaxy is found.
}
\label{fig.fll}
\end{figure}

\citet{Kim2010} showed that young ($<$ 5 Gyr) ellipticals host a larger fraction of
luminous X-ray sources than old ellipticals.
With the young and old elliptical samples (Section \ref{sec.break}),
we define the fraction ($F_{LX}$) of luminous X-ray sources (within the $D_{25}$ ellipse) following \citet{Kim2010},
\begin{equation}
F_{LX} = N(L_{X} > 5\times10^{38}\ {\rm erg~s^{-1}})/N(L_{X} > 10^{38}\ {\rm erg~s^{-1}}).
\end{equation}
The luminous X-ray source fractions ($F_{LX}$) are 0.32$\pm$0.04 (68 out of 211) and 0.18$\pm$0.02 (199 out of 1085) for young
and old ellipticals, respectively.
Therefore, the number of luminous X-ray sources is higher
by a factor of $\sim$ 1.8 in the young sample compared to the old sample, similar to that ($\sim$ 2) of \citet{Kim2010}.
In addition, no dependency of $F_{LX}$ on the stellar luminosity of the galaxy is found (Figure \ref{fig.fll}b), which is in agreement with \citet{Kim2010}.

\begin{figure}
\plotone{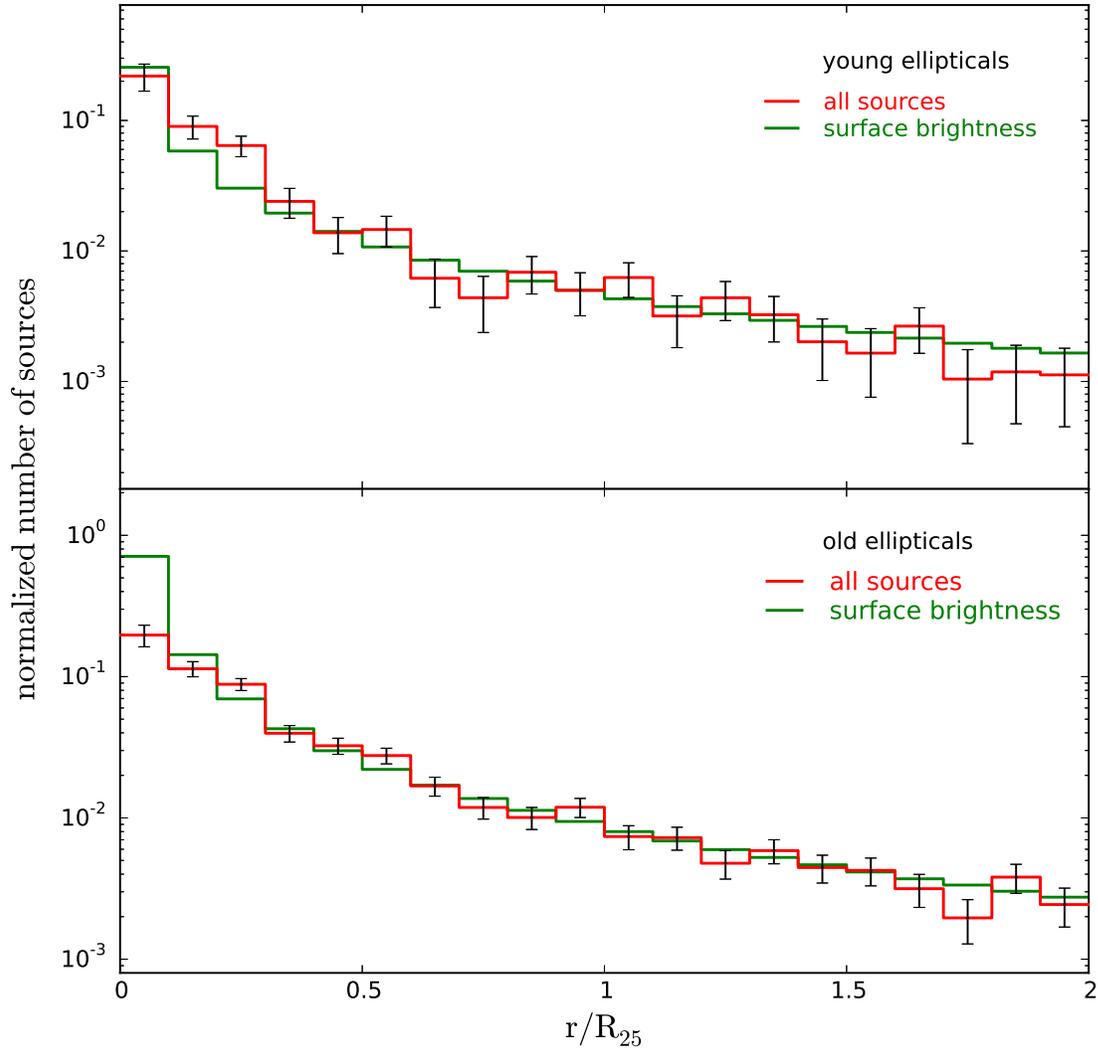}
\caption{The radial distribution of normalized number of sources for young and old elliptical galaxies.
The green line reprensents the radial distribution of B-band surface brightness, 
while the red line represents the radial distribution of normalized number of X-ray sources.
}
\label{fig.yoradius}
\end{figure}

A further examination, about the radial distribution of X-ray sources in young and old ellipticals,
is given in Figure \ref{fig.yoradius}.
As shown in \citet{Brassington2008, Brassington2009},
the radial distribution of X-ray sources follows that of the optical light,
especially for old ellipticals.
No clear discrepancy is found for the young and old elliptical samples.
Although young early-type galaxies contain more bright field LMXBs and flatter field-LMXB XLFs,
there is no significant difference of the (GC-LMXB and field-LMXB) combined XLFs between
young and old early-type galaxies \citep{Lehmer2004}.

\subsection{Stellar Mass and SFR Indicators}

Previous studies of nearby passive early-type galaxies and star-forming late-type galaxies
have shown that the X-ray point-source emission from old LMXBs and younger HMXBs
correlates well with galaxy stellar mass and SFR, respectively \citep{Colbert2004}.
The total number of LMXBs and their combined luminosity
are proportional to the stellar mass of the host galaxy,
and the X-ray luminosity of a galaxy due to LMXBs can be used as an independent
stellar mass indicator \citep{Gilfanov2004};
on the other hand, the number of HMXBs and their collective luminosity scale with the SFR,
therefore HMXBs are a good tracer of the recent star-formation activity in the host galaxy
\citep{Bauer2002,Grimm2003,Gilfanov2004a,Gilfanov2004b,Lehmer2008,Mineo2012,Mineo2014b,Mineo2014c}.
This is explained by that at high X-ray energies (2--10 keV),
the emission power from young stars and hot interstellar gas decreases steeply,
while HMXBs start to dominate the galaxy-wide X-ray intensity 
and therefore correlate strongly with SFR \citep{Persic2002,Persic2004,Persic2007,Ranalli2003,Lehmer2010}.

\begin{figure}
\plottwo{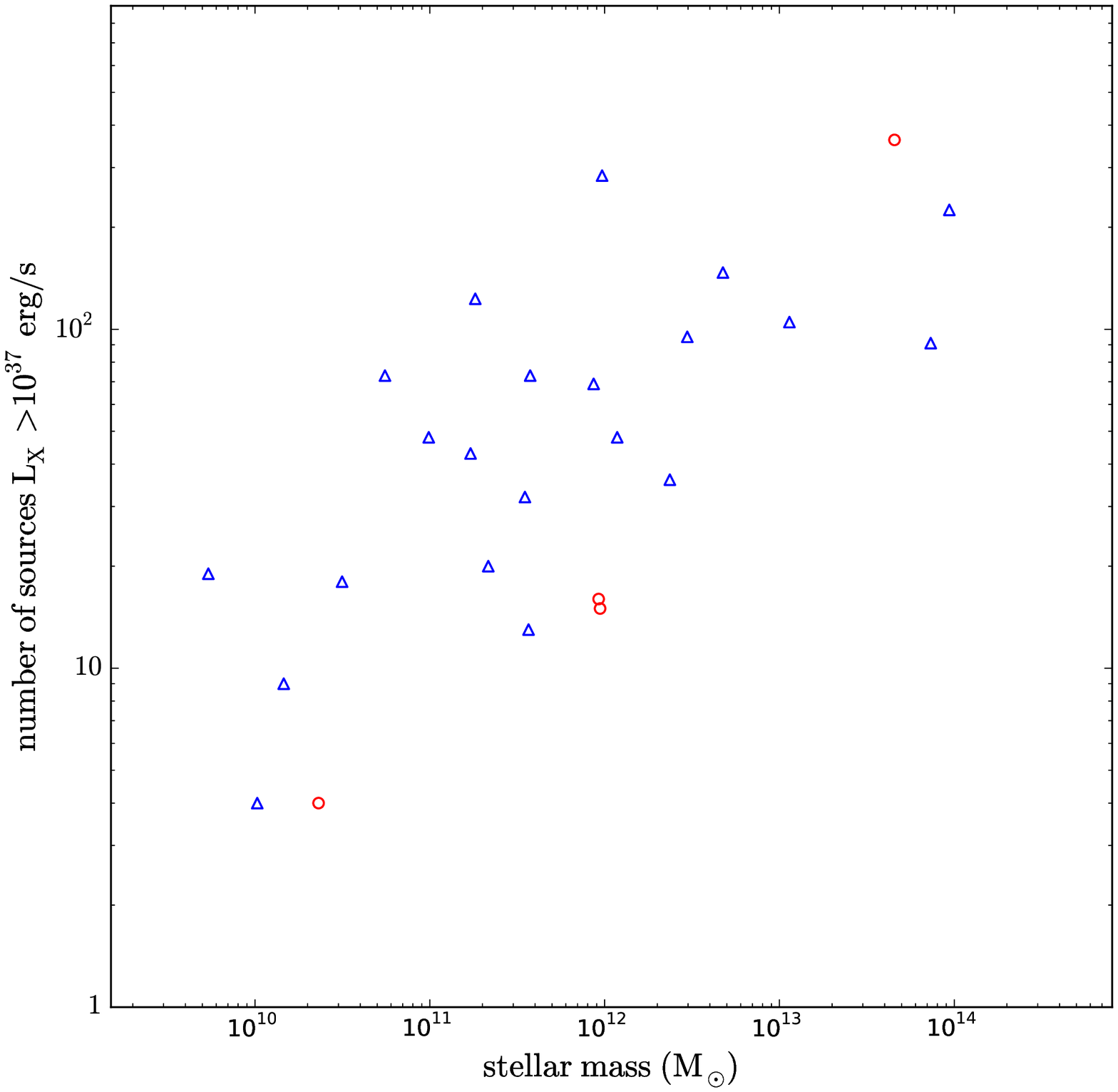}{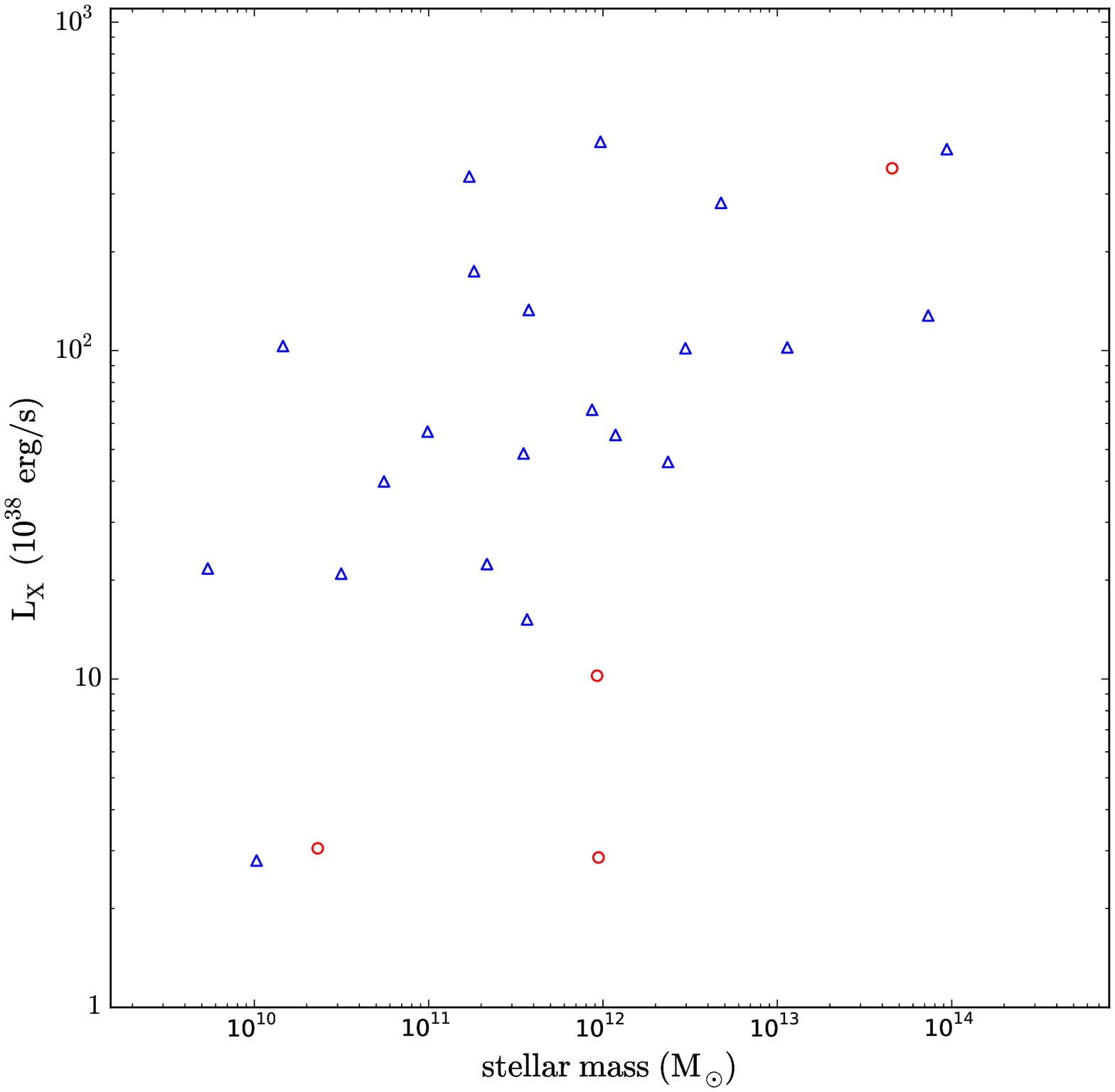}
\caption{(a) Number of sources with luminosities $L_X~>~10^{37}$ erg s$^{-1}$ versus stellar mass. 
The data for galaxies are shown by circles and triangles for early- and late-type galaxies respectively. 
(b) The collective X-ray luminosity verse stellar mass.
A clear trend can be seen, but the distribution is more diffuse than that reported by Gilfanov (2004).
}
\label{fig.massVSlx}
\end{figure}

\begin{figure}
\plottwo{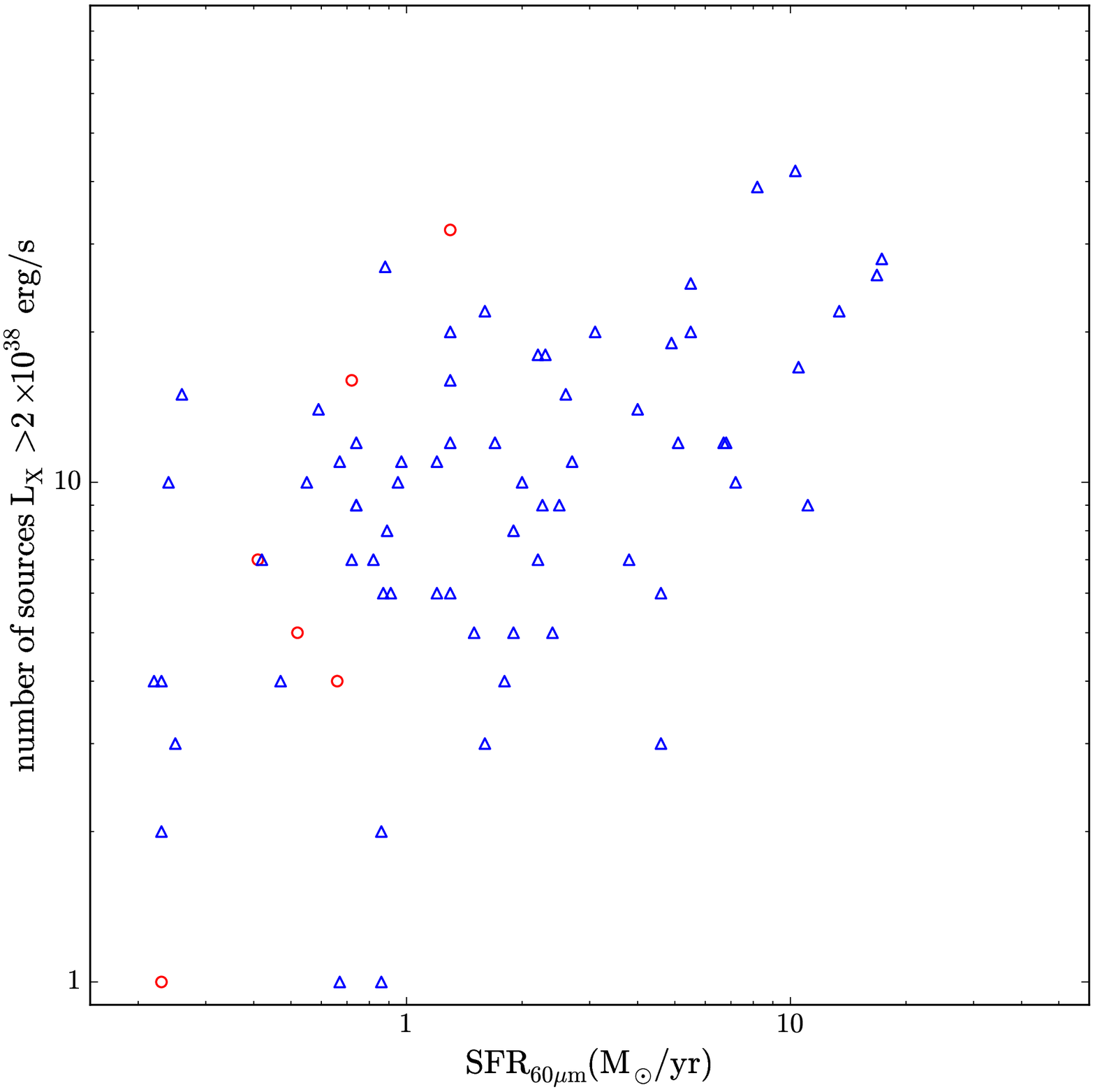}{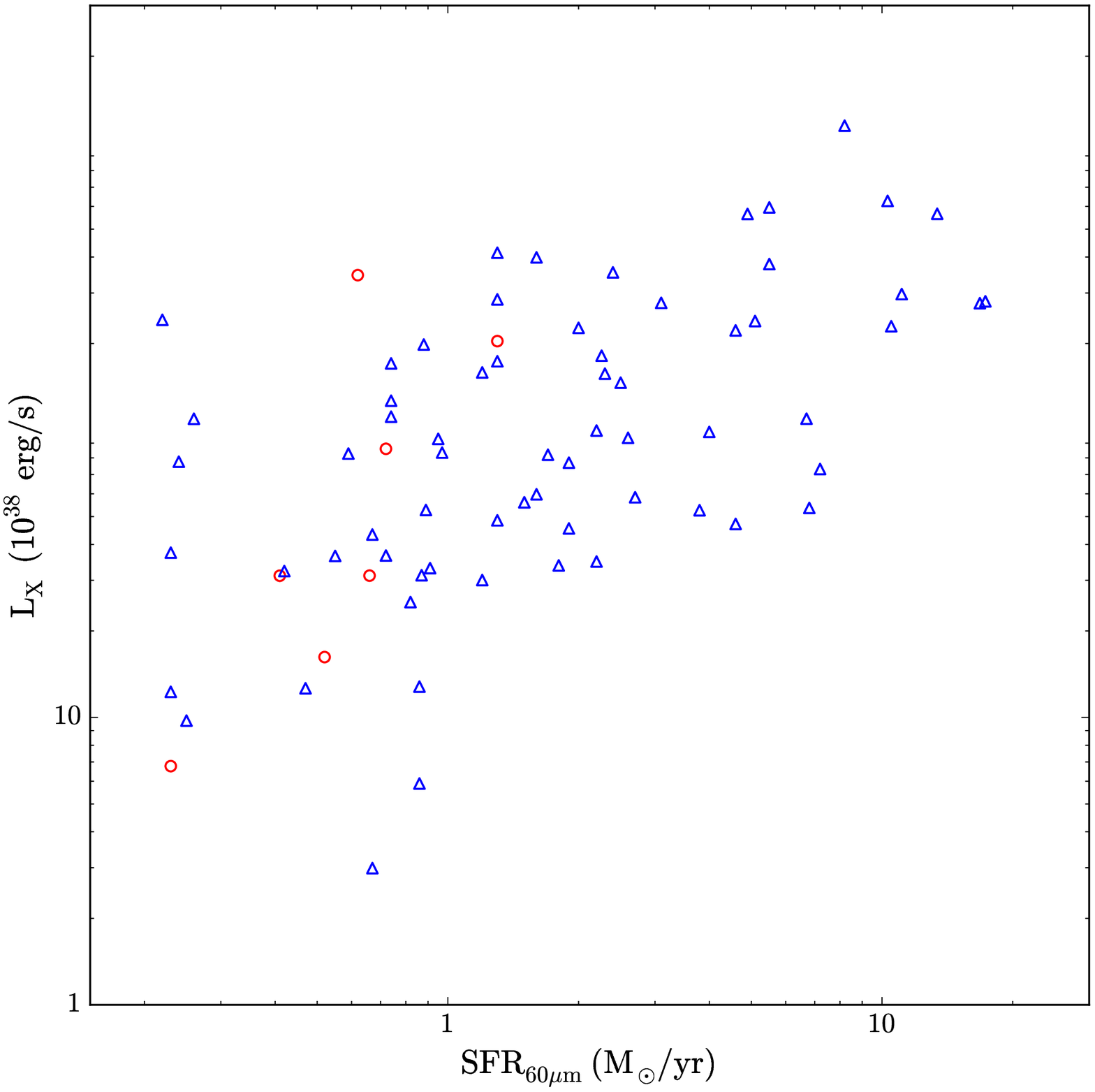}
\caption{(a) Number of sources with luminosities exceeding $2\times10^{38}$ erg s$^{-1}$ versus SFR for galaxies. 
The data for galaxies are shown by circles and triangles for early- and late-type galaxies respectively. 
(b) The collective X-ray luminosity verse SFR.
The trend is clear, but more diffuse than those from previous studies (e.g., Gilfanov et al. 2004a).
}
\label{fig.sfrVSlx}
\end{figure}

Here we make a simple examination on the population of X-ray binaries and its relation to
the stellar mass and SFR of the host galaxy.
The masses of galaxies in our sample is
collected from \citet{Tully2015}, which are calculated from the $K$-band luminosity with
assumed light to mass conversion factor. Galaxies with detection thresholds
below $10^{37}$ erg s$^{-1}$ are selected, and sources with luminosities $L_X~>~10^{37}$ erg s$^{-1}$
are chosen
following \citet{Gilfanov2004}. However, early-type galaxies typically
have higher detection threshold/completeness limit, therefore few early-type galaxies are selected.
Figure \ref{fig.massVSlx} displays the number of sources with luminosities $L_X~>~10^{37}$ erg s$^{-1}$
and their collective X-ray luminosity versus stellar mass. A clear trend can be seen,
but the distribution is more diffuse than that reported by \citet{Gilfanov2004}.
This could be due to two reasons: (1) the masses of the inner part of galaxies (e.g., galaxy nucleus) are not excluded,
while nuclear X-ray sources have been excluded; (2) more late-type galaxies are included in the sample,
which contains a number of HMXBs and therefore causes some scatter.
Figure \ref{fig.sfrVSlx} displays the number of sources with luminosities $L_X~>~2\times10^{38}$ erg s$^{-1}$
and their collective X-ray luminosity versus SFR. Only galaxies with SFR above 0.2 $M_{\odot}$/yr are plotted.
The trend is clear, but more diffuse than those from previous studies \citep[e.g.,][]{Gilfanov2004a}.
These AGNs (e.g., LINERS, Seyfert galaxies) in our sample may partly contribute to the diffuse distribution
\citep{Ranalli2003,Lehmer2010}.

\subsection{ULXs in Elliptical Galaxies}

\begin{figure}
\plottwo{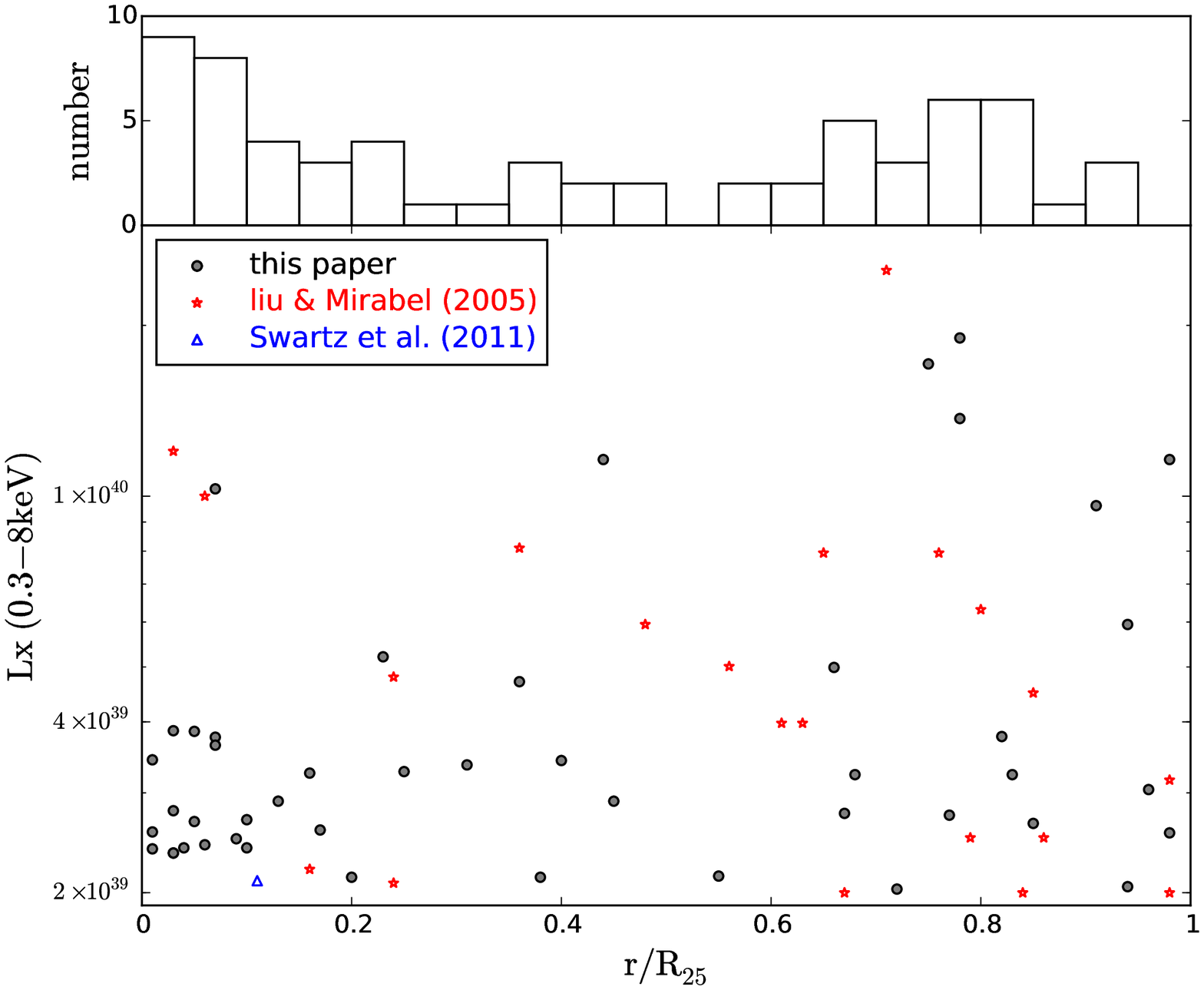}{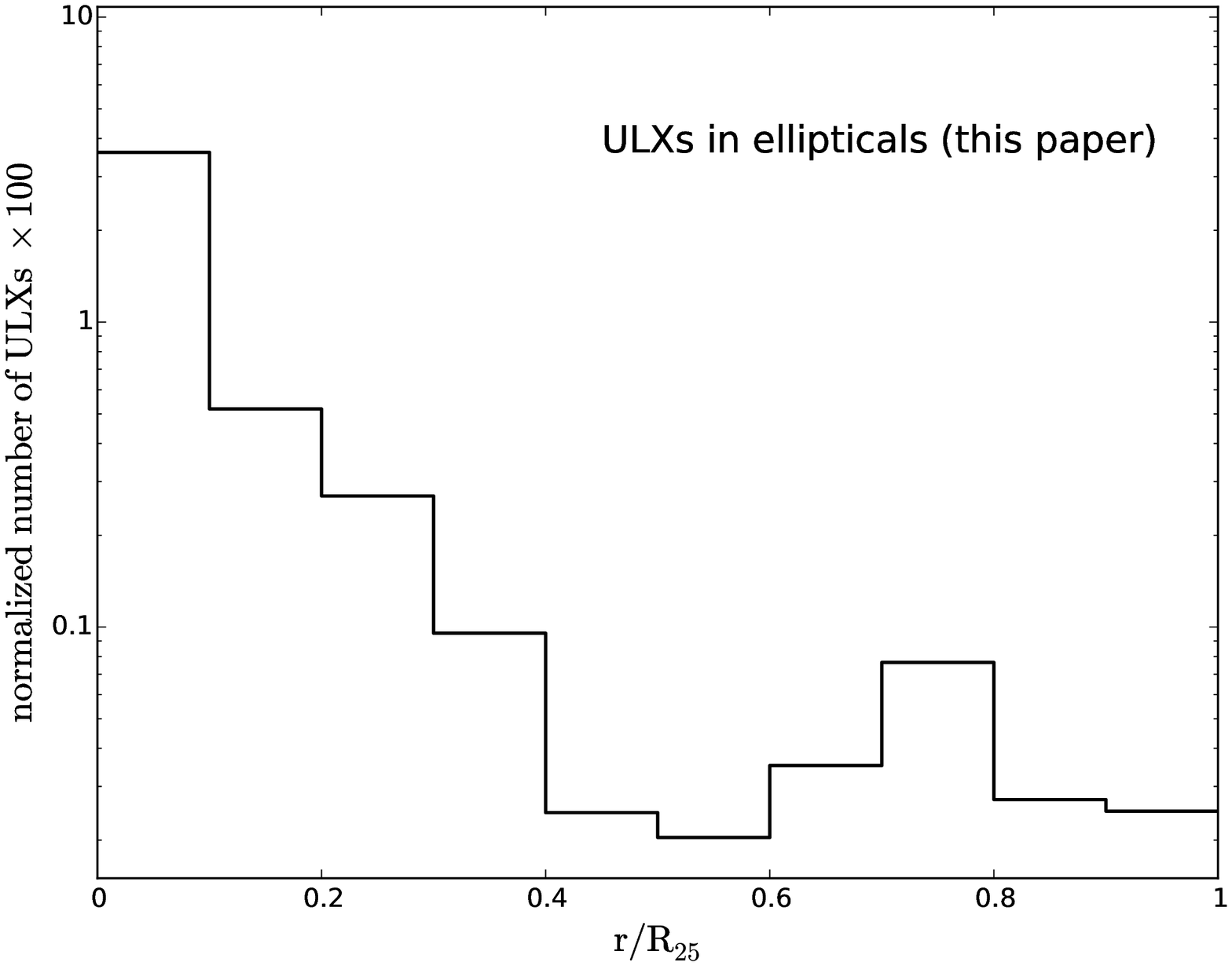}
\caption{(a) The radial distribution of ULXs in elliptical galaxies detected in this paper and previous studies \citep{Liu2005, Swartz2011}.
The ULXs have a wide distribution in galaxies,
although more ULXs above $10^{39}$ erg s$^{-1}$ are detected between 0.7 $D_{25}$ and $D_{25}$ isophote.
(b) The radial distribution of normalized number of ULXs for elliptical galaxies.
It is clear that a second peak is located out of the 0.5 $D_{25}$ isophote.
}
\label{fig.ulx}
\end{figure}

This study, with its large number of galaxies and X-ray sources, also reveals
some highly significant new results.
The radial distribution of the ULXs detected in elliptical galaxies in this paper and previous studies
\citep{Liu2005, Swartz2011} is presented in Figure \ref{fig.ulx}.
The ULXs have a wide distribution in galaxies,
although more ULXs above $10^{39}$ erg s$^{-1}$ are detected between 0.7 $D_{25}$ and $D_{25}$ isophote.
The normalized radial distribution (Figure \ref{fig.ulx}b) shows a second peak located out of
the 0.5 $D_{25}$ isophote, which may indicate a different population compared with the ULXs
in the inner parts of galaxies.

Previous studies suggested that there are very few ULXs in early-type galaxies,
e.g., $0.1\pm0.1$ ULXs per surveyed early-type galaxy in a $ROSAT$ HRI survey of
nearby galaxies \citep{Liu2006}, and most ULXs above $2\times10^{39}$ erg s$^{-1}$ in early-type galaxies are
probably foreground or background objects not physically associated with the
host galaxies \citep{Irwin2004}.
Recently, \citet{Swartz2011} reported a ULX rate of 0.23 ULXs per elliptical galaxy, but
only on a $2\sigma$ confidence level.
This study, with the foreground/background objects and galactic nuclear sources
carefully removed, finds a total of 49 ULXs above $2\times10^{39}$ erg s$^{-1}$ from
the stitched survey of 130 early-type galaxies.
Given 17.3 foreground/background objects predicted for the stitched survey of
these galaxies, the ULX rate is $0.24\pm0.05$ per surveyed early-type galaxy,
which is in good agreement with \citet{Swartz2011};
this is five  (4$\sim$7) times lower than the rate for late-type galaxies, but
nonetheless a $5\sigma$ non-zero result.
There is still a significant population of ULXs above $4\times10^{39}$ erg s$^{-1}$,
with 17 ULXs and 7.3 foreground/background objects, and a ULX rate of
$0.07\pm0.03$ ULXs per surveyed early-type galaxy.
Most of these ULXs reside in elliptical galaxies, which has a ULX rate 3-9
times higher than lenticular galaxies, consistent with a much steeper and lower
XLF for lenticular galaxies as shown in Figure \ref{fig.12}b.
Given the old ages of the stellar populations in early-type galaxies, these
ULXs must have rather old and low mass companions to their primary black holes,
similar to the ULXs located in the bulges of early-type spirals \citep{Swartz2009}.
Such systems have low accretion rates, and are most likely soft X-ray
transients with small duty cycles for outbursts \citep{King2002}.

\begin{figure}
\plottwo{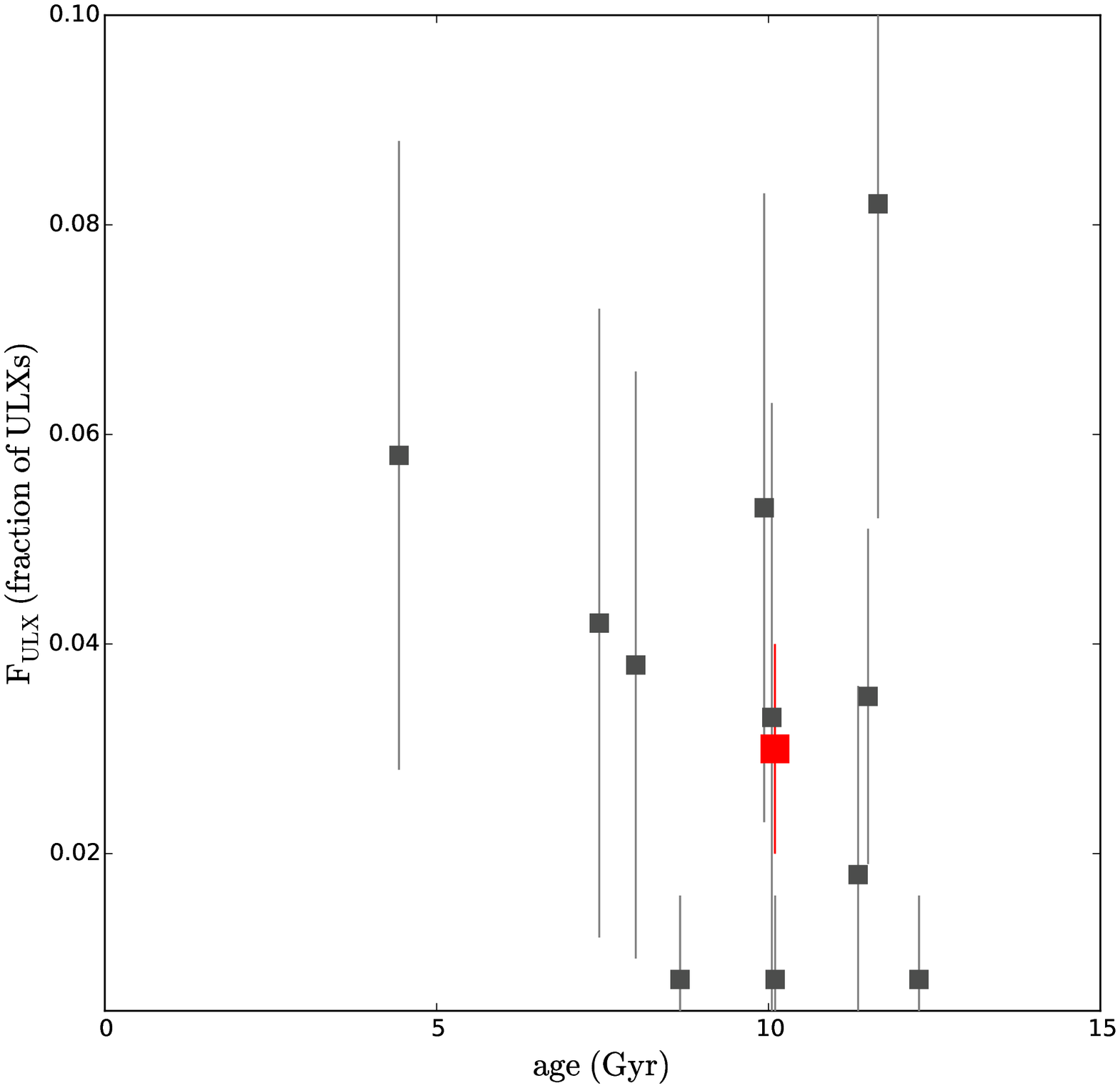}{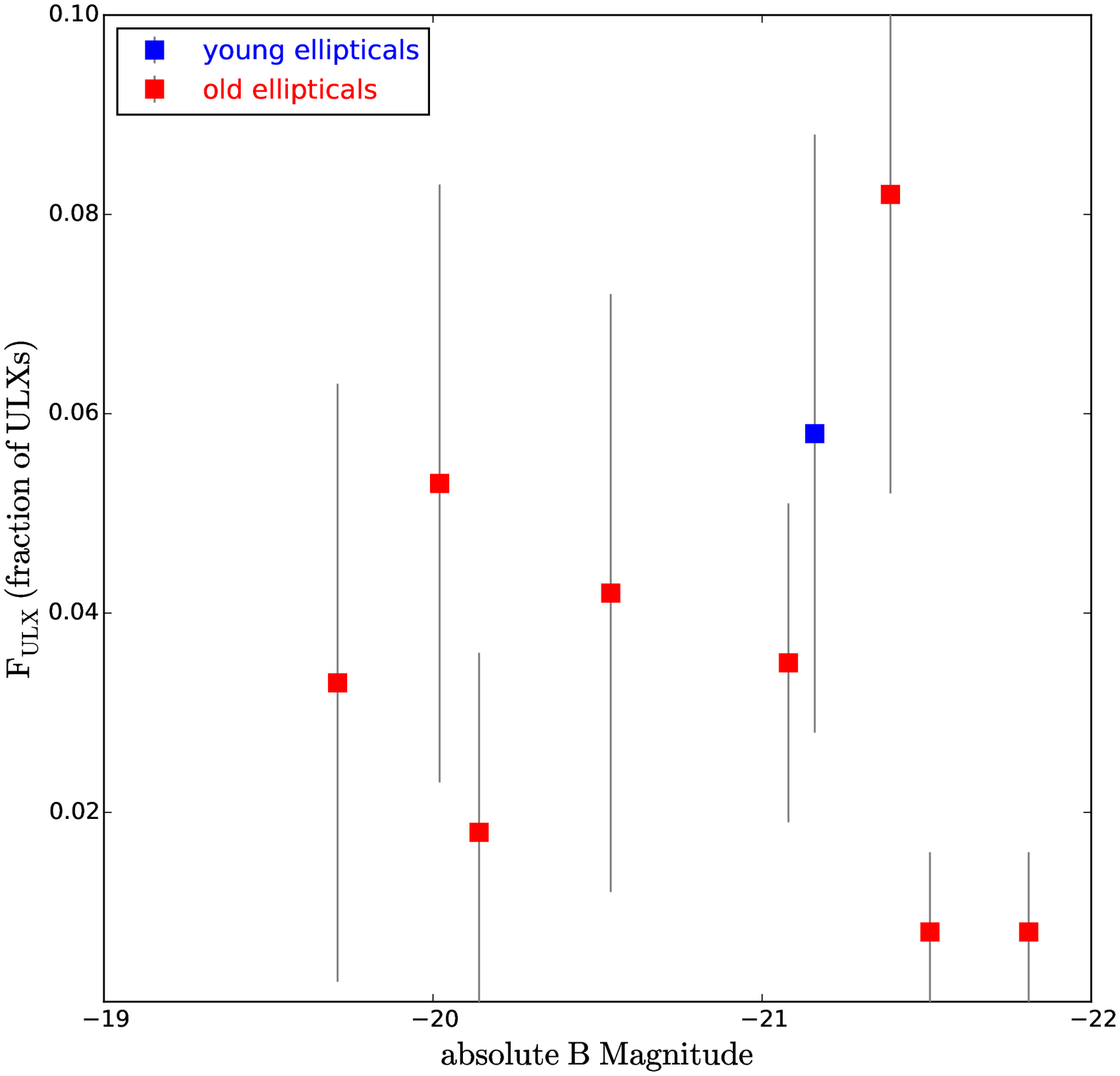}
\caption{Fraction ($F_{ULX}$) of luminous ($L_{X} > 2\times10^{39}$ erg s$^{-1}$) X-ray sources
against age and absolute $B$-band magnitude.
A weak relation is seen between $F_{ULX}$ and age, while 
no dependency of $F_{ULX}$ on the stellar luminosity of the galaxy is found.
}
\label{fig.fllULX}
\end{figure}

An examination of the age dependency of ULXs is also performed using the young and old elliptical examples.
The fraction ($F_{ULX}$) of ULXs (within the $D_{25}$ ellipse) is defined as,
\begin{equation}
F_{ULX} = N(L_{X} > 2\times10^{39}\ {\rm erg~s^{-1}})/N(L_{X} > 10^{38}\ {\rm erg~s^{-1}}).
\end{equation}
\citet{Kim2010} showed that young ($<$ 5 Gyr) elliptical galaxies
hosts more ($\sim$ 5 times) ULX type LMXBs ($L_X > 2\times10^{39}$ erg s$^{-1}$)
than old ($>$ 7 Gyr) elliptical galaxies.
The ULX fractions ($F_{ULX}$) are 0.06$\pm$0.03 (3 out of 52) and 0.03$\pm$0.01 (19 out of 763) for young
and old ellipticals, respectively.
Compared to \citet{Kim2010}, a weaker relation is seen between $F_{ULX}$ and age (Figure \ref{fig.fllULX}a).
Also, no dependency of $F_{ULX}$ on the stellar luminosity of the galaxy is found (Figure \ref{fig.fllULX}b).

\subsection{ULX Classification from A New XLF Break}

The multitude of ULXs from this survey has enabled detailed studies of XLFs at
the high luminosity end.
In a previous study of the X-ray point sources in nearby galaxies with
$ROSAT$/HRI, the XLF in late-type galaxies is shown to be a smooth power-law with
a break at $10^{40}$ erg s$^{-1}$ \citep{Liu2006}.  Such a break was also present in
a study of X-ray point sources in star forming galaxies \citep{Grimm2003}.
This break, if real, suggests that ULXs below $10^{40}$ erg s$^{-1}$ belong to the
same population of stellar mass black hole binaries,
while the handful ULXs above $10^{40}$ erg s$^{-1}$ belong to another population of
IMBHs with different binary formation mechanisms.
However, this break may be an artifact caused  by the scarcity of ULXs above
$10^{40}$ erg s$^{-1}$ in these studies.
Indeed, with 75 ULXs above $10^{40}$ erg s$^{-1}$ from this survey, the XLFs extend
smoothly to $4\times10^{40}$ erg s$^{-1}$ without any breaks at $10^{40}$ erg s$^{-1}$ for
the all-galaxy sample, the late-type galaxy sample, and the spiral-galaxy
sample, confirming that the XLF break at $10^{40}$ erg s$^{-1}$ from previous studies
are simply caused by small number of ULXs above $10^{40}$ erg s$^{-1}$.
There is a break around $4\times10^{40}$ erg s$^{-1}$ in the new XLFs from this survey
as shown in Figure \ref{fig.9}a.
Again, this break may be the dividing line between the population of the
stellar mass black hole binaries and another population of IMBH binaries.
Note that this break luminosity would correspond to massive stellar black holes
\citep[as massive as 30 $M_\odot$ as in the case of IC 10 X-1;][]{Silverman2008}
with mildly super-Eddington radiation \citep{King2009}, while the best IMBH
candidate HLX-1 can have luminosities as low as a few $\times10^{40}$ erg s$^{-1}$
\citep{Farrell2009}.
This break may also be an artifact caused by the small number (three) of ULXs
above $4\times10^{40}$ erg s$^{-1}$ in this survey.
To test whether the break is real, we are carrying out another {\it Chandra}/ACIS
survey of $\sim1800$ galaxies within 200 Mpc, the results of which will be
reported in another paper.

\section{SUMMARY}

The properties of XLFs and the ULX populations are studied with uniform
procedures for different galaxy samples in this work utilizing the recently
finished {\it Chandra} ACIS survey of X-ray point sources in nearby galaxies.
This study makes use of the largest sample of extragalactic X-ray point sources
within 343 nearby galaxies, an order of magnitude increase in number as
compared to other XLF and ULX studies.
As an archive study, the observations used were from different programs with
different exposure times.
However, all observations were reduced and analyzed with uniform procedures,
and the detection threshold can be well quantified for X-ray sources with
different OAAs and background levels as described in Section 2.1.
The foreground stars and background AGN/QSOs projected into the nearby galaxies
are seriously contaminating the samples of ordinary X-ray binaries in these
galaxies, especially so for a large survey like this one.
In this work, such contamination is meticulously estimated and removed  with
both the log$N$-log$S$ relations and the scaled source counts from the blank fields
void of nearby galaxies.
In addition, galactic nuclear sources are identified for the surveyed galaxies,
and removed to construct clean samples of ordinary X-ray binaries.

With the surveyed galaxies, we have studied the XLFs and ULX properties across different host galaxy types,
and confirm with good statistics that the XLF slope flattens from lenticular ($\alpha\sim1.50\pm0.07$)
to elliptical ($\sim1.21\pm0.02$), to spirals ($\sim0.80\pm0.02$) of S0/a-Sa ($\sim0.98\pm0.03$), Sab-Sb ($\sim0.78\pm0.03$),
Sbc-Sc ($\sim0.70\pm0.04$), Scd-Sd ($\sim0.54\pm0.04$) to Sdm-Sm ($\sim0.34\pm0.17$) subtypes, to peculiars ($\sim0.55\pm0.30$),
and to irregulars ($\sim0.26\pm0.10$).
The XLF break dividing the neutron star and black hole binaries is also confirmed,
albeit at quite different break luminosities for different types of galaxies.
The higher break luminosity than previous studies for early-type galaxies may
be due to the presence of binaries with massive neutron stars or He-enriched neutron stars,
since these systems could be existing in young early-type galaxies in our sample,
while in previous studies, few young early-type galaxies were included.
A radial dependency is found for ellipticals, with a flatter XLF slope for sources located between $D_{25}$ and 2$D_{25}$.
The radial distribution of luminous sources ($10^{39} \leq L_{X} < 10^{49}$ erg s$^{-1}$) shows different feature with fainter ones,
further suggesting the XLF slopes in the out region of early-type galaxies are dominated by GC LMXBs.
The age dependency is confirmed that young ($<$ 5 Gyr) ellipticals host a larger fraction of
luminous X-ray sources than old ellipticals, which is consistent with previous studies.
The relations between the X-ray population and galaxy stellar mass and SFR are also confirmed,
although they are more diffuse than those from previous studies.

This study shows that the ULX rate in early-type galaxies is $0.24\pm0.05$ ULXs per surveyed galaxy,
a $5\sigma$ non-zero result, confirming the existence of ULXs in early-type galaxies.
With a large number of ULXs in this survey,
the XLF break at $10^{40}$ erg s$^{-1}$ seen in previous studies is proved to be an artifact caused by small number statistics.
It is found that the XLF for late-type galaxies extends smoothly until it drops abruptly around $4\times10^{40}$ erg s$^{-1}$.
If this break at $4\times10^{40}$ erg s$^{-1}$ is real and not due to low number statistics, it will suggest that ULXs below $4\times10^{40}$ erg s$^{-1}$
belong to the stellar black hole population possibly including 30 $M_\odot$ black holes with mildly super-Eddington radiation,
while the handful ULXs above $4\times10^{40}$ erg s$^{-1}$ belong to another population of IMBHs with different binary formation mechanisms.

\acknowledgements
We especially thank the anonymous referee for his/her thorough report and helpful comments
and suggestions that have significantly improved the paper.
This research has made use of data obtained from the $Chandra$ Data Archive.
The authors acknowledge support from the National Science Foundation of China under grants
NSFC-11273028 and NSFC-11333004, and support from the National Astronomical Observatories,
Chinese Academy of Sciences under the Young Researcher Grant.

\appendix

\section{LIGHT PROFILES OF SURVEYED GALAXIES}

The light profile for a galaxy is computed from the total blue light
$\pounds_B$ and the effective radius $R_e$ that encloses 50\% of the total
light.
The effective radii are taken from RC3 for two thirds of the survey galaxies,
with an average $R_e$ of 0.15$D_{25}$, while for the rest galaxies
without $R_e$ from RC3, the $R_e$ is assumed as 0.15$D_{25}$.
The blue magnitudes of these survey galaxies are drawn from the RC3 catalog, and
converted to blue light $\pounds_B$ in unit of $L_\odot$ with $M_{B\odot} =
5.46$ mag \citep{Liu2006}.

The light profiles are expressed by different forms for early-type galaxies and late-type galaxies.
For the former one, it is usually described by the de
Vaucouleurs $R^{1/4}$ law, $I(R) = I_e 10^{3.33-3.33(R/R_e)^{1/4}}$, with $R$
as the elliptical radius and  $I_e$ as the surface brightness at the effective radius $R_e$.
The light profile can therefore be computed as
$I(R) = {\pounds_B \over 7.22\pi R_e^2} 10^{3.33-3.33(R/R_e)^{1/4}}$,
with the total light expressed as $\pounds_B = 7.22\pi R_e^2 I_e$.
For the latter one, the light profile can be decomposed into a de Vaucouleurs
bulge in the form of $I(R) = I_e 10^{3.33-3.33(R/r_e)^{1/4}}$,  and an
exponential disk in the form of $I(R) = I_0 e^{-R/h}$, where $h$ is the scale
height, and $I_0$ is the surface brightness at the galactic center. The total
light of the disk can be expressed as $2\pi h^2 I_0$.
The relative prominence of the two components, which changes with the Hubble type T
and the bulge-to-disk ($B/D$) ratios,
is taken from \citet{Graham2001} and interpolated if necessary.
In general, the relative prominence decreases toward later galaxy types
(from 25\% at T=0 to 1\% at T=10).
The $r_e/h$ ratio is quite constant despite the variation in bulge-to-disk ratios,
and it is adopted as 0.2 following \citet{Graham2001}.
%
With the assumption that the effective radius $R_e$ from RC3 encloses 50\% of the
disk light, we find $h = R_e/1.7$, $r_e = R_e/8.5$. This is
reasonable for most surveyed galaxies with $B/D \ll 1$, while for galaxies
with nonnegligible bulges, the $r_e$ and $h$ may be slightly overestimated.
The $I_0$ and $I_e$ can be computed from $\pounds_B$ and $R_e$, with
$I_0 = {\pounds_B \over 1+B/D} {1.7^2 \over 2\pi R_e^2}$,
and $I_e = {\pounds_B \over D/B+1 } {8.5^2 \over 7.22\pi R_e^2}$.
Finally, the light profile for late-type galaxies can be calculated with
$I(R) = I_e 10^{3.33-3.33(R/r_e)^{1/4}} + I_0 e^{-R/h}$.

\clearpage

\clearpage

\clearpage

\clearpage

\setlength{\tabcolsep}{2pt}

\begin{deluxetable}{c|cccc|ccc|ccc|cccccc}
\tablewidth{7in}
\tabletypesize{\scriptsize}
\tablecaption{X-ray Point Source Populations in 343 Nearby Galaxies}
\tablehead{
	\colhead{Galaxy} & \colhead{$N_{obs}$} & \colhead{$N^a_{src}$} & \colhead{$N^t_X$} & \colhead{$f_{Cov}$} & \colhead{$\lg L_{X,lim}$} & \colhead{$N_X$} & \colhead{$N_{b/f}$} & \colhead{$N^a_{U}$} & \colhead{$n_{U}$} & \colhead{$n_{b/f}$} & \colhead{GType} & \colhead{$L_B$} & \colhead{GArea} & \colhead{Dist.} & \colhead{SFR$_{60\mu{\rm m}}$} & \colhead{activity} \\
    \colhead{(1)} &  \colhead{(2)}  &  \colhead{(3)} &  \colhead{(4)} & \colhead{(5)} &  \colhead{(6)}  & \colhead{(7)} &  \colhead{(8)}  &  \colhead{(9)} & \colhead{(10)} &  \colhead{(11)} & \colhead{(12)} & \colhead{(13)}  & \colhead{(14)}  &  \colhead{(15)} & \colhead{(16)} & \colhead{(17)}
}

\startdata

NGC5813 & 1 & 20 & 19 & 1.00 & 38.57 & 19 & 1.88 & 5 & 4.57 & 0.43 & elliptical & 4.3e+10 & 9.87 & 32.21 & $<$0.12 &  LINER \\
NGC1407 & 1 & 73 & 72 & 1.00 & 38.38 & 66 & 3.43 & 5 & 4.49 & 0.51 & elliptical & 6.7e+10 & 15.40 & 28.84 & 0.052 & - \\
NGC1399 & 11 & 129 & 117 & 1.00 & 37.98 & 117 & 8.58 & 5 & 2.53 & 0.47 & elliptical & 4.1e+10 & 35.11 & 19.95 & $<$0.16 & - \\
NGC3557 & 1 & 8 & 7 & 0.93 & 39.02 & 7 & 1.29 & 3 & 2.26 & 0.74 & elliptical & 1.6e+11 & 9.48 & 45.71 & 0.23 & - \\
NGC4494 & 2 & 18 & 18 & 1.00 & 38.13 & 17 & 2.14 & 2 & 1.89 & 0.11 & elliptical & 2.7e+10 & 13.29 & 17.06 & 0.028 &  LINER \\
IC1459 & 1 & 48 & 47 & 0.99 & 38.22 & 47 & 4.83 & 2 & 1.41 & 0.59 & elliptical & 6.1e+10 & 15.64 & 29.24 & 0.18 &  LINER \\
NGC3125 & 1 & 2 & 2 & 1.00 & 38.42 & 2 & 0.03 & 1 & 1.00 & 0.00 & elliptical & 1.2e+09 & 0.63 & 11.5 & 0.3 & - \\
NGC3379 & 6 & 69 & 43 & 1.00 & 37.22 & 43 & 7.28 & 1 & 0.95 & 0.05 & elliptical & 1.5e+10 & 20.28 & 10.57 & $<$0.026 &  LINER \\
\enddata

\tablecomments{The columns are:
(1) Galaxy name;
(2) number of observations;
(3) number of X-ray point sources detected above $4\sigma$ in all observations;
(4) number of sources detected in the stitched deep survey;
(5) the surveyed area A as a fraction of $D_{25}$ isophotal area;
(6) the limiting $L_X$ at which the surveyed blue light is 50\% of the total
blue light surveyed;
(7) number of sources detected above $L_{X,lim}$;
(8) the predicted foreground/background sources for $L>L_{X,lim}$;
(9) the number of ULXs found in all observations;
(10) the number of ULXs detected in the deep survey minus the predicted
foreground/background sources for $L>2\times10^{39}$ erg s$^{-1}$;
(11) the predicted foreground/background sources for $L>2\times10^{39}$ erg s$^{-1}$;
(12) the galaxy type;
(13) the blue light content of the galaxy in unit of $L_\odot$;
(14) the $D_{25}$ isophotal area in unit of arcmin$^2$;
(15) the galaxy distance;
(16) the SFR in $M_\odot$/yr estimated based on the flux
density at $60\mu$m observed by IRAS;
(17) the star-forming and nuclear activities. }

\end{deluxetable}

\clearpage

\setlength{\tabcolsep}{2pt}

\begin{deluxetable}{lrrr|rrrr|rrrcc|rrrcc|rrrcc}
\rotate
\tablewidth{9.2in}
\tabletypesize{\scriptsize}
\tablecaption{ULX Rates for Different Survey Galaxy Samples}
\tablehead{
   \colhead{}  & \colhead{}  & \colhead{}  & \colhead{} & \colhead{} & \colhead{} & \colhead{}  & \colhead{}  & \multicolumn{5}{c}{$L_X>2\times10^{39}$ erg s$^{-1}$} & \multicolumn{5}{c}{$L_X>4\times10^{39}$ erg s$^{-1}$} & \multicolumn{5}{c}{$L_X>1\times10^{40}$ erg s$^{-1}$} \\
   \colhead{Sample} & \colhead{$N_{gal}$} & \colhead{$N^a_{src}$} & \colhead{$N^a_{U}$} & \colhead{Area} & \colhead{$L_B$} & \colhead{$N_{X}$} & \colhead{$N_{b/f}$} & \colhead{$n_{U}$} & \colhead{$n_{G}$} & \colhead{$n_{b/f}$} & \colhead{$U/G$} & \colhead{$U/L_B$} & \colhead{$n_{U}$} & \colhead{$n_{G}$} & \colhead{$n_{b/f}$} & \colhead{$U/G$} & \colhead{$U/L_B$} & \colhead{$n_{U}$} & \colhead{$n_{G}$} & \colhead{$n_{b/f}$} & \colhead{$U/G$} & \colhead{$U/L_B$} \\
	\colhead{(1)} &  \colhead{(2)}  &  \colhead{(3)} &  \colhead{(4)} & \colhead{(5)} &  \colhead{(6)}  & \colhead{(7)} &  \colhead{(8)}  &  \colhead{(9)} & \colhead{(10)} &  \colhead{(11)} & \colhead{(12)} & \colhead{(13)}  & \colhead{(14)}  &  \colhead{(15)} & \colhead{(16)} & \colhead{(17)}  & \colhead{(18)}  &  \colhead{(19)} & \colhead{(20)}  & \colhead{(21)} & \colhead{(22)}  & \colhead{(23)}
}

\startdata

all & 343 & 5920 & 282 & 9.9e+07 & 247.4 & 4970 & 1170.56 & 218 & 117 & 33.1 & 0.54$\pm$0.04 & 0.75$\pm$0.06 & 98 & 91 & 14.1 & 0.24$\pm$0.03 & 0.34$\pm$0.04 & 31 & 66 & 4.0 & 0.08$\pm$0.02 & 0.11$\pm$0.02 \\
early & 130 & 2311 & 66 & 2.1e+07 & 129.1 & 1945 & 259.86 & 49 & 58 & 17.3 & 0.24$\pm$0.05 & 0.25$\pm$0.05 & 17 & 47 & 7.5 & 0.07$\pm$0.03 & 0.07$\pm$0.03 & 6 & 33 & 2.1 & 0.03$\pm$0.02 & 0.03$\pm$0.02 \\
late & 213 & 3609 & 216 & 7.8e+07 & 118.3 & 3025 & 910.7 & 169 & 59 & 15.7 & 0.72$\pm$0.06 & 1.30$\pm$0.11 & 81 & 44 & 6.6 & 0.35$\pm$0.04 & 0.63$\pm$0.08 & 25 & 33 & 2.0 & 0.11$\pm$0.02 & 0.19$\pm$0.04 \\
elliptical & 55 & 1509 & 47 & 8.7e+06 & 81.4 & 1299 & 123.59 & 34 & 28 & 10.5 & 0.43$\pm$0.11 & 0.29$\pm$0.07 & 10 & 21 & 4.5 & 0.10$\pm$0.06 & 0.07$\pm$0.04 & 5 & 15 & 1.2 & 0.07$\pm$0.04 & 0.05$\pm$0.03 \\
lenticular & 75 & 802 & 19 & 1.2e+07 & 47.6 & 646 & 136.27 & 15 & 30 & 6.8 & 0.11$\pm$0.05 & 0.17$\pm$0.08 & 7 & 26 & 3.0 & 0.05$\pm$0.04 & 0.08$\pm$0.06 & 1 & 18 & 0.8 & 0.00$\pm$0.01 & 0.00$\pm$0.02 \\
spiral & 187 & 3380 & 190 & 7.5e+07 & 113.7 & 2865 & 874.4 & 154 & 56 & 14.9 & 0.74$\pm$0.07 & 1.22$\pm$0.11 & 70 & 42 & 6.2 & 0.34$\pm$0.04 & 0.56$\pm$0.07 & 20 & 32 & 1.9 & 0.10$\pm$0.02 & 0.16$\pm$0.04 \\
irregular & 21 & 173 & 8 & 2.9e+06 & 2.3 & 124 & 31.52 & 7 & 1 & 0.2 & 0.32$\pm$0.13 & 3.00$\pm$1.17 & 5 & 1 & 0.1 & 0.23$\pm$0.11 & 2.17$\pm$0.99 & 2 & 0 & 0.0 & 0.09$\pm$0.07 & 0.87$\pm$0.63 \\
peculiar & 5 & 56 & 18 & 3.3e+05 & 2.4 & 36 & 4.75 & 8 & 2 & 0.6 & 1.48$\pm$0.57 & 3.13$\pm$1.20 & 6 & 1 & 0.3 & 1.15$\pm$0.49 & 2.43$\pm$1.04 & 3 & 1 & 0.1 & 0.58$\pm$0.35 & 1.24$\pm$0.73 \\
\enddata
\tablecomments{The columns are:
(1) name of the galaxy survey sample;
(2) the number of surveyed galaxies;
(3) the number of point sources above $4\sigma$ detected from all observations;
(4) the number of ULXs detected from all observations;
(5) the surveyed sky area in pixel$^2$ from the stitched deep survey of individual galaxies;
(6) the surveyed blue light content in unit of $10^{10} L_\odot$;
(7) the total number of X-ray point sources detected in the stitched survey;
(8) the number of foreground/background objects expected for the survey;
(9) the number of ULXs ($L_X>2\times10^{39}$ erg s$^{-1}$) in the survey;
(10) the number of galaxies with at least one ULX ($L_X>2\times10^{39}$ erg s$^{-1}$) in the survey;
(11) the number of predicted contaminating sources with $L_X>2\times10^{39}$ erg s$^{-1}$;
(12) the net number of ULXs ($L_X>2\times10^{39}$ erg s$^{-1}$) per surveyed galaxy and its error;
(13) the net number of ULXs ($L_X>2\times10^{39}$ erg s$^{-1}$) per $10^{10} L_\odot$ blue light and its error;
(14) the number of ULX ($L_X>4\times10^{39}$ erg s$^{-1}$) in the survey;
(15) the number of galaxies with at least one ULX ($L_X>4\times10^{39}$ erg s$^{-1}$) in the survey;
(16) the number of predicted contaminating sources with $L_X>4\times10^{39}$ erg s$^{-1}$;
(17) the net number of ULXs ($L_X>4\times10^{39}$ erg s$^{-1}$) per surveyed galaxy and its error;
(18) the net number of ULXs ($L_X>4\times10^{39}$ erg s$^{-1}$) per $10^{10} L_\odot$ blue light and its error;
(19) the number of ULX ($L_X>10^{40}$ erg s$^{-1}$) in the survey;
(20) the number of galaxies with at least one ULX ($L_X>10^{40}$ erg s$^{-1}$) in the survey;
(21) the number of predicted contaminating sources with $L_X>10^{40}$ erg s$^{-1}$;
(22) the net number of ULXs ($L_X>10^{40}$ erg s$^{-1}$) per surveyed galaxy and its error;
(23) the net number of ULXs ($L_X>10^{40}$ erg s$^{-1}$) per $10^{10} L_\odot$ blue light and its error.}

\end{deluxetable}

\clearpage

\setlength{\tabcolsep}{4pt}

\begin{deluxetable}{lccc|rcc|rrcccc}
\rotate
\tablewidth{8.4in}
\tabletypesize{\scriptsize}
\tablecaption{Single and Broken Power Fits to Differential Curves for Net Non-nuclear Sources}
\tablehead{
   \colhead{}  & \colhead{}  & \colhead{}  & \colhead{} & \multicolumn{3}{c}{single power-law} & \multicolumn{6}{c}{broken power-law} \\
   \colhead{Sample} & \colhead{$\lg L_{X1}$} & \colhead{$\lg L_{X2}$} & \colhead{$N_{bin}$} & \colhead{$\chi^2_1$} & \colhead{A} & \colhead{$\alpha$} & \colhead{$\chi^2_2$} & \colhead{$prob$} & \colhead{A} & \colhead{$\alpha$} & \colhead{$L_b$} & \colhead{$\alpha_2$} \\
 \colhead{(1)} &  \colhead{(2)}  &  \colhead{(3)} &  \colhead{(4)} & \colhead{(5)} &  \colhead{(6)}  & \colhead{(7)} &  \colhead{(8)}  &  \colhead{(9)} & \colhead{(10)} &  \colhead{(11)} & \colhead{(12)} & \colhead{(13)}
}

\startdata

all & 38.02 & 40.47 & 51 & 86.78 & 0.2 $\pm$ 2.2\% & 0.98 $\pm$ 0.01 & 48.84 & 0.0498 & 0.21 $\pm$ 2.2\% & 0.51 $\pm$ 0.06 & 38.42 $\pm$ 0.02 & 1.10 $\pm$ 0.02 \\
early & 38.13 & 40.13 & 42 & 78.95 & 0.15 $\pm$ 3.3\% & 1.33 $\pm$ 0.02 & 35.53 & 0.0149 & 0.2 $\pm$ 3.2\% & 1.02 $\pm$ 0.03 & 38.95 $\pm$ 0.01 & 2.29 $\pm$ 0.18 \\
late & 37.87 & 40.47 & 54 & 67.81 & 0.21 $\pm$ 3.2\% & 0.78 $\pm$ 0.02 & 51.16 & 0.3187 & 0.24 $\pm$ 3.1\% & 0.61 $\pm$ 0.03 & 38.80 $\pm$ 0.02 & 0.96 $\pm$ 0.05 \\
elliptical & 38.02 & 40.13 & 44 & 119.46 & 0.17 $\pm$ 3.6\% & 1.21 $\pm$ 0.02 & 40.73 & 0.0055 & 0.19 $\pm$ 3.5\% & 0.68 $\pm$ 0.11 & 38.65 $\pm$ 0.02 & 1.80 $\pm$ 0.03 \\
lenticular & 38.27 & 39.77 & 32 & 39.65 & 0.087 $\pm$ 8.6\% & 1.50 $\pm$ 0.07 & 27.66 & 0.3412 & 0.1 $\pm$ 8.3\% & 0.96 $\pm$ 0.10  & 38.87 $\pm$ 0.02 & 2.74 $\pm$ 0.10 \\
spiral & 37.87 & 40.47 & 54 & 67.54 & 0.21 $\pm$ 3.2\% & 0.80 $\pm$ 0.02 & 50.62 & 0.3079 & 0.25 $\pm$ 3.2\% & 0.64 $\pm$ 0.02 & 38.90 $\pm$ 0.04 & 1.02 $\pm$ 0.06 \\
irregular & 36.87 & 39.83 & 61 & 32.57 & 0.33 $\pm$ 6.6\% & 0.26 $\pm$ 0.10 \\
peculiar & 38.42 & 40.27 & 39 & 12.61 & 0.32 $\pm$ 4.1\% & 0.55 $\pm$ 0.30 \\
\enddata

\tablecomments{The columns are:
(1) name of the galaxy survey sample;
(2) the low luminosity end for the XLF curve;
(3) the high luminosity end for the XLF curve;
(4) the number of bins ($\delta \lg L = 0.05$) for the XLF curve;
(5) the resulted $\chi^2$ for the single power-law fit;
(6) the A parameter and error in percentage normalized by the surveyed blue light for the sample;
(7) the power-law slope and error;
(8) the resulted $\chi^2$ for the broken power-law fit;
(9) the probability for the broken power-law fit is no better than the single power-law;
(10) the A parameter and error in percentage normalized by the surveyed blue light for the sample;
(11) the power-law slope and error;
(12) the luminosity at which the power-law breaks;
(13) the second power-law slope and error.}

\end{deluxetable}


\begin{thebibliography}{}

\bibitem[Annibali et al.(2007)]{Annibali2007} Annibali, F., Bressan, A., Rampazzo, R.,
Zeilinger, W.~W., \& Danese, L.\ 2007, \aap, 463, 455

\bibitem[Bauer et al.(2002)]{Bauer2002} Bauer, F.~E., Alexander, D.~M., Brandt, W.~N., et al.\ 2002, \aj, 124, 2351

\bibitem[Bell \& De Jong(2001)]{Bell2001} Bell, E., \& De Jong, R., 2001, ApJ, 550, 212

\bibitem[Bildsten \& Deloye(2004)]{Bildsten2004} Bildsten, L., \& Deloye, C. J. 2004, ApJ, 607, L119

\bibitem[Brassington et al.(2008)]{Brassington2008} Brassington, N.~J.,
Fabbiano, G., Kim, D.-W., et al.\ 2008, \apjs, 179, 142

\bibitem[Brassington et al.(2009)]{Brassington2009} Brassington, N.~J.,
Fabbiano, G., Kim, D.-W., et al.\ 2009, \apjs, 181, 605

\bibitem[Colbert et al.(2004)]{Colbert2004} Colbert, E.~J.~M., Heckman, T.~M., Ptak, A.~F.,
Strickland, D.~K., \& Weaver, K.~A.\ 2004, \apj, 602, 231

\bibitem[de Vaucouleurs et al.(1991)]{Vaucouleurs1991} de Vaucouleurs, G., de Vaucouleurs, A., Corwin, H., et al.
1991, {\it Third Reference Catalogue of Bright Galaxies} Vol. I, Vol. II, Vol. III (New York: Springer)

\bibitem[Evans et al.(2010)]{Evans2010} Evans, I.~N., Primini, F.~A.,
Glotfelty, K.~J., et al.\ 2010, \apjs, 189, 37-82

\bibitem[Fabbiano(1989)]{Fabbiano1989} Fabbiano, G.\ 1989, \araa, 27, 87

\bibitem[Fabbiano(2005)]{Fabbiano2005} Fabbiano, G.\ 2005, Science, 307, 533

\bibitem[Fabbiano(2006)]{Fabbiano2006} Fabbiano, G. 2006, ARA\&A, 44, 323

\bibitem[Fabbiano \& White(2006)]{FW2006} Fabbiano, G., \& White, N. E. 2006,
 in Compact Stellar X-ray Sources in Normal Galaxies, ed. W. H. G. Lewin, \& M. van der Klis
(Cambridge: Cambridge Univ. Press), 475

\bibitem[Fabbiano et al.(2001)]{Fabbiano2001} Fabbiano, G., Zezas, A., \& Murray, S.~S.\ 2001, \apj, 554, 1035

\bibitem[Farrell et al.(2009)]{Farrell2009} Farrell, S., Webb, N., Barret, D., et al. 2009, Nature, 460, 73

\bibitem[Fragos et al.(2008)]{Fragos2008} Fragos, T., Kalogera, V., Belczynski, K., et al.\ 2008, \apj, 683, 346-356

\bibitem[Freedman et al.(1994)]{Freedman1994} Freedman, W.~L., Hughes, S.~M., Madore, B.~F., et al.\ 1994, \apj, 427, 628 

\bibitem[Giacconi et al.(2001)]{Giacconi2001} Giacconi, R., Rosati, P., Tozzi, P., et al. 2001, ApJ, 551, 624

\bibitem[Gilfanov(2004)]{Gilfanov2004} Gilfanov, M.\ 2004, \mnras, 349, 146

\bibitem[Gilfanov et al.(2004a)]{Gilfanov2004a} Gilfanov, M., Grimm, H.-J., \& Sunyaev, R.\ 2004, \mnras, 347, L57

\bibitem[Gilfanov et al.(2004b)]{Gilfanov2004b} Gilfanov, M., Grimm, H.-J., \& Sunyaev, R.\ 2004,
Nuclear Physics B Proceedings Supplements, 132, 369

\bibitem[Gilfanov et al.(2004c)]{Gilfanov2004c} Gilfanov, M., Grimm, H.-J., \& Sunyaev, R.\ 2004, \mnras, 351, 1365

\bibitem[Graham(2001)]{Graham2001} Graham, A., 2001, AJ, 121, 820

\bibitem[Grimm et al.(2002)]{Grimm2002} Grimm, H.-J., Gilfanov, M., \& Sunyaev, R. 2002, A\&A, 391, 923

\bibitem[Grimm et al.(2003)]{Grimm2003} Grimm, H.-J., Gilfanov, M., \& Sunyaev, R. 2003, MNRAS, 339, 793

\bibitem[Hasinger et al.(1993)]{Hasinger1993}
Hasinger, G., Burg, R., Giacconi, R., et al.\ 1993, \aap, 275, 1


\bibitem[Heger \& Woosley(2002)]{Heger2002} Heger, A., \& Woosley, S.~E.\ 2002, \apj, 567, 532

\bibitem[Humphrey \& Buote(2008)]{Humphrey2008} Humphrey, P.~J., \& Buote, D.~A.\ 2008, \apj, 689, 983-1004

\bibitem[Irwin(2005)]{Irwin2005} Irwin, J.~A.\ 2005, \apj, 631, 511

\bibitem[Irwin et al.(2004)]{Irwin2004} Irwin, J., Bregman, J., \& Athey, A. 2004, ApJ, 601, 143

\bibitem[Ivanova \& Kalogera(2006)]{Ivanova2006} Ivanova, N., \&  Kalogera, V. 2006, ApJ, 636, 985

\bibitem[Jarrett et al.(2003)]{Jarrett2003} Jarrett, T.~H., Chester, T., Cutri, R., Schneider, S.~E., \& Huchra, J.~P.\
2003, \aj, 125, 525

\bibitem[Jeltema et al.(2003)]{Jeltema2003} Jeltema, T.~E., Canizares, C.~R., Buote, D.~A., \& Garmire, G.~P.\
2003, \apj, 585, 756

\bibitem[Kilgard et al.(2002)]{Kilgard2002} Kilgard, R., Kaaret, P., Krauss, M., et al. 2002, ApJ, 573, 138

\bibitem[Kim et al.(2004)]{Kim2004} Kim, D.-W., Cameron, R., Drake, J., et al. 2004, ApJS, 150, 19

\bibitem[Kim \& Fabbiano(2003)]{KF2003} Kim, D.-W., \& Fabbiano, G. 2003, ApJ, 586, 826

\bibitem[Kim \& Fabbiano(2004)]{KF2004} Kim, D.-W., \& Fabbiano, G. 2004, ApJ, 611, 846

\bibitem[Kim \& Fabbiano(2010)]{Kim2010} Kim, D.-W., \& Fabbiano, G.\ 2010, \apj, 721, 1523

\bibitem[Kim et al.(2009)]{Kim2009} Kim, D.-W., Fabbiano, G., Brassington, N.~J., et al.\ 2009, \apj, 703, 829

\bibitem[Kim et al.(2006)]{Kim2006} Kim, D.-W., Fabbiano, G., Kalogera, V., et al. 2006, ApJ, 652, 1090

\bibitem[King(2002)]{King2002} King, A. R.\ 2002, \mnras, 335, L13

\bibitem[King(2004)]{King2004} King, A.~R.\ 2004, \mnras, 347, L18

\bibitem[King(2009)]{King2009} King, A. R. 2009, MNRAS, 393, L41

\bibitem[King et al.(2001)]{King2001} King, A.~R., Davies, M.~B., Ward,
M.~J., Fabbiano, G., \& Elvis, M.\ 2001, \apjl, 552, L109

\bibitem[Kong et al.(2003)]{Kong2003} Kong, A., Di Stefano, R., Garcia, M., \& Greiner, J., 2003, ApJ, 585, 298

\bibitem[Kundu et al.(2007)]{Kundu2007} Kundu, A., Maccarone, T.~J., \& Zepf, S.~E.\ 2007, \apj, 662, 525

\bibitem[Lehmer et al.(2010)]{Lehmer2010} Lehmer, B.~D., Alexander, D.~M., Bauer, F.~E., et al.\ 2010, \apj, 724, 559

\bibitem[Lehmer et al.(2014)]{Lehmer2004} Lehmer, B.~D., Berkeley, M., Zezas, A., et al.\ 2014, \apj, 789, 52

\bibitem[Lehmer et al.(2008)]{Lehmer2008} Lehmer, B.~D., Brandt, W.~N., Alexander, D.~M., et al.\
2008, \apj, 681, 1163-1182


\bibitem[Liu(2011)]{Liu2011} Liu, J. 2011, ApJS, 192, 10

\bibitem[Liu et al.(2006)]{Liu2006} Liu, J., Bregman, J., \& Irwin, J. 2006, ApJ, 642, 171


\bibitem[Liu \& Mirabel(2005)]{Liu2005} Liu, Q.~Z., \& Mirabel, I.~F.\ 2005, \aap, 429, 1125

\bibitem[Long \& van Speybroeck(1983)]{Long1983} Long, K.~S., \& van Speybroeck, L.~P.\ 1983,
Accretion-Driven Stellar X-ray Sources, ed. W. H. G. Lewin \& E. P. J. van den Heuvel
(Cambridge: Cambridge Univ. Press), 117

\bibitem[Makishima et al.(2000)]{Makishima2000} Makishima, K., Kubota, A.,
Mizuno, T., et al.\ 2000, \apj, 535, 632

\bibitem[McDonald et al.(2011)]{McDonald2011} McDonald, M., Courteau, S., Tully, R.~B., \& Roediger, J.\
2011, \mnras, 414, 2055

\bibitem[Miller \& Colbert(2004)]{Miller2004} Miller, M. C., \& Colbert, E. 2004, IJMPD, 13, 1

\bibitem[Mineo et al.(2014a)]{Mineo2014a} Mineo, S., Fabbiano, G., D'Abrusco, R., et al.\ 2014, \apj, 780, 132

\bibitem[Mineo et al.(2014b)]{Mineo2014b} Mineo, S., Gilfanov, M., Lehmer, B.~D.,
Morrison, G.~E., \& Sunyaev, R.\ 2014, \mnras, 437, 1698

\bibitem[Mineo et al.(2012)]{Mineo2012} Mineo, S., Gilfanov, M., \& Sunyaev, R.\ 2012, \mnras, 419, 2095

\bibitem[Mineo et al.(2014c)]{Mineo2014c} Mineo, S., Rappaport, S., Levine, A., et al.\ 2014, \apj, 797, 91

\bibitem[Mineo et al.(2013)]{Mineo2013} Mineo, S., Rappaport, S., Steinhorn, B., et al.\ 2013, \apj, 771, 133

\bibitem[Mushotzky et al.(2000)]{Mushotzky2000} Mushotzky, R., Cowie, L., Barger, A., \& Arnaud, K.
2000, Nature, 404, 459

\bibitem[Paolillo et al.(2011)]{Paolillo2011} Paolillo, M., Puzia, T.~H., Goudfrooij, P., et al.\ 2011, \apj, 736, 90

\bibitem[Peacock \& Zepf(2016)]{Peacock2016} Peacock, M.~B., \& Zepf, S.~E.\ 2016, \apj, 818, 33

\bibitem[Persic \& Rephaeli(2002)]{Persic2002} Persic, M., \& Rephaeli, Y.\ 2002, \aap, 382, 843

\bibitem[Persic \& Rephaeli(2007)]{Persic2007} Persic, M., \& Rephaeli, Y.\ 2007, \aap, 463, 481

\bibitem[Persic et al.(2004)]{Persic2004} Persic, M., Rephaeli, Y., Braito, V., et al.\ 2004, \aap, 419, 849

\bibitem[Podsiadlowski et al.(2002)]{Podsiadlowski2002} Podsiadlowski, P., Rappaport, S., \& Pfahl, E.~D.\
2002, \apj, 565, 1107

\bibitem[Postnov \& Kuranov(2005)]{Postnov2005} Postnov, K.~A., \& Kuranov, A.~G.\ 2005, Astronomy Letters, 31, 7

\bibitem[Primini et al.(2011)]{Primini2011} Primini, F.~A., Houck, J.~C.,
Davis, J.~E., et al.\ 2011, \apjs, 194, 37

\bibitem[Ranalli et al.(2003)]{Ranalli2003} Ranalli, P., Comastri, A., \& Setti, G.\ 2003, \aap, 399, 39

\bibitem[Rice et al.(1988)]{Rice1988} Rice, W., Lonsdale, C, Soifer, B., et al. 1988, ApJS, 68, 91

\bibitem[Roberts \& Warwick(2000)]{Roberts2000} Roberts, T.~P., \& Warwick, R.~S.\ 2000, \mnras, 315, 98

\bibitem[Rosa-Gonz{\'a}lez et al.(2002)]{Rosa-Gonzalez2002} Rosa-Gonz{\'a}lez, D., Terlevich, E., \& Terlevich, R.,
2002, MNRAS, 332, 283

\bibitem[S{\'a}nchez-Bl{\'a}zquez et al.(2006)]{Sanchez2006} S{\'a}nchez-Bl{\'a}zquez, P., Gorgas, J.,
Cardiel, N., \& Gonz{\'a}lez, J.~J.\ 2006, \aap, 457, 809

\bibitem[Sarazin et al.(2000)]{Sarazin2000} Sarazin, C. L., Irwin, J. A., \& Bregman, J. N. 2000, ApJ, 544, L101

\bibitem[Sarazin et al.(2001)]{Sarazin2001} Sarazin, C.~L., Irwin, J.~A., \& Bregman, J.~N.\ 2001, \apj, 556, 533

\bibitem[Sarazin et al.(2003)]{Sarazin2003} Sarazin, C. L., Kundu, A., Irwin, J. A., et al. 2003, ApJ, 595, 743


\bibitem[Sell et al.(2011)]{Sell2011} Sell, P., Pooley, D., Zezas, A., et al. 2011, ApJ, 735, 26

\bibitem[Silverman \& Filippenko(2008)]{Silverman2008} Silverman, J., \& Filippenko, A. 2008, ApJ, 678, 17

\bibitem[Swartz et al.(2008)]{Swartz2008} Swartz, D., Soria, R., \& Tennant, A. 2008, ApJ, 684, 282

\bibitem[Swartz et al.(2011)]{Swartz2011} Swartz, D.~A., Soria, R., Tennant, A.~F., \& Yukita, M.\ 2011, \apj, 741, 49

\bibitem[Swartz et al.(2009)]{Swartz2009} Swartz, D.~A., Tennant, A.~F., \& Soria, R.\ 2009, \apj, 703, 159

\bibitem[Terlevich \& Forbes(2002)]{Terlevich2002} Terlevich, A.~I., \& Forbes, D.~A.\ 2002, \mnras, 330, 547

\bibitem[Thomas et al.(2005)]{Thomas2005} Thomas, D., Maraston, C., Bender, R., \& Mendes de Oliveira, C.\
2005, \apj, 621, 673

\bibitem[Trager et al.(2000)]{Trager2000} Trager, S.~C., Faber, S.~M., Worthey, G., \& Gonz{\'a}lez, J.~J.\
2000, \aj, 120, 165

\bibitem[Tully(2015)]{Tully2015} Tully, R.~B.\ 2015, \aj, 149, 171




\bibitem[Voss et al.(2009)]{Voss2009} Voss, R., Gilfanov, M., Sivakoff, G.~R., et al.\ 2009, \apj, 701, 471


\bibitem[Zezas \& Fabbiano(2002)]{Zezas2002} Zezas, A., \& Fabbiano, G.\ 2002, \apj, 577, 726

\bibitem[Zezas et al.(1999)]{Zezas1999} Zezas, A.~L., Georgantopoulos, I., \& Ward, M.~J.\ 1999, \mnras, 308, 302


\end{thebibliography}
\end{document}